\documentclass[aps,superscriptaddress,onecolumn,10pt,prx]{revtex4-2}

\usepackage{amsmath,mathtools,amsthm,amssymb}
\usepackage{tabularx}
\usepackage{tabularray}

\usepackage{color}
\usepackage[dvipsnames]{xcolor}
\usepackage{bbold}
\usepackage{enumitem}
\usepackage{physics}
\usepackage{comment}
\usepackage{amsmath}
\usepackage{soul}
\usepackage{array}
\usepackage{graphics}
\usepackage{wrapfig}
\graphicspath{{figures/}}
\usepackage{biolinum}
\usepackage{bbm}
\usepackage{dsfont}
\usepackage{mathrsfs}
\usepackage{mathdots}
\usepackage{enumitem}
\usepackage{pifont}
\definecolor{myrefcolor}{rgb}{0.067,0.5,0.5}
\definecolor{myurlcolor}{rgb}{0.1,0,0.9}

\usepackage[a4paper, total={6.3in, 9.4in}]{geometry}

\usepackage[
    breaklinks,
    pdftex,
    colorlinks=true,
    linkcolor=myrefcolor,
    citecolor=myrefcolor,
    urlcolor=myrefcolor
]{hyperref}

\usepackage[capitalize]{cleveref}

\newcommand{\pur}[0]{\mathrm{Pur}}

\newtheorem*{theorem*}{Theorem}
\newtheorem*{corollary*}{Corollary}
\newtheorem*{definition*}{Definition}
\newcounter{thm}
\newtheorem{theorem}[thm]{Theorem}
\newtheorem{lemma}[thm]{Lemma}

\newtheorem{proposition}{Proposition}
\newtheorem{definition}{Definition}

\newtheorem{corollary}{Corollary}

\theoremstyle{remark}
\newtheorem{remark}{Remark}

\DeclareMathOperator{\stab}{STAB}
\usepackage{algorithm}
\usepackage{algpseudocode}

\DeclareMathOperator{\de}{d\!}
\DeclareMathOperator{\sym}{sym}
\DeclareMathOperator{\cl}{Cl}

\DeclareMathOperator{\diag}{diag}

 \newcommand{\even}{\mathrm{Even}(\mathbb{F}_{2}^{k\times m})}
\newcommand{\symf}{\mathrm{Sym}(\mathbb{F}_2^{m\times m})}

\newcommand{\haar}[0]{\operatorname{Haar}}

\definecolor{airforceblue}{rgb}{0.36, 0.54, 0.66}
\newcommand{\be}{\begin{equation}\begin{aligned}\hspace{0pt}}
\newcommand{\bbb}{\begin{equation*}\begin{aligned}}
\newcommand{\ee}{\end{aligned}\end{equation}}
\newcommand{\eee}{\end{aligned}\end{equation*}}

\allowdisplaybreaks
\makeatletter
\renewcommand{\subsubsection}{\@startsection{subsubsection}{3}{0pt}%
  {1.5ex plus 1ex minus .2ex}%
  {1ex plus .2ex}%
  {\bfseries}}
\makeatother
\newcommand{\nocontentsline}[3]{}
\let\origcontentsline\addcontentsline
\newcommand\stoptoc{\let\addcontentsline\nocontentsline}
\newcommand\resumetoc{\let\addcontentsline\origcontentsline}

\begin{document}

\title{Non-Clifford Cost of Random Unitaries}


\author{Lorenzo Leone}\email{loleone@unisa.it}
\affiliation{Dipartimento di Ingegneria Industriale, Università degli Studi di Salerno, Via Giovanni Paolo II, 132, 84084 Fisciano (SA), Italy}

\affiliation{Dahlem Center for Complex Quantum Systems, Freie Universit\"at Berlin, 14195 Berlin, Germany}

\author{Salvatore F.~E. Oliviero}\email{s.oliviero@fu-berlin.de}
\affiliation{NEST, Scuola Normale Superiore and Istituto Nanoscienze, Consiglio Nazionale delle Ricerche, Piazza dei Cavalieri 7, IT-56126 Pisa, Italy}
\affiliation{Dahlem Center for Complex Quantum Systems, Freie Universit\"at Berlin, 14195 Berlin, Germany}

\author{Alioscia Hamma}\email{alioscia.hamma@unina.it}

\affiliation{Dipartimento di Fisica `Ettore Pancini', Universit\`a degli Studi di Napoli Federico II, Via Cintia 80126,  Napoli, Italy}
\affiliation{INFN, Sezione di Napoli, Italy}

\author{Jens Eisert}\email{jense@fu-berlin.de}
\affiliation{Dahlem Center for Complex Quantum Systems, Freie Universit\"at Berlin, 14195 Berlin, Germany}

\author{Lennart Bittel}\email{l.bittel@fu-berlin.de}
\affiliation{Dahlem Center for Complex Quantum Systems, Freie Universit\"at Berlin, 14195 Berlin, Germany}

\begin{abstract}
Recent years have enjoyed a strong interest in exploring properties and applications of random quantum circuits. In this work, we explore the ensemble of \( t \)-doped Clifford circuits on \( n \) qubits, consisting of Clifford circuits interspersed with \( t \) single-qubit non-Clifford gates. We establish rigorous convergence bounds towards unitary \( k \)-designs, revealing the intrinsic cost in terms of non-Clifford resources in various flavors. First, we analyze the \( k \)-th order frame potential, which quantifies how well the ensemble of doped Clifford circuits is spread within the unitary group. We prove that a quadratic doping level, \(t=\tilde{\Theta}(k^2)\), is both necessary and sufficient to approximate the frame potential of the full unitary group. As a consequence, we refine existing upper bounds on the convergence of the ensemble towards state \( k \)-designs. Second, we derive tight bounds on the convergence of \( t \)-doped Clifford circuits towards relative-error \( k \)-designs, showing that \( t = \tilde{\Theta}(nk) \) is both necessary and sufficient for the ensemble to form a relative \( \varepsilon \)-approximate \( k \)-design. Similarly, $t=\tilde{\Theta}(n)$ is required to generate pseudo-random unitaries. All these results highlight that generating random unitaries is extremely costly in terms of non-Clifford resources, and that such ensembles fundamentally lie beyond the classical simulability barrier. Additionally, we analyze high-order doped-Clifford Weingarten functions to derive analytic expressions for the twirling operator over the ensemble of random doped Clifford circuits, and we establish their asymptotic behavior in relevant regimes.
\end{abstract}

\maketitle

\tableofcontents

\section{Introduction}

Randomness is a central and powerful tool in quantum physics, with a wealth of applications spanning quantum cryptography~\cite{ambainis_smaapp2004,hayden_ransta2004,kretschmer_quapru2021,morimae_quacom2022,ananth_cryfro2022}, quantum algorithms~\cite{sen_ranmea2006, brandao_expspe2013, boixo_characterizing_2018,arute_quasup2019,bouland_complexity_2019}, quantum sensing~\cite{kueng_lowran2017,kimmel_pharet2017,kueng_distinguishing_2016,oszmaniec_ranbos2016}, benchmarking and certification
\cite{Eisert_2020,dankert_exact_2009,Magesan_2011,magniAnticoncentrationCliffordCircuits2025} and quantum communication~\cite{devetak_thepri2005,devetak_relqua2004,groisman_quacla2005,abeyesinghe_themot2009,dupuis_oneshot2014,szehr_decuni2013,horodecki_parqua2005,horodecki_quasta2007,nakata_onesho2021,wakakuwa_onesho2023}. The use of random unitary operators is also prevalent in providing a coarse-grained description of the evolution of complex quantum systems, yielding insights into the spectra of complex Hamiltonians and fundamental processes and features such as quantum thermalization~\cite{popescu_entanglement_2006,linden_quantum_2009,delrio_relthe2016,kaneko_chaofm2020,christian_review,ippoliti_solmod2022}, the holographic principle~\cite{hayden_black_2007,nakata_blaasc2023,RandomHamiltonians,
nakata2025computationalcomplexityunitarystate,oliviero_unscrambling_2024,leone_learning_2024}, notions of quantum scrambling~\cite{sekino_fast_2008,lashkari_fast_2013,maldacena_bound_2016,roberts_chaos_2017,cotler_emequa2023,mark_maxent2024,pilatowsky_hilerg2024}, and quantum thermodynamics~\cite{Munson2025Mar}. Given the ubiquity of quantum randomness in both applications and rather fundamental topics, understanding its precise origin constitutes 
a fundamental question.


In its simplest form, quantum randomness can be achieved by drawing unitary matrices uniformly at random according to the Haar measure over the unitary group 
\( \mathcal{U}_n 
= U(2^n)\) on $n$ qubits. 
However, for most applications in many-body physics (for large \( n \)), constructing Haar random unitaries is 
practically infeasible, for the circuit complexity becoming prohibitively
large. A simple counting argument reveals that an exponential number (in \( n \)) of elementary gates is needed to approximate a Haar random unitary with constant precision. However, for many applications, sampling Haar random unitaries is too much to ask for anyway. For most practical purposes,
one can be content with constructing an ensemble of unitaries \( \mathcal{E} \)  reproducing up to the first $k$ moments of the distribution, leading to the concept of \emph{unitary \( k \)-designs}~\cite{emerson_pseudorandom_2003,Gross_2007,haferkamp2020QuantumHomeopathyWorks,brandao_local_2016,Harrow2009Random2-designs,roberts_chaos_2017,harrow_approximate_2018,laracuente2024approximateunitarykdesignsshallow}.
An even weaker---and conceptually somewhat differently motivated---request is that of computational pseudo-randomness, that is, indistinguishably from true randomness given computational constraints. 
Both concepts have proven to be highly fruitful in a wealth of applications,
both practically and conceptually \cite{Ji_2018, metger2024pseudorandomunitariesnonadaptivesecurity, schuster2025randomunitariesextremelylow, ma2024constructrandomunitaries, haug2024pseudorandomunitariesrealsparse}. 
Recently, there has been a breakthrough in Refs.~\cite{schuster2025randomunitariesextremelylow,laracuente2024approximateunitarykdesignsshallow} with the introduction of sets generating unitary \( k \)-designs and pseudo-random unitaries with the 
minimal possible depths, i.e., scaling logarithmically with the number of qubits \( n \). This represents a significant reduction compared to the exponential scaling of gate complexity for Haar random unitaries.

Despite the existence of extremely shallow circuits capable of generating pseudo-random unitaries, in many quantum hardware architectures 
involving early quantum computers, resource costs stem not only from the \textit{number} of gates but also from the \textit{types} of gates used. One of the main bottlenecks in quantum error correction is a consequence of the \emph{Eastin-Knill theorem}~\cite{Eastin_2009}, which asserts that no universal gate set can be implemented transversely---that is, without incurring spatial overhead. As a result, the best that a fault-tolerant, gate-based quantum computation scheme can achieve is to use only a single non-transversal gate which, when combined with a transversal gate set, enables universal quantum computation. A common approach is to use transversal Clifford operations, which are available in many stabilizer \emph{quantum error-correcting codes}~\cite{gottesman_stabilizer_1997}, and to invest resources in implementing a single non-Clifford gate---making use of what is called \emph{magic}---that is sufficient to promote the Clifford group to a universal gate set.


In this context, the challenge becomes generating unitary designs using the fewest possible non-Clifford gates, since Clifford operations, being transversal, can be effectively considered \textit{free}. Notably, the \emph{Clifford group} alone is sufficient to generate exact unitary $3$-designs~\cite{webb_clifford_2016,zhu_multiqubit_2017}, a highly useful property for learning algorithms~\cite{scott_optimizing_2008} and benchmarking~\cite{roth_recovering_2018, wallman_randomized_2014, wallman_randomized_2018}. 
However, it falls short in producing higher-order designs~\cite{zhu2016cliffordgroupfailsgracefully}. Consequently, non-Clifford operations become the critical resource required for achieving higher-order unitary designs, while their usage must be minimized to reduce the associated overhead.

\begin{figure}
    \centering
    \includegraphics[width=1\linewidth]{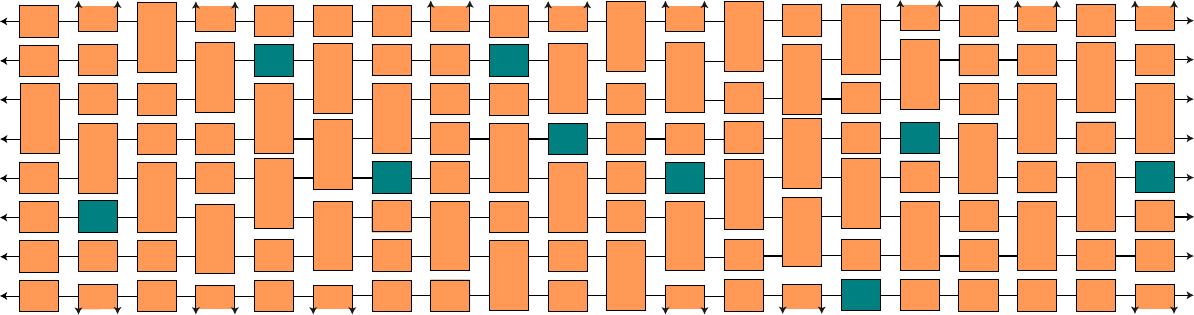}
    \caption{Pictorial representation of $t$-doped Clifford circuits. The orange gates are Clifford gates, while the teal gates are single qubit non-Clifford gates. Hence, non-Clifford gates are interleaved within a circuit that predominantly consists of feasible Clifford operations.}
    \label{fig:enter-label}
\end{figure}

In this work, we explore the generation of unitary designs using \emph{\( t \)-doped Clifford circuits}---deep Clifford circuits, which we regard as free resources, augmented with \( t \) single-qubit resource-intensive non-Clifford gates (as illustrated in \cref{fig:enter-label})---with a primary focus on determining the minimal use of non-Clifford resources required to generate unitary $k$-designs. 
The main question we seek to answer in this 
work is the following.
\begin{quote}
    \textit{
   How well `spread' is the set of doped Clifford circuits and what relative error design does it achieve as a function of the doping $t$?
    }
\end{quote}
By developing a general framework for computing averages over doped Clifford circuits---an approach of independent interest---this work provides detailed answers to both of the aforementioned questions. 

We remark that addressing this question, concerning the quantum randomness attained by doped Clifford circuits, also has implications for scrambling and for quantum advantage experiments~\cite{roberts_chaos_2017,deshpande2025peakedquantumadvantageusing}, where Clifford+T architectures play a crucial role.

First, we establish tight bounds on the \( k \)-th frame potential ---a scalar quantity that measures how uniformly an ensemble of unitaries is spread over the unitary group--- as a function of the doping parameter \( t \). Specifically, we show that a quadratic doping level, \( t = \Theta(k^2 + \log \varepsilon^{-1}) \), is both necessary and sufficient to approximate the frame potential within an additive error \( \varepsilon \) relative to the frame potential of Haar-random unitaries. While one might have expected a linear scaling of the doping with respect to the degree \( k \) of the frame potential, this result reveals that generating quantum randomness using doped Clifford circuits incurs an intrinsic cost in terms of non-Clifford resources. This convergence bound has a direct application in quantifying the scrambling capabilities of doped Clifford circuits; notably, only a polylogarithmic amount of doping is both necessary and sufficient to achieve maximal scrambling.

Second, we establish tight bounds on the convergence of the \( t \)-doped Clifford ensemble towards unitary \( k \)-designs. We show that a doping level scaling linearly in both the number of qubits \( n \) and the degree \( k \), i.e., \( t = \tilde{\Theta}(kn) \) (up to logarithmic factors), is \textit{necessary and sufficient} for the ensemble to form a \( k \)-design up to a relative approximation error \( \varepsilon \), even for values of $\varepsilon$ exponentially large in $nk$. While the main result of Ref.~\cite{haferkamp2020QuantumHomeopathyWorks} shows that a doping level independent of the number of qubits $n$ is sufficient to produce a $k$-design with additive approximation error, our result demonstrates that generating a $k$-design with relative approximation error is necessarily costly in terms of non-Clifford resources. Similarly, we show that $t=\tilde{\Theta}(n)$ are necessary and sufficient to construct pseudo-random unitaries. The stark contrast between these two results lies in the notion of approximation: additive-error designs guarantee indistinguishability from Haar-random unitaries (up to precision \( \varepsilon \)) only under a restricted class of protocols, whereas relative-error designs, as well as pseudo-random unitaries, ensure indistinguishability under \emph{any} quantum protocols.
In the following sections, we carefully explain our results and place them in context with previously known bounds.

\begin{table}[h!]
  \centering
  \begin{tabular}{@{}|l|c|c|@{}}
    \hline
    & \textbf{ non-Clifford cost} & \textbf{Operational meaning} \\
    \hline
    Additive unitary $k$-design~\cite{haferkamp2020QuantumHomeopathyWorks,zhang2025designsmagicaugmentedcliffordcircuits} &     \textcolor{black}{$O(k^4+k\log\varepsilon^{-1})$}    & {Secure against non-adaptive $k$-query experiments} \\
   Relative unitary $k$-designs~\cite{haferkamp2020QuantumHomeopathyWorks}[$\star$]&
   \textcolor{black}{$\Theta(nk)$} & {Secure against any $k$-query quantum experiment} \\
  Pseudorandom unitaries~\cite{schuster2025randomunitariesextremelylow}[$\star$] &  \textcolor{black}{$\tilde{\Theta}(n)$} &
    {Secure against any poly-time quantum experiment} \\
    Unitary frame potential~[$\star$] & \textcolor{black}{${\tilde{\Theta}}(k^2+\log\varepsilon^{-1})$} & {Spreading on the unitary group and OTOCs} \\
    State $k$-designs~[$\star$] & \textcolor{black}{$O(k^2+\log \varepsilon^{-1})$} & Secure against any $k$-query quantum experiment \\
    \hline
  \end{tabular}
    \caption{Summary of results on unitary design constructions and their associated non-Clifford costs. Results (or portions thereof) proven in this work are marked with [$\star$].}\label{table1}
\end{table}

\subsection{Overview of the results}\label{sec:overview}
Let us provide an informal overview of our results, also summarized in \cref{table1}. Our model of $t$-doped Clifford circuits consists of randomly drawn Clifford circuits interspersed with single-qubit random gates. As we discuss later, the specific choice of `doping gates' is not crucial to our analysis. In this paper, we mainly examine two commonly used choices: random diagonal single-qubit gates and Haar-random single-qubit gates. Both yield qualitatively similar results. However, the techniques we develop are independent of the particular choice of doping.

\medskip

\subsubsection{The spreading of doped Clifford circuits via frame potential}
One of the question we address in this paper is how spread is the ensemble of doped Clifford circuits with respect to the amount of doping, quantified by the number $t$ of non-Clifford gates inserted. The frame potential measures whether \( \mathcal{E} \) is ‘spread’ or ‘concentrated’ within the full unitary group \( \mathcal{U}_n \). Specifically, the frame potential of order \( k \) measures how well the finite collection of unitaries
\( \mathcal{E}^{\otimes k} \coloneqq \{ U^{\otimes k} : U \in \mathcal{E} \} \) is uniformly distributed on \( \mathcal{U}_n^{\otimes k} \), and it is defined as
\be\label{introframepotential}
\mathcal{F}_{\mathcal{E}}^{(k)} \coloneqq \mathbb{E}_{U,V \in \mathcal{E}} |\mathrm{tr}(UV^{\dagger})|^{2k}.
\ee

As the name suggests, the frame potential can also be thought of as a `repulsive force': we can think of the unitaries in \( \mathcal{E} \) as particles on the unitary group
subject to a repulsive force proportional to \( |\mathrm{tr}(UV^{\dagger})|^k \), and \cref{introframepotential} gives the potential of the configuration~\cite{Gross_2007}. 
In the extreme case where the ensemble consists of only one element with probability one, the forward evolution perfectly cancels the backward evolution, leading to the maximal possible value of the frame potential. In contrast,
if the ensemble \( \mathcal{E} \) is sufficiently spread, then two independently drawn samples \( U \) and \( V \) are likely to be very different, meaning that \( |\mathrm{tr}(UV^{\dagger})| \) will be very small. As expected, the most spreading ensemble is the ensemble equipped with the uniform measure over the full unitary group, the Haar measure: in fact, this is the equilibrium configuration exactly minimizing the frame potential~\cite{Gross_2007}.
It holds that for any ensemble of unitaries, we have \( \mathcal{F}_{\mathcal{E}}^{(k)} \ge \mathcal{F}_{\haar}^{(k)} \).
We quantify the spreading of an ensemble $\mathcal{E}$ with the difference 
$\mathcal{F}_{\mathcal{E}}^{(k)}-\mathcal{F}_{\haar}^{(k)}$, which is positive, and zero if and only if $\mathcal{E}$ is an (exact) $k$-design.

Let us first consider the ensemble of unitaries constituted by Clifford operators, belonging to the Clifford group. Such ensemble is discrete and it features a minimum distance between its elements. As such, it  is not sufficiently spread across the unitary group, 
\begin{equation}
\mathcal{F}_{\cl}^{(k)}-\mathcal{F}_{\haar}\ge \exp(\Omega(k^2)), 
\end{equation}
i.e., the Clifford group gets further away, increasing $k$, in a quadratic exponential fashion. The ultimate reason for this behavior is that the frame potential of subgroups of the unitary groups is given by the dimension of the commutant, which, for the Clifford group, increases quadratic-exponentially in $k$~\cite{bittel2025completetheorycliffordcommutant}. The relevant question then becomes how much doping is necessary for a sufficiently uniform spreading of the doped Clifford ensemble across the unitary group.

The first central result of this work resolves around a tight bound on the frame potential of the ensemble of doped Clifford circuits.

\begin{theorem*}[Frame potential convergence. Informal of \cref{th:framepotentialconvergence}]
    Let $n\ge \Omega(k^2+t\log k)$. There exists two constants $A,B=\Theta(1)$ such that the difference between the frame of $t$-doped Clifford circuits $\mathcal{F}_{t}^{(k)}$ and the Haar frame potential $\mathcal{F}_{\haar}^{(k)}$ is upper and lower bounded as
\be\label{introeq1}
2^{\Omega(k^2- At\log k)}\le \mathcal{F}_{t}^{(k)}-\mathcal{F}_{\haar}^{(k)}\le 2^{O(k^2-Bt)}.
\ee
\end{theorem*} 
There are two worth-noticing aspects around the result summarized in \cref{introeq1}. First, the upper bound is completely independent from the number of qubits $n$ of the quantum circuit. However, the result of generating system-size independent additive-error designs with $t$-doped Clifford circuits was established in the seminal paper~\cite{haferkamp2020QuantumHomeopathyWorks}, and it is well-understood in the literature. The second, and novel, aspect of \cref{introeq1} is that the bounds showcase a quadratic dependence of $t$ upon the design order $k$. As the order of the frame potential grows, the convergence requires a quadratic amount of doping. Indeed, to have $\mathcal{F}_{t}^{(k)}-\mathcal{F}_{\haar}^{(k)}=\varepsilon$ then a necessary and sufficient amount of doping (up to log factors) is $t=\Theta(k^2+\log\varepsilon^{-1})$.  The result can be understood by the following hand-wavy argument: the ultimate responsible for this convergence bound is the quadratic-exponential growth of the commutant of the Clifford group with repsect to $k$ agaist the factorial growing of the unitary group commutant~\cite{bittel2025completetheorycliffordcommutant}. Indeed, the doping can effectively `kill' an exponential fraction (in $t$) of operators within the commutant of the Clifford group, resulting in $\exp(k^2)\exp(-t)$ scaling. 
The above result already highlights an inherent substantial cost---in terms of non-Clifford resources---associated with generating higher-order designs through doping, a point that will be further explored in the following sections.
\medskip

\subsubsection{Implications for approximate state $k$-designs} Our result in \cref{introeq1} has implications in the context of unitary $k$-designs, particularly to \textit{state $k$-designs}. Given an ensemble of unitaries, it is natural to consider the ensemble of states $\mathcal{S} = \{U\ket{0} \mid U \in \mathcal{E}\}$ induced by $\mathcal{E}$ when acting on the trivial state vector $\ket{0}$. The ensemble of unitaries $\mathcal{E}$ generates an $\varepsilon$-approximate state $k$-design if and only if $\|\Psi_{\mathcal{E}}^{(k)}- \Psi_{\haar}^{(k)}\|_1\le\varepsilon$, where we define 
\begin{equation}
\Psi_{\mathcal{E}}^{(k)} \coloneqq \mathbb{E}_{U\sim\haar}(U\ketbra{0}U^{\dag})^{\otimes k} 
\end{equation}
as the average state of the given ensemble. 
A key application of our main result in \cref{introeq1} is in establishing convergence bounds on the ability of doped Clifford circuits to generate approximate state $k$-designs. 
\begin{theorem*}[Doped Clifford circuits producing state $k$-designs. See  \cref{lem:relativeframepotentialandframepotentialtdoped,cor:statekdesignconvergence}]
Let $n\ge\Omega(k^2+t\log k)$. Let $\Psi_{t}^{(k)}$ the average $k$-copy state produced by the ensemble of $t$-doped Clifford circuits, and let $\Psi_{\haar}^{(k)}$ the Haar $k$-copy state. Then
\be
 \|\Psi_{\haar}^{(k)}-\Psi_{t}^{(k)}\|_1 \le  \sqrt{\frac{3}{2}\frac{\mathcal{F}_{t}^{(k)}-\mathcal{F}_{\haar}^{(k)}}{\mathcal{F}_{\haar}}},\label{eqintro2}
\ee
which, combined with \cref{introeq1}, implies that a quadratic doping level $t = \Omega(k^2 + \log \varepsilon^{-1})$ suffices for doped Clifford circuits to produce an $\varepsilon$-approximate state $k$-design. 
\end{theorem*}

This result constitutes an improvement over previously established bounds. In particular, the best known bound prior to this work follows from the stronger results of Ref.~\cite{haferkamp_random_2022}, which show that a doping level of \( t = \Omega(\log^2 k (k^4 + k \log \varepsilon^{-1})) \) is sufficient to achieve a unitary \( k \)-design up to an additive error \( \varepsilon \). 

Moreover, \cref{eqintro2} provides an operational interpretation of the frame potential. While frame potentials have become a central tool for analyzing quantum randomness and have inspired connections across a broad range of fields---such as circuit complexity lower bounds~\cite{jian2022lineargrowthcircuitcomplexity,Haferkamp2022May} and information scrambling~\cite{roberts_chaos_2017}---\cref{eqintro2} directly links the frame potential to the rigorous notion of approximate state \( k \)-designs.


\medskip

\subsubsection{Applications to quantum scrambling}
Scrambling is the phenomenon by which local information—i.e., information encoded in a local subsystem—is dispersed, as a result of quantum evolution, into nonlocal degrees of freedom. Consequently, information is said to be scrambled if no local quantum measurement can recover the initial information. Quantum scrambling is thus a hallmark of chaotic and complex quantum dynamics, making it one of the most intriguing features of quantum evolutions.

Since scrambling is a property of the quantum evolution itself, diagnosing scrambling behavior comes with an intrinsic challenge: to decide whether a unitary $U$ scrambles information, one would in principle need to verify that its action on all possible localized quantum states renders the information unrecoverable. A more practical bottom-up approach is instead to check whether quantum information can be locally recovered via local quantum measurements, which, without loss of generality, can be taken to be Pauli measurements (as these form a complete operator basis). At the operator level, diagnosing quantum scrambling can then be achieved through exotic correlation functions between local operators. In particular, given a unitary $U$, one can define $k$-order Out-of-Time-Ordered Correlation (OTOC) functions~\cite{kitaev_hidden_2014}. Consider two $k$-tuples of Pauli operators $A_1,\ldots, A_k$ and $B_1,\ldots, B_k$, and define $\tilde{B}_i \coloneqq U^{\dagger} B_i U$. The $k$-OTOC is then defined as
\begin{align}
    \operatorname{OTOC}_k(U,\{A\},\{B\}) \coloneqq \frac{1}{d}\tr(A_{1}\tilde{B}_1\cdots A_k\tilde{B}_k)\,.
\end{align}
The careful reader will note that the behavior of $\operatorname{OTOC}_k(U,\{A\},\{B\})$ can strongly depend on the choice of the tuples of observables $\{A\}, \{B\}$. To eliminate this spurious dependence, a common approach is to average over all Pauli observables: $\operatorname{OTOC}_k(U) \coloneqq \mathbb{E}_{\{A\},\{B\}}\operatorname{OTOC}_k(U,\{A\},\{B\})\,$. 

A fundamental benchmark of scrambling behavior is given by Haar-random unitary dynamics, which maximally disperse local information into nonlocal degrees of freedom. A natural figure of merit is therefore $\delta\operatorname{OTOC}_k(U)$, defined as the relative difference between $\operatorname{OTOC}_k(U)$ and the Haar-random average $\mathbb{E}_{U\sim\mathcal{U}_n}\operatorname{OTOC}_k(U)$.

Combining the convergence bound on the frame potential discussed above with the findings of Ref.~\cite{roberts_chaos_2017}, which relate the frame potential of unitary ensembles to OTOCs, we can estimate the average scrambling behavior of doped Clifford circuits.
\begin{corollary*}[Scrambling in $t$-doped Clifford circuits. Informal of \cref{cor:scrambling}] 
Let $n \ge \Omega(k^2 + t\log k)$ and $c > 0$. The relative difference between the average $\operatorname{OTOC}_k(U)$ of a $t$-doped Clifford circuit and that of Haar-random unitaries is negligibly small if and only if $t = \tilde{\Theta}(k^2 + \log^{1+c} n)$.
\end{corollary*}

These results establish both upper and lower bounds, revealing the ultimate capabilities and limitations of scrambling in doped Clifford circuits. The conclusion is noteworthy: only a polylogarithmic number of additional non-Clifford gates is sufficient to reproduce scrambling behavior that is indistinguishable—up to negligible relative error—from Haar-like scrambling. In sharp contrast, as we will later demonstrate, relative $k$-designs (and pseudorandom unitaries) require an extensive $\Omega(n)$ number of non-Clifford gates. This underscores a fundamental distinction: while relative designs entail a linear overhead, reproducing complex scrambling dynamics demands only a polylogarithmic number of non-Clifford resources. These findings are akin to those of Ref.~\cite{magniAnticoncentrationCliffordCircuits2025}, which showed that reproducing the Porter–Thomas distribution of measurement statistics for Haar-random states in the computational basis requires only a polylogarithmic number of non-Clifford gates. This suggests that the two properties, namely \textit{anticoncentration} and \textit{scrambling}, are closely related.

Before concluding this section, one might argue that $\operatorname{OTOC}_k(U)$ inherently washes out the locality aspect of scrambling, since averaging over the full Pauli group includes highly nonlocal measurements. However, because we are concerned with comparing deep unitaries—such as doped Clifford circuits and Haar-random unitaries—locality becomes irrelevant in this regime.

\subsubsection{The twirling operator over doped Clifford circuits} On a more technical note, proving the tight convergence bounds for $t$-doped Clifford circuits relies on a formal understanding of the twirling operator over this ensemble. In particular, given an ensemble of unitaries $\mathcal{E}$, the \emph{twirling superoperator} 
(also commonly referred to as the 
$k$-th \emph{moment superoperator}) is defined as
\be\label{eqintrotwirling}
\Phi_{\mathcal{E}}^{(k)}(O)\coloneqq\mathbb{E}_{U\in\mathcal{E}}[U^{\otimes k}OU^{\dag\otimes k}]\,.
\ee
If the ensemble $\mathcal{E}$ forms a unitary $k$-design, then the twirling superoperator coincides with the one over the entire unitary group, i.e.,\ $\Phi_{\haar}^{(k)}(O)\coloneqq\mathbb{E}_{U\in\mathcal{U}_n}[U^{\otimes k}OU^{\dag\otimes k}]$. In \cref{sec:dopedcliffordhaaraverage}, we characterize the twirling operator over the doped Clifford circuits, regardless of the particular choice of doping gates, in terms of the elements $\Omega$ of the Clifford commutant, 
\be\label{introinformaltwirling}
\Phi_{t}^{(k)}(O) =\Phi_{\haar}^{(k)}(O)+\sum_{\Omega,\Omega'}[\Delta^{t}\mathcal{W}^{-1}]_{\Omega,\Omega'}\tr(O\Omega)\Omega'
\ee  
the so-called \textit{Pauli monomials}, introduced in Ref.~\cite{bittel2025completetheorycliffordcommutant}.
Here, the components of the matrix $\Delta^t\mathcal{W}^{-1}$ are referred to as \textit{doped-Clifford Weingarten functions}: the matrix $\mathcal{W}^{-1}$ encodes the Clifford Weingarten functions, introduced and studied in Refs.~\cite{leone_quantum_2021,oliviero_transitions_2021,bittel2025completetheorycliffordcommutant}, while the matrix $\Delta$ captures the influence of doping in the circuit. In \cref{sec:asymptoticdopedcliffordweingarten}, we derive the asymptotic scalings of these functions for $k^2 = O(n)$, which allows for simpler analytic expressions. For the doping architectures analyzed in this paper, one has $\|\Delta\|_{\infty}\le C<1$, with $C=\Theta(1)$ being a constant. \cref{introinformaltwirling} illustrates the competition between the Clifford contribution to the twirling, which gets exponentially suppressed with increasing doping $t$, and the gradual emergence of the twirling operator $\Phi_{\haar}^{(k)}(\cdot)$ over the full unitary group. From~\cref{introinformaltwirling}, it immediately follows 
that whenever $\|\Delta\|_{\infty}<1$,
\be\label{eq:informalconvergence}
\lim_{t \to \infty} \Phi_{t}^{(k)}(O)=\Phi_{\haar}^{(k)}(O)\,,
\ee
the only remaining question being the scaling competition between $t$, $k$ and $n$. The convergence towards Haar random depends on the matrix $\Delta$. In particular, the crux of the lower bound in \cref{introeq1} resolves around a relation between the trace of $\Delta$ and the \textit{spectral form factor} of the ensemble of non-Clifford gates used to dope the Clifford group and promote it to universality (see \cref{lem:lowerbound}). The relationship between the convergence rate of doped Clifford circuits and spectral form factors is intriguing due to their widespread presence across various areas of quantum information. Originally introduced in the context of Random Matrix Theory \cite{bhatia_matrix_1997} as a tool to measure correlations between the energy eigenvalues of quantum systems, spectral form factors have found numerous applications in quantum information theory due to their ability to model complex systems in a simple and effective way. In particular, they have become fundamental tools to characterize chaotic dynamics \cite{cotler_chaos_2017, liu_spectral_2018, leone_isospectral_2021, oliviero_random_2021, gharibyan_onset_2018}, thermalization and ergodicity \cite{winer2023reappearancethermalizationdynamicslatetime}, and  also found applications in number theory \cite{Keating2000RandomMT} and quantum transport \cite{Imry1986ActiveTC}.


\subsubsection{Convergence of doped Clifford circuits to relative unitary designs} The second question we address in this paper concerns the convergence of $t$-doped Clifford circuits to unitary $k$-designs. While \cref{eq:informalconvergence} guarantees that, with an appropriately chosen measure, $t$-doped Clifford circuits always converge to exact unitary designs in the limit of infinite doping, the fundamental question that remains is how much doping $t$ is necessary and sufficient for the ensemble to be effectively considered a unitary $k$-design on $n$ qubits. 

There are essentially two 
common notions of approximate unitary $k$-designs for an ensemble $\mathcal{E}$: $\varepsilon$-approximate unitary designs in terms of \textit{additive} error and in terms of \textit{relative} error. The first notion ensures that any quantum algorithm, with resolution $\varepsilon$, making $k$-fold queries to $U^{\otimes k}$ cannot distinguish $\mathcal{E}$ from unitaries drawn uniformly from the Haar measure. The second, stronger notion guarantees that even quantum algorithms making $k$ (possibly adaptive) queries cannot distinguish $\mathcal{E}$ from the Haar-random unitaries.  We refer the reader to \cref{sec:unitarykdesign} for a formal introduction; in the following, we focus on the stronger notion of approximate $k$-designs in relative error. In particular, an ensemble $\mathcal{E}$ of unitaries is a relative $\varepsilon$-approximate unitary $k$-design if and only if
\be\label{re:main}
(1-\varepsilon)\Phi_{\haar}^{(k)}\le\Phi_{t}^{(k)}\le (1+\varepsilon)\Phi_{\haar}^{(k)}
\ee
where $\Phi-\Phi'\ge 0$ iff $\Phi-\Phi'$ is a completely positive map. 
Thanks to the characterization of the twirling operator of doped Clifford circuits in \cref{introinformaltwirling}, we establish the following convergence bounds to relative approximate $k$-designs.

\begin{theorem*}[Optimal convergence to relative unitary designs] Let \( \delta > 0 \), \( 0 \le \varepsilon \le 2^{n\frac{k}{2}(1 - \delta)} \) and $k\le 2^{\frac{n}{2}}$. Then, for the ensemble of \( t \)-doped Clifford circuits $\mathcal{C}_t$, the following results hold.
\begin{enumerate}[label=(\roman*)]
    \item For $k\le \delta n/2$, the ensemble $\mathcal{C}_t$ cannot form a relative \( \varepsilon \)-approximate unitary \( k \)-design unless 
    \be
t=\tilde{\Omega}(nk)\,\,\,\text{(\cref{th:lowerboundrelativeerrorunitarydesigns})}.
    \ee
\item For $k=O(\sqrt{n})$, the ensemble $\mathcal{C}_t$ does form an $\varepsilon$-approximate $k$-design with $t=\tilde{\Omega}(nk)$ (\cite{haferkamp2020QuantumHomeopathyWorks}, see \cref{prop:haferkamp}).
\item For $k>8n\delta^{-1}$, the ensemble $\mathcal{C}_t$ cannot form a relative \( \varepsilon \)-approximate unitary \( k \)-design unless 
    \be
t=\Omega(k) \,\,\,\text{(\cref{th:alternativelowerbound})}.
\ee
\end{enumerate}
In particular, item (i) and (ii) together implies an optimal convergence bound towards relative $\varepsilon$-approximate $k$-designs with $t=\tilde{\Theta}(nk)$, in the regime $k=O(\sqrt{n})$. 
\end{theorem*}

The theorem above implies that constructing relative approximate designs in the context of doped Clifford circuits is highly resource-intensive---even when a relative error exponentially large in $nk$ is allowed. In particular, when $k = O(n)$, the number of non-Clifford gates required scales linearly with both the tensor power $k$ and the number of qubits $n$ (and it is also saturated for $k=O(\sqrt{n})$). Similarly, when $k = \Omega(n)$, the number of non-Clifford gates must grow at least linearly with $k$. In the intermediate regime $\Omega(n) \leq k \leq O(n^2)$, where the first bound provides a tighter constraint than the second, it follows that at least $t = \Omega(n^2)$ non-Clifford gates are necessary, since any $(k+1)$-design is also at least a $k$-design. This result strictly generalizes the finding of Ref.~\cite{leone_quantum_2021}, where a linear number (in $n$) of non-Clifford gates was shown to be necessary and sufficient to reproduce an $\varepsilon$-approximate 4-design in relative error.

At this point, the reader may wonder how this result reconciles with the observed convergence of the frame potential that appears independent of $n$, as well as with the `homeopathic' convergence in additive-error designs reported in Ref.~\cite{haferkamp2020QuantumHomeopathyWorks}. The key lies in the notion of approximation: while a doping level independent of $t$ is sufficient for the convergence of the frame potential, and for achieving $\varepsilon$-approximate state designs and unitary designs in \emph{additive} error, it does not preclude the existence of a quantum algorithm that, with $k$ queries to $U$ and resolution $\sim\varepsilon$, can distinguish the ensemble from the Haar measure. This leads to the stronger notion of unitary designs in \emph{relative} error. The theorem above shows that such distinguishing algorithms do exist unless the doping is extremely high---specifically, linear in $nk$. This highlights the fact that an ensemble of unitaries which is indistinguishable from Haar random unitaries up to the $k$ moment must necessarily lie beyond the classical simulability threshold, which is set by $t = O(\log n)$~\cite{bravyi_improved_2016,guMagic2025}.

\medskip

\subsection{Remarks on $t$-doped Clifford circuits generating pseudo-random unitaries}
As anticipated, a weaker alternative to generating relative $\varepsilon$-approximate unitary $k$-designs is the construction of pseudo-random unitaries~\cite{ji_pseudorandom_2018}. A set of unitaries $\mathcal{U}$ is called pseudo-random if no polynomial-time quantum algorithm can distinguish whether a unitary was sampled uniformly from $\mathcal{U}$ or from the Haar measure on the full unitary group. In contrast with unitary $k$-design, pseudo-random unitaries demand indistinguishability only against polynomial-time adversaries.

Although not the primary focus of this work, we briefly examine the construction of pseudo-random unitaries in the framework of $t$-doped Clifford circuits. We consolidate existing results from the literature to determine the optimal level of doping that is both necessary and sufficient for generating pseudo-random unitaries.

\begin{proposition}[Pseudo-random \( t \)-doped Clifford circuits]\label{prop1:pseudo}
Any pseudo-random set of unitaries must involve at least $\Omega(n)$ non-Clifford gates. Conversely, there exists a pseudo-random set constructed using $\tilde{O}(n)$ (up to log factors) non-Clifford gates.
\end{proposition}
The proof of the above proposition is entirely based on the results of Refs.~\cite{grewal_efficient_2023,schuster2025randomunitariesextremelylow}, and is provided in \cref{sec:pseudo-randomess} for completeness.
While we have shown that generating relative-error unitary \( k \)-designs requires doping with \( t = \tilde{\Theta}(nk) \) non-Clifford gates, the proposition above demonstrates that a significantly lower doping level of \( t = \tilde{\Theta}(n) \) is both necessary and sufficient for generating pseudo-random unitaries. Although the weaker requirement of indistinguishability from Haar-random unitaries by polynomial-time quantum adversaries allows for more favorable scaling of \( t \), the fact that it still grows linearly with system size underscores the inherent cost of randomness generation in terms of non-Clifford resources, a recurring theme throughout this work.

\medskip

\subsection{Discussion, outlook, and open questions}\label{sec:discussion}
In this work, we study how well doped Clifford circuits spread over the unitary group and establish tight bounds on both their frame potential and the approximation of state $k$-designs. Additionally, we
carefully analyze the precise conditions under which the ensemble of doped Clifford circuits constitutes a relative-error unitary $k$-design, a natural and, at the same time, strong sense of approximation. The recurring theme in our results is that generating randomness is extremely demanding in terms of non-Clifford resources, and, therefore, substantially challenge the fault-tolerant implementation of unitary $k$-designs. On the other hand, a positive interpretation of the present results is to see them as an encouragement for efforts aimed at producing quantum circuits outside the realm of classical simulability.

The first main result of this work is that a quadratic amount of doping, in terms of the degree $k$, is both necessary and sufficient for the ensemble to approximate the Haar frame potential. To prove this, we introduce new technical tools for averaging over $t$-doped Clifford circuits from a mathematical perspective, including the construction of doped-Clifford Weingarten functions and an analysis of their asymptotic scaling in $n$. These tools are of independent theoretical interest and are aimed at increasing the portfolio of rigorous methods to study random processes in quantum physics.

Our second main result is a linear lower bound in both $n$ and $k$ for doped Clifford circuits to form a relative $\varepsilon$-approximate $k$-design. Specifically, we show that the ensemble cannot constitute a relative design unless $t = \Omega(nk)$. This lower bound is optimal in the regime $k=O(\sqrt{n})$: the seminal work of Haferkamp et al.~\cite{haferkamp_random_2022} establishes a matching upper bound, demonstrating that doped Clifford circuits do yield relative designs when $t = O(nk)$. 
Seen in this light,
our results establish the optimality of the construction from Ref.~\cite{haferkamp_random_2022}, showing that a linear amount of doping, $t = \tilde{\Theta}(nk)$, is both necessary and sufficient for achieving relative-error unitary designs. Beyond this regime, we show that $\Omega(k)$ many non-Clifford gates are necessary for constructing relative unitary $k$-designs. 

While our results provide a conclusive analysis of the convergence of the frame potential and unitary $k$-designs in the relative error setting in the regime $k=O(\sqrt{n})$, they leave 
important open questions regarding other fundamental aspects of unitary $k$-designs, which we explicitly invite further investigation into.

First, convergence guarantees for $t$-doped Clifford circuits forming relative $\varepsilon$-approximate $k$-designs---potentially matching our lower bounds---are currently lacking in the regime $k =\Omega(\sqrt{n})$. 
Second, we explicitly study Clifford circuits doped with non-Clifford unitaries. However, the same question is also relevant---and, in fact, some of our results carry over directly---to scenarios where the doping is implemented via suitable non-Clifford mid-circuit measurements rather than unitaries, as in what are known as \emph{dynamical quantum circuits}~\cite{PhysRevX.9.031009}. While the results of Ref.~\cite{oliviero_transitions_2021} imply convergence to a relative $\varepsilon$-approximate $4$-design using $\tilde{\Theta}(n)$ mid-circuit measurements, an analysis of higher-order designs remains lacking.

Third, finally, the frame potential captures the difference between the moment operators associated with an ensemble of unitaries in the $2$-norm. Similarly, approximation in relative error for the twirling channel provides a strong indistinguishability notion, even against adversaries capable of adaptive queries to copies of the unitaries. However, alternative approximation measures may be more appropriate and physically plausible in certain contexts. In particular, the diamond distance between the twirling operator defined in~\cref{eqintrotwirling} and the Haar twirling operator has operational significance analogous to the trace distance used in state designs, as formalized through a generalization of the Holevo-Helstrom theorem. Our results do not yield strong implications for this notion of approximation. For this reason, they are also neither contradicting nor 
incompatible in spirit to the findings of Ref.\ \cite{haferkamp2020QuantumHomeopathyWorks}.
Although relative-error designs constitute a robust and practically useful framework---often outperforming additive-error designs---an intriguing open question remains: does the quadratic scaling observed in the convergence of the frame potential also arise in the convergence of additive-error $k$-designs? Or is this scaling merely a consequence of the spreading properties captured by the frame potential?

While these questions are primarily mathematical in nature, they also invite conceptual investigations into which precise sense of approximation is most natural in what context, and they will be the subject of fruitful future research.


\subsection{Structure of this work}
The remainder of this work is organized as follows. We begin in \cref{sec:preliminaries} by setting the notation and introducing key concepts that are instrumental for proving our main results. In \cref{sec:haaraveragecliff}, we briefly review the main findings concerning the Haar measure over the Clifford group---a critical component of our approach---while referring the reader to Ref.~\cite{bittel2025completetheorycliffordcommutant} for a comprehensive introduction to the topic. In \cref{sec:dopedcliffordhaaraverage}, we define the ensemble of $t$-doped Clifford circuits, introduce an appropriate measure over this ensemble, and present a systematic method for computing averages using doped-Clifford Weingarten functions---a generalization of the Clifford-Weingarten functions from Ref.~\cite{bittel2025completetheorycliffordcommutant}. We also derive the asymptotic behavior of these functions in the regime $k = O(\sqrt{n})$. Finally, in \cref{sec:mainresults}, we leverage the framework developed in the preceding sections to prove the main results discussed in \cref{sec:overview}.

\section{Preliminaries}\label{sec:preliminaries}
In this section, we briefly introduce the tools that are necessary to introduce and prove the main results in this work.
Let $\mathcal{H}$ be the Hilbert space of $n$ qubits and $d=2^n$ its dimension. We consider $k$ copies of it $\mathcal{H}^{\otimes k}$. We denote as $\mathcal{B}(\mathcal{H})$ the set of operators acting on $\mathcal{H}$, and $\mathcal{U}_n$ denotes the unitary group on $n$ qubits.  
We use the notation $[n]$ to denote the set $[n]:=\{1,\dots,n\}$, where $n\in\mathbb{N}$.
The finite field $\mathbb{F}_2$ consists of the elements $\{0, 1\}$ with addition and multiplication defined modulo 2. For $x\in\mathbb{F}_{2}^{k}$, we denote $|x|$ the Hamming weight of $x$. The space of $k \times m$ binary matrices over $\mathbb{F}_2$ is denoted by $\mathbb{F}_2^{k \times m}$. The set \( \mathrm{Sym}(\mathbb{F}_2^{m\times m}) \) denotes the set of all symmetric \( m \times m \) matrices over the finite field \( \mathbb{F}_2 \), having a null diagonal. That is,
\[
\mathrm{Sym}(\mathbb{F}_2^{m\times m}) \coloneqq \left\{ M \in \mathbb{F}_2^{m \times m} : M^T = M\,,\, M_{j,j}=0\,\,\forall j\in[m] \right\}.
\]
Similarly, we define set $\even$ as the set of binary matrices $V\in\mathbb{F}_{2}^{k\times m}$ with $m\in[k]$ with column vectors $V_{i}^{T}$ of even Hamming weight
\be
\mathrm{Even}(\mathbb{F}_{2}^{k\times m})\coloneqq\{V\in\mathbb{F}_{2}^{k\times m}\,:\, |V^{T}_{i}|=0\mod 2\,\,\forall i\in[m]\}.
\ee
\subsection{Pauli operators}\label{sec:paulioperators}
The entire discussion of this work is focusing on capturing properties of quantum circuits consisting on $n$ qubits. To this purpose, let us introduce the qubit Pauli matrices $\{I,X,Y,Z\}$ as
\be 
I=
\begin{pmatrix}
    1 & 0 \\
    0 & 1
\end{pmatrix}\,,\quad
X=
\begin{pmatrix}
    0 & 1 \\
    1 & 0
\end{pmatrix}\,,\quad Y=
\begin{pmatrix}
    0 & -i  \\
    i &  0
\end{pmatrix}\,,
\quad Z=
\begin{pmatrix}
    1 & 0 \\
    0 & -1 
\end{pmatrix}\,.
\ee 
The Pauli group on \( n \)-qubits is defined as the \( n \)-fold tensor product of the set of Pauli matrices, multiplied by a phase factor from \( \{\pm 1, \pm i\} \). We denote the Pauli group modulo phases as \( \mathbb{P}_n \coloneqq \{I, X, Y, Z\}^{\otimes n} \).
Let us define the function 
\begin{align}\label{def:chidef}
\chi(A, B) \coloneqq \frac{1}{d} \tr(A B A^\dagger B^\dagger),
\end{align}
where \(A\) and \(B\) are operators. The function \(\chi(A, B)\) satisfies the symmetry properties:
\begin{align}
\label{eq:trivialpropCHI}
\chi(A, B) = \chi(\phi_A A, \phi_B B), \quad \chi(A, B) = \chi(B^\dagger, A) = \chi(A^\dagger, B^\dagger),
\end{align}
where $\phi_A, \phi_B \in \{-1, 1, -i, +i\}$ and the latter follows from the cyclicity of the trace. Note that, the following property holds:
\(\chi(P, Q)\) can be written as
\begin{align}
        \chi(P, Q) = 
        \begin{cases} 
        1, & \text{if } [P, Q] = 0, \\ 
        -1, & \text{if } \{P, Q\} = 0,
        \end{cases}
\end{align}
and we have that \( PQP = \chi(P, Q) Q \).

A Clifford operator is a unitary operator that maps any Pauli operator to another Pauli operator under conjugation. This family of unitaries is extremely important in quantum information theory, many-body physics~\cite{PhysRevLett.134.150403}, and it plays a dominating role in quantum error correction.

\begin{definition}[Clifford group]
    The set of Clifford operators forms a group, called the Clifford group $\mathcal{C}_n$, which is a subgroup of the unitary group $\mathcal{U}_n$, defined as the normalizer of the Pauli group $\mathbb{P}_n$.  
Concretely, this means that  
\[
\mathcal{C}_n = \{ U \in \mathcal{U}(2^n) \mid U P U^{\dagger} \in \mathbb{P}_n, \, \forall P \in \mathbb{P}_n \}.
\]
It is straightforward to verify that this leads to the group property of $\mathcal{C}_n$.
\end{definition}

\begin{definition}[Stabilizer states] Stabilizer states are pure quantum states obtained by $\ket{0}^{\otimes n}$ via the action of a Clifford unitary operator $C\in\mathbb{C}_n$. The set of stabilizer states on $n$ qubits will be denoted as $\stab_n$.
\end{definition}

\subsection{Unitary $k$-designs}\label{sec:unitarykdesign}

Unitary designs are sets of unitaries that approximate Haar-random unitaries by reproducing the expectation values of low-order polynomials. In this section, we briefly introduce the relevant concepts and refer the reader to Ref.~\cite{brandao_local_2016} for a more in-depth overview.

We denote as $\de U$ the Haar measure over the unitary group, which is the only left/right invariant measure on $\mathcal{U}_n$. 

\begin{definition}[$k$-fold channel]\label{def:kfoldchannel} Let $O\in\mathcal{B}(\mathcal{H})$ be a operator. The $k$-fold channel (or twirling) is defined as
\be\label{eq:kfoldchannelhaar}
\Phi_{\haar}^{(k)}(O)\coloneqq\int\de U\,U^{\otimes k}OU^{\dag\otimes k}.
\ee
\end{definition}
A general recipe to compute the $k$-fold channel associated to the unitary group is via the Weingarten functions~\cite{weingarten_asymptotic_1978}.
\begin{lemma}[Weingarten calculus] Let $\boldsymbol{\Lambda}$ the $k!\times k!$ matrix with components $\boldsymbol{\Lambda}_{\pi, \sigma}=\tr(T_{\pi}T_{\sigma})$, then for any $O\in\mathcal{B}(\mathcal{H})$ the $k$-fold twirling in \cref{eq:kfoldchannelhaar} reads
\be
\Phi_{\haar}^{(k)}=\sum_{\pi,\sigma}(\boldsymbol{\Lambda}^{-1})_{\pi,\sigma}\tr(T_{\pi}O)T_{\sigma}
\ee
where the components $(\boldsymbol{\Lambda}^{-1})_{\pi, \sigma}$ are called Weingarten functions.
\end{lemma}

\begin{lemma}[Asymptotics of Weingarten functions~\cite{weingarten_asymptotic_1978}] \label{asymptoticweingarten} Let $\boldsymbol{\Lambda}$ be the $k!\times k!$ matrix with components $\boldsymbol{\Lambda}_{\pi, \sigma}=\tr(T_{\pi}T_{\sigma})$. Then, in the limit of large $d$, the components of the inverse matrix $\boldsymbol{\Lambda}^{-1}$ behave as
\be
(\boldsymbol{\Lambda}^{-1})_{\pi,\pi}&=&d^{-k}+O(d^{-(k+2)}),\\
(\boldsymbol{\Lambda}^{-1})_{\pi, \sigma}&=&O(d^{-(k+|\pi\sigma|)}),\quad\pi\neq\sigma.
\ee    
\end{lemma}
The above asymptotics are extremely useful for various applications, see also Ref.~\cite{liu_entanglement_2018}.
Let us introduce the concept of unitary \( k \)-designs. Denoting \( \mathcal{E} \subseteq \mathcal{U}_n \) as an ensemble of unitaries, either discrete or continuous, equipped with a measure \( \mathrm{d}\mu(U) \), one can define the \( k \)-fold channel of \( \mathcal{E} \) on \( O \in \mathcal{B}(\mathcal{H}) \) simply as
\begin{equation}\label{eq:k-foldoperator}
\Phi_{\mathcal{E}}^{(k)}(O) \coloneqq \int_{\mathcal{E}} \mathrm{d} \mu(U) \, U^{\otimes k} O U^{\dag \otimes k}\,.
\end{equation}
This operator is also commonly referred to as the $k$-th moment superoperator.
Whenever \( \Phi_{\mathcal{E}}^{(k)}(O) \) agrees with the \( k \)-fold channel on the full unitary group, we say that \( \mathcal{E} \) is (an exact) 
unitary \( k \)-design.
\begin{definition}[Unitary $k$-design] Let $\mathcal{E}\subseteq\mathcal{U}_n$ be an ensemble of unitaries equipped with a measure $\de\mu(U)$. $\mathcal{E}$ is a unitary $k$-design iff
\be
\Phi_{\mathcal{E}}^{(k)}(O)=\Phi_{\haar}^{(k)}(O)\label{unitarydesigndefinition}
\ee
for any $O\in\mathcal{B}(\mathcal{H})$. 
\end{definition}

Let us now introduce one of the central objects of this work in the context of unitary $k$-design: the \textit{frame potential} \cite{Gross_2007}. 
The frame potential is a scalar function, which can determine whether an ensemble $\mathcal{E}$ is a $k$-design or not. The frame potential can be seen as a `repulsive force': The global minimum of the (unitary) frame potential corresponds to an exact unitary design. 
The frame potential is defined as
\be\label{def:framepotential}
\mathcal{F}_{\mathcal{E}}^{(k)}\coloneqq\int_{\mathcal{E}}\de\mu(U)\de\mu(V)|\tr(UV^{\dag})|^{2k}.
\ee
\begin{lemma}[Unitary $k$-design via the frame potential]\label{lem:framepotential} Let $\mathcal{E}$ be an ensemble of unitaries equipped with the measure $\de\mu(U)$. Then
\be
\mathcal{F}_{\mathcal{E}}^{(k)}\ge\mathcal{F}^{(k)}_{\haar},
\ee
with equality if and only if $\mathcal{E}$ is a $k$-design. Moreover, 
$\mathcal{F}^{(k)}_{\haar}=k!$.
\end{lemma}
Let us now define the concept of approximate unitary $k$-designs. There are essentially two distinctly different
but, at the same time, common definitions of approximate unitary $k$-designs.

\begin{definition}[$\varepsilon$-approximate unitary design in additive error]\label{def:unitarydesignadditiveerror} Let $\varepsilon>0$. Let $\mathcal{E}$ be an ensemble of unitaries equipped with the measure $\de\mu(U)$. $\mathcal{E}$ is an $\varepsilon$-approximate unitary design in additive error iff
\be
\|\Phi_{\mathcal{E}}-\Phi_{\haar}\|_{\diamond}\le\varepsilon
\ee
where $\Phi_{\mathcal{E}}$ (resp.\ $\Phi_{\haar}$) is the $k$-fold channel defined in \cref{def:kfoldchannel} and $\|\cdot\|_{\diamond}$ denotes the diamond norm.
\end{definition}
The notion of $\varepsilon$-approximate unitary designs in additive error has the operational meaning that the state produced by any quantum algorithm that queries $U^{\otimes k}$ when is sampled from $\mathcal{E}$ or according to the Haar measure is at most $\varepsilon$ close in trace distance. However, recently, the concept of relative error is becoming increasingly prominent in the context of the study of unitary $k$-designs \cite{low_pseudorandomness_2010,haferkamp_random_2022,schuster2025randomunitariesextremelylow}.

\begin{definition}[$\varepsilon$-approximate unitary design in relative error]\label{def:unitarydesignrelativeerror} Let $\varepsilon>0$. Let $\mathcal{E}$ be an ensemble of unitaries equipped with the measure $\de\mu(U)$. $\mathcal{E}$ is an $\varepsilon$-approximate unitary design in relative error iff
\be
\label{re}
(1-\varepsilon)\Phi_{\haar}\le \Phi_{\mathcal{E}}\le (1+\varepsilon)\Phi_{\haar}
\ee
where $\Phi\le \Phi'$ denotes that $\Phi'-\Phi$ is a completely positive map.
\end{definition}
The operational meaning of relative error design in \cref{def:unitarydesignrelativeerror} is much stronger than the additive error in \cref{def:unitarydesignadditiveerror}. This is indeed referred to as the `gold standard' in Ref.\ \cite{schuster2025randomunitariesextremelylow}, making reference to Ref.\ \cite{brandao_local_2016}: if $\mathcal{E}$ is a unitary design in relative error then the state produced by any quantum algorithm that makes $k$ queries to $U$ (which can be adaptive queries even) when $U$ is sampled from $\mathcal{E}$ or from the Haar measure is at most $2\varepsilon$-close in trace distance (for in-depth discussions, see, e.g.,  Refs.~\cite{low_pseudorandomness_2010,brandao_local_2016}
and, for a modern perspective, Ref.\ \cite{schuster2025randomunitariesextremelylow}). As a matter of fact, \cref{def:unitarydesignrelativeerror} implies \cref{def:unitarydesignadditiveerror} (up to a factor of $2$), while the converse is not true unless exponential factors are in the error:

\begin{lemma}[Additive 
vs.\ relative error. Lemma 3 in Ref.~\cite{brandao_local_2016}]\label{lem:additivetorelativeerror} Let $\varepsilon>0$. Let $\mathcal{E}$ be an ensemble of unitaries equipped with the measure $\de\mu(U)$. If $\mathcal{E}$ is $\varepsilon$-approximate $k$ design in additive error, then it is a $d^{2k}\varepsilon$-approximate $k$-design in relative error. Conversely, if $\mathcal{E}$ is an $\varepsilon$-approximate $k$-design in relative error, then $\mathcal{E}$ is an $2\varepsilon$-approximate $k$-design in additive error. 
\end{lemma}

Lastly, let us introduce the concept of $k$-expander, tightly related to the relative approximate $k$-design. We refer to Ref.~\cite{brandao_local_2016} for more details. Let us define the $k$-th moment operator
\begin{equation}
M^{(k)}_{\mathcal{E}}\coloneqq\mathbb{E}_{U\in \mathcal{E}} (U^{\otimes k} \otimes \bar{U}^{\otimes k}),
\end{equation}
where $\bar{U}$ denotes 
the complex conjugate of $U$. Let $M^{(k)}_{\mathrm{ Haar}}$ be the equivalent for the Haar measure. 

Then a $\delta$-approximate $k$-expander $\mathcal{E}$ is defined as~\cite{brandao_local_2016,brandao_expspe2013,haferkamp_random_2022,EpsilonNet}
\be
\|M^{(k)}_{\mathcal{E}}-M_{\haar}^{(k)}\|_{\infty}\le \delta\,.
\ee
As anticipated, there is a intimate relation between $\delta$-approximate $k$-expander and relative $\varepsilon$-approximate $k$-design, as the following lemma, due to Ref.~\cite{brandao_local_2016}, summarizes.

\begin{lemma}[$k$-expander and relative $k$-designs, Lemma 4 in Ref.~\cite{brandao_local_2016}]\label{lem:spectralgap}  
Denote with $G_\mathcal{E}$ the smallest $\varepsilon$ for which $\mathcal{E}$ forms an $\varepsilon$-approximate $k$-design in relative error, so that~\cref{re} holds true. Then
\begin{equation}\label{con}
2^{-nk/2-1}
\| M^{(k)}
- 
M^{(k)}_{\mathrm{ Haar}}\|_\infty
\leq 
G_\mathcal{E} \leq 2^{2kn}
\| M^{(k)}
- 
M^{(k)}_{\mathrm{ Haar}}\|_\infty.
\end{equation}
\end{lemma}

\subsection{State $k$-designs and state frame potential}\label{sec:projectivedesigns}

In this section, we introduce the concept of \textit{state $k$-design}, also called \textit{projective $k$-design}. This concept is closely related to the one of unitary design but restricted on pure states (i.e., rank one projectors). For the remainder of this work, we indicate as $\ket{\psi}$ state vectors and $\psi$ their density matrices, i.e., quantum states.

\begin{definition}[Projective $k$-design] Consider a set of state vectors $\mathcal{S}=\{\ket{\psi}\}$, equipped with a measure $\de\mu(\psi)$. Defining
\be\label{eq:projectivekdesignstate}
\Psi_{\mathcal{S}}^{(k)}\coloneqq\int\de\mu(\psi)\,\psi^{\otimes k}\,,
\ee
then $\mathcal{S}$ is a projective $k$-design iff $\Psi_{\mathcal{S}}^{(k)}=\Psi_{\haar}^{(k)}$, where $\Psi_{\haar}^{(k)}\coloneqq\Phi_{\haar}^{(k)}(\psi^{\otimes k})$.
\end{definition}

\begin{proposition}[Haar twirling]
For any pure state $\psi$, the Haar twirling over $\psi^{\otimes k}$ reads
\be
\Phi_{\haar}^{(k)}(\psi^{\otimes k})=\frac{\Pi_{\sym}}{\tr(\Pi_{\sym})}
\ee
where $\Pi_{\sym}=\frac{1}{k!}\sum_{\pi\in S_k}T_{\pi}$ is the projector onto the symmetric subspace of $\mathcal{H}^{\otimes k}$, $S_{k}$ is the symmetric group of order $k$ and $T_{\pi}$ are representation of permutation operators $\pi\in S_k$. 
\end{proposition}

Given an ensemble $\mathcal{E}$ of unitary operators, $\mathcal{E}$ induces an ensemble of states when applied on a reference state vector $\ket{0}$, i.e., $\mathcal{S}_{\mathcal{E}}:=\{U\ket{0}\,, U\in\mathcal{E}\}$ such that $|\mathcal{S}_{\mathcal{E}}|\le |\mathcal{E}|$. We have that if $\mathcal{E}$ is unitary $k$-design, then $\mathcal{S}_{\mathcal{E}}$ is a state $k$-design. The converse, naturally, does not hold. However, if $\mathcal{E}$ is not a unitary $k$-design, it can happen that, with a accurate choice of the reference state vector $\ket{0}$ in the definition of $\mathcal{S}_{\mathcal{E}}$, then  $\mathcal{S}_{\mathcal{E}}$ is state $k$-design. One relevant example is given by the Clifford orbit of quantum state. In particular, Ref.~\cite{zhu2016cliffordgroupfailsgracefully} has shown the existence of fiducial quantum state vectors $\ket{\psi}$ promoting the Clifford orbit of $\ket{\psi}$ to a state $4$-design. Similarly, Ref.~\cite{lamiQuantumStateDesigns2025}has shown that ensembles of matrix product states provide good fiducial vectors for state $4$-designs as their stabilizer entropies attain universal values.

Similarly to the case of unitary design, there is a notion of 
a \textit{state frame potential} for projective $k$-design, which corresponds to the \emph{purity}  
\begin{equation}
\pur(\rho)\coloneqq\tr(\rho^2) 
\end{equation}
of the state vector $\Psi_{S}^{(k)}$ in \cref{eq:projectivekdesignstate}.
Similarly, to the frame potential in \cref{lem:framepotential}
\cite{Gross_2007}, also the state frame potential provides a necessary and sufficient condition for state $k$-designs.

\begin{lemma}[State design via frame potential]\label{lem:designviastateframepotential} Let $\mathcal{S}$ be an ensemble of states equipped with the measure $\de\mu(\psi)$. Then
\be
\pur(\Psi_{\mathcal{E}}^{(k)})\ge \pur(\Psi_{\haar}^{(k)}),
\ee
with equality if and only if $\mathcal{S}$ is a projective $k$-design. 
Moreover, 
\begin{equation}
\pur(\Psi_{\haar}^{(k)})=\frac{1}{\tr(\Pi_{\sym})}, 
\end{equation}
where $\tr(\Pi_{\sym})=\binom{d+k-1}{k}.$
\end{lemma}
%
We can define a notion of $\varepsilon$-approximate projective design in trace distance.

\begin{definition}[$\varepsilon$-approximate state design in trace distance]\label{def:approximatestatekdesigns} Let $\varepsilon>0$. Let $\mathcal{S}$ be an ensemble of pure states equipped with a measure $\de\mu(\psi)$. $\mathcal{S}$ is an $\varepsilon$-approximate state $k$-design iff
\be\label{eq:approximatestatekdesigns}
\|\Psi_{\mathcal{E}}^{(k)}-\Psi_{\haar}^{(k)}\|_{1}\le\varepsilon.
\ee
\end{definition}
The notion of $\varepsilon$-approximate state $k$-design in trace distance has a clear operational meaning. Through the Holevo-Helstrom theorem, $k$-copies of a state drawn uniformly at random from the set $\mathcal{S}$ cannot be dististinguished from a Haar random state with probability greater than $\frac{1}{2}+\varepsilon$ if $\mathcal{S}$ is an $\varepsilon$-approximate $k$-design.

The state frame potential, being related to the purity of the state vector $\Psi_{\mathcal{E}}$, is amaneable for practical computation and also provides a useful bounds to the less computable notions of approximate state designs \cref{def:approximatestatekdesigns}. For this purpose, it is useful to define the \textit{relative frame potential} of an ensemble $\mathcal{S}$ of states, and relate it to \cref{def:approximatestatekdesigns}.
\begin{definition}[Relative frame potential]\label{def:relativeframepotential} Let $\mathcal{S}$ be an ensemble of states equipped with the measure $\de\mu(\psi)$. The relative frame potential is defined as
\be
\mathcal{R}_{\mathcal{S}}^{(k)}\coloneqq\frac{\pur(\Psi_{\mathcal{S}}^{(k)})-\pur(\Psi_{\haar}^{(k)})}{\pur(\Psi_{\haar}^{(k)})}.
\ee
\end{definition}
\begin{lemma}[Bounds via state relative frame potential]
 Let $\mathcal{S}$ be an ensemble of pure states equipped with a measure $\de\mu(\psi)$, and $\Psi_{\mathcal{S}}^{(k)}$ be the mixed state associated with $\mathcal{S}$. Let $\mathcal{R}_{\mathcal{S}}^{(k)}$ be the relative frame potential of the ensemble $\mathcal{S}$. Then the general bounds 
    \be\label{eq:bound1stateframepotential}
\|\Psi_{\mathcal{S}}^{(k)}-\Psi_{\haar}^{(k)}\|_{1}\le \sqrt{\mathcal{R}_{\mathcal{S}}^{(k)}}
    \ee
    hold. In particular, \cref{eq:bound1stateframepotential} implies that if the relative state frame potential $\mathcal{R}_{\mathcal{S}}^{(k)}\le \varepsilon^2$ then $\mathcal{S}$ is an $\varepsilon$-approximate state design in trace distance. 
    \begin{proof}
The bound in \cref{eq:bound1stateframepotential} descends from a useful property of Schatten $p$-norms, i.e., given $A\in\mathcal{B}(\mathcal{H})$, then $\|A\|_1\le\sqrt{\operatorname{rank}(A)}\|A\|_2$. Hence, we can readily apply it to \cref{eq:approximatestatekdesigns}, 
to get
\be
\|\Psi_{\mathcal{S}}^{(k)}-\Psi_{\haar}^{(k)}\|_{1}\le\sqrt{\operatorname{rank}(\Psi_{\mathcal{S}}^{(k)}-\Psi_{\haar}^{(k)})}\|\Psi_{\mathcal{S}}^{(k)}-\Psi_{\haar}^{(k)}\|_2.
\ee
From \cref{lem:designviastateframepotential}, we know that $\|\Psi_{\mathcal{S}}^{(k)}-\Psi_{\haar}^{(k)}\|_2^2=\pur(\Psi_{\haar}^{(k)})-\pur(\Psi_{\mathcal{S}}^{(k)})$. Let us now show that 
\be
\operatorname{rank}(\Psi_{\mathcal{S}}^{(k)}-\Psi_{\haar}^{(k)})\le \operatorname{rank}(\Psi_{\haar}^{(k)}). 
\ee
First, we have that $\operatorname{rank}(\Psi_{\haar}^{(k)})=\operatorname{rank}(\Pi_{\sym})$. Then, since $T_{\pi}\ket{\psi^{\otimes k}}=\ket{\psi^{\otimes k}}$ for any $\psi$, we also have that $\Pi_{\sym}\Psi_{\mathcal{S}}^{(k)}=\Psi_{\mathcal{S}}^{(k)}$, i.e., $\Pi_{\sym}$ acts identically on $\Psi_{\mathcal{S}}^{(k)}$, from which it immediately follows that $\operatorname{rank}(\Psi_{\mathcal{S}}^{(k)}-\Psi_{\haar}^{(k)})\le \operatorname{rank}(\Psi_{\haar}^{(k)})$. Recalling that $\Psi_{\haar}^{(k)}$ is proportional to a projector, i.e.,  $\Pi_{\sym}$, the statement follows. The validity of \cref{eq:bound1stateframepotential} then follows. 
    \end{proof}
\end{lemma}
\subsection{Haar average over the Clifford group}\label{sec:haaraveragecliff}
In this section, we summarize the findings of Ref.~\cite{bittel2025completetheorycliffordcommutant} regarding the average over the Clifford group, as it will be instrumental for the proof of our main result. 
Let us first define the set of \textit{reduced Pauli monomials}, which will be referred to as the set of Pauli monomials for brevity.

\begin{definition}[Pauli monomials]\label{def:paulimonomials}
Let \( k,m \in \mathbb{N} \), $V\in\mathrm{Even}(\mathbb{F}_{2}^{k\times m})$ with independent column vectors and $M\in\mathrm{Sym}(\mathbb{F}_{2}^{m\times m})$. A Pauli monomial, denoted as \( \Omega(V, M) \in \mathcal{B}(\mathcal{H}^{\otimes k}) \), is defined as
\begin{align}
\Omega(V, M) \coloneqq \frac{1}{d^m} \sum_{\boldsymbol{P} \in \mathbb{P}_n^m}  
P_1^{\otimes v_1} P_2^{\otimes v_2} \cdots P_m^{\otimes v_m}
\left( \prod_{\substack{i, j \in [m] \\ i < j}} \chi(P_i, P_j)^{M_{i,j}} \right),
\end{align}
where  $ \chi(P_i, P_j)$ is defined in \cref{def:chidef}, and we denote a string of Pauli operators in $\mathbb{P}_{n}$ as $\boldsymbol{P}\coloneqq P_1,\ldots, P_k$. We define the set of reduced Pauli monomials as
\be
\mathcal{P}\coloneqq\{\Omega(V,M)\,|\, V\in\even\,:\,\rank(V)=m\,,\,M\in\symf\,,\,m\in[k-1]\}\,. 
\ee
Furthermore, we define $\mathcal{P}_U\coloneqq\mathcal{P}\cap\mathcal{U}_{nk}$ the set of reduced Pauli monomial which are also unitary and $\mathcal{P}_P$ the set of reduced Pauli monomials proportional to projectors, formally defined as
\be
\mathcal{P}_P\coloneqq\{\Omega(V,0)\,|\, V\in \even\,:\, |V_i^{T}|=0\mod4\,,\, V_i^T\cdot V_j^T=0\mod2\,,\, i,j\in[m]\}.
\ee
\end{definition}

\begin{lemma}[Relevant properties of Pauli monomials]\label{lem:relevantpropertiespaulimonomials} The following facts hold~\cite{bittel2025completetheorycliffordcommutant}:
\begin{itemize}
    \item The commutant of the Clifford group is spanned by $\mathcal{P}$.
    \item For $n\le k-1$, $\mathcal{P}$ contains linearly independent operators.
    \item  $
|\mathcal{P}|=\prod_{i=0}^{k-2}(2^i+1) $, and the following bounds holds $2^{\frac{k^2-3k-1}{2}} \le \vert \mathcal P \vert \le 2^{\frac{k^2-3k+12}{2}}$.    
    \item For $\Omega\in \mathcal{P}$, then $\Omega=\omega^{\otimes n}$ (i.e., factorizes on qubits), where
    \be
\omega=\frac{1}{2^m} \sum_{\boldsymbol{P} \in \{I,X,Y,Z\}^m}  
P_1^{\otimes v_1} P_2^{\otimes v_2} \cdots P_m^{\otimes v_m}\times \left( \prod_{\substack{i, j \in [m] \\ i < j}} \chi(P_i, P_j)^{M_{i,j}} \right).
    \ee
\item $\Omega^{\dag}=\Omega^{T}$ for all $\Omega\in\mathcal{P}$.
\item $\Omega=\Omega_U\Omega_P$ where $\Omega_U\in\mathcal{P}_U$, $\Omega_P\in\mathcal{P}_P$ for all $\Omega\in\mathcal{P}$.
\item $\|\Omega\|_1=\tr(\Omega_P(V,0))=d^{k-\rank(V)}$ where $\Omega_P(V,0)\in\mathcal{P}_P$, for  any $\Omega\in\mathcal{P}$.
\item $\tr(\Omega\rho^{\otimes k})\le1$ for any state $\rho$ and any $\Omega\in\mathcal{P}$.
\end{itemize}
\end{lemma}

\begin{lemma}[Twirling over the Clifford group~\cite{bittel2025completetheorycliffordcommutant}]\label{sec:cliffordweingartencalculus} Consider the Clifford group $\mathcal{C}_n$. The $k$-fold channel of $\mathcal{C}_n$, denoted as $\Phi_{\cl}(\cdot)$, on an operator $O$ reads
\be
\Phi_{\cl}(O)=\sum_{\Omega,\Omega'\in\mathcal{P}}(\mathcal{W}^{-1})_{\Omega,\Omega'}\tr(\Omega^{\dag} O)\Omega'
\ee
where $\mathcal{P}$ is the set of Pauli monomials in \cref{def:paulimonomials}, and $\mathcal{W}^{-1}$ are the Clifford-Weingarten functions introduced in~\cite{bittel2025completetheorycliffordcommutant}, obtained by inverting the Gram-Schmidt matrix $\mathcal{W}_{\Omega,\Omega'}\coloneqq\tr(\Omega^{\dag}\Omega')$. 
\end{lemma}

\begin{lemma}[Average stabilizer state~\cite{gross_schurweyl_2019}]\label{lem:averagestabstate} Let $\sigma\in\stab_n$, then 
\be
\Phi_{\cl}^{(K)}(\sigma^{\otimes k})=\mathbb{E}_{\sigma}\sigma^{\otimes k}=\frac{1}{Z_n}\sum_{\Omega\in\mathcal{P}}\Omega=\frac{1}{Z_n}\sum_{\Omega_P\in\mathcal{P}_P}c_{\Omega}\Pi_U\Omega_P\Pi_U
\ee
where $\Pi_U\coloneqq\frac{1}{|\mathcal{P}_U|}\sum_{\Omega\in\mathcal{P}_U}\Omega$ and $Z_n=2^n\prod_{i=0}^{k-2}(2^n+2^i)$. The coefficient $c_{\Omega}$ is determined by the following identity 
\be
\sum_{\Omega\in\{\Omega_U\Omega_P\Omega_{U'}\,:\, \Omega_U,\Omega_{U'}\in\mathcal{P}_U\}}\Omega_P=c_{\Omega}\Pi_U\Omega_P\Pi_U
\ee
which is the multiplicity of the projection. The state frame potential, coinciding with its purity, is given by $\pur(\Phi_{\cl}^{(K)}(\sigma^{\otimes k}))=\frac{|\mathcal{P}|}{Z_n}$.
\end{lemma}

In what follows, we summarize the asymptotic results  proven in Ref.~\cite{bittel2025completetheorycliffordcommutant}, that will be particularly useful to derive the expression of the twirling operator for $t$-doped Clifford circuits.

\begin{lemma}[Asymptotics of Clifford-Weingarten functions~\cite{bittel2025completetheorycliffordcommutant}]\label{lem:asymptoticweingarten}
Let $\mathcal{W}$ the Gram-Schmidt matrix, and let $\mathcal{W}^{-1}$ be its inverse. Let $n\ge k^2-3k+13$. Then the following properties hold.
\begin{enumerate}[label=(\roman*)]
    \item $\mathcal{W}$ is symmetric.
    \item $1\le \mathcal{W}_{\Omega,\Omega'}\le d^{k-1} ,\quad \Omega\neq \Omega'\in\mathcal{P}$.
    \item $ \mathcal{W}_{\Omega,\Omega}= d^k,\quad  \Omega\in \mathcal{P}$.
    \item $d^k-|\mathcal{P}|d^{k-1}\le\lambda(\mathcal{W})\le d^k+|\mathcal{P}|d^{k-1}$ where $\lambda(\mathcal{W})$ is any eigenvalue of $\mathcal{W}$.
    \item $\det(\diag\mathcal{W})\left(1-\frac{|\mathcal{P}|^2}{d}\right)\le \det(\mathcal{W})\le \det(\diag\mathcal{W})\left(1+\frac{2|\mathcal{P}|^2}{d}\right)$, if $n\ge k^2-3k+13$.
    \item The following asymptotics hold for Clifford-Weingarten functions,
       \be
   \left|(\mathcal{W}^{-1})_{\Omega,\Omega}-\frac{1}{d^k}\right|&\le \frac{6|\mathcal{P}|^2}{d^{k+1}}\,,\\
    |(\mathcal{W}^{-1})_{\Omega,\Omega'}|&\le \frac{5|\mathcal{P}|^2}{d^{k+1}}.
    \ee
\end{enumerate}
\end{lemma}


\section{Twirling operator over the doped-Clifford group}\label{sec:dopedcliffordhaaraverage}
In this section, we define and compute the twirling operator for \( t \)-doped Clifford unitaries. \( t \)-doped Clifford unitaries are generated by applying Clifford gates augmented with at most \( t \) non-Clifford gates, see \cref{fig:enter-label}. It is well known that enhancing the Clifford group with any gate outside the Clifford set yields a universal gate set. Studying \( t \)-doped Clifford unitaries thus serves as a framework for studying the transition to the universality regime by controlling the degree of non-Cliffordness introduced in the unitary transformation.  

Before proceeding, we emphasize an important distinction: while the twirling operators for both the unitary and Clifford groups can be simply expressed as elements of the commutant of the unitary (or Clifford) group, \( t \)-doped Clifford unitaries do not form a group (as can be readily verified). Consequently, we cannot employ the Haar measure, which is the uniform measure over groups. Instead, we must define an appropriate measure over \( t \)-doped Clifford circuits, noting that this choice is not unique.

\subsection{Doped Clifford measures}

\begin{definition}[$t$-doped Clifford measure]\label{def:dopedrandomclifford} Denote $\mathcal{C}_n^{(t)}$ the ensemble of $t$-doped Clifford circuits, defined as
\be
\mathcal{C}_n^{(t)}\coloneqq\left\{U_t\,:\, U_t=\left(\prod_{i=1}^{t}C_iK_i\right)C_0\,,\, C_i\in\mathcal{C}_n\,\,\forall i\in[0,t],\, K_i\in\mathcal{U}_1\,\,\forall i\in[1, t]\right\}
\ee
then a random $t$-doped Clifford circuit is defined according to the following measure, denoted as $\mu(U_t)$
\be
\mu(U_t)=\mu_{\cl}\star \mu_{1}\star\mu_{\cl}\cdots\star\mu_{\cl}
\ee
where $\star$ denotes the convolution between measures, while $\mu_{\cl}$ is the uniform measure over the $n$-qubit Clifford group, while $\mu_{1}$ is \textit{any} measure over the single qubit unitary group. Notice that $\mu_1$ can also be deterministically a single gate. Let $f(U_t)$ be any function of $U_t\in \mathcal{C}_{n}^{(t)}$, we denote the average over $C_{n}^{(t)}$ as
\be
\int \de U_t f(U_t)\coloneqq\int_{\cl} \de C_t\int_{\mu_1}\de\mu_1(K_t)\cdots \int_{\cl}\de C_0 f(U_t)
\ee
where $\de C\coloneqq \de\mu_{\cl}(C)$. Given $O\in\mathcal{B}(\mathcal{H}^{\otimes k})$, the twirling operator $\Phi_{t}^{(k)}(O)$ reads
\be
\Phi_{t}^{(k)}(O)\coloneqq\int\de U_t U_{t}^{\otimes k}OU^{\dag\otimes k}_t.
\ee
\end{definition}
Before explicitly computing the twirling operator through Weingarten calculus, let us give an alternative expression that we can use later. 
We can parametrize single qubit unitaries $K_i$ (up to an irrelevant phase) with a vector $\vec{\alpha}=(\alpha_x,\alpha_y,\alpha_z)$ as $K_i(\vec{\alpha})=e^{i\vec{\alpha}\cdot \vec{\sigma}}$ where $\vec{\sigma}=(X,Y,Z)$. Hence the measure $\de\mu_1(K_i)=\de\vec{\alpha}$ can be conveniently thought of as a measure on the vector $\vec{\alpha}$. That said, we have the following alternative expression.
\be\label{eq:alternativeexpressioncliffordtwirling}
\Phi_{t}^{(k)}(O)=\mathbb{E}_{\vec{P}_i}\int\de\vec{\alpha_1}\cdots \de\vec{\alpha_t}U(\alpha_t)\cdots U(\alpha_1)\Phi_{\cl}^{(k)}(O)U^{\dag}_{1}(\alpha_1)\cdots U_{t}^{\dag}(\alpha_t)
\ee
where $U_{i}(\alpha_i)=e^{i\vec{\alpha}\cdot\vec{P}_i}$ where $\vec{P}_i=(P_{i,x},iP_{i,x}P_{i,z},P_{i,z})$ and $P_x,P_z$ are a pair of anticommuting Pauli randomly choosen from the Pauli group $\mathbb{P}_n$ (modulo identity), and $\mathbb{E}_{\vec{P}_i}$ denotes the uniform average over this choice. The above expression follows directly from the fact that random Clifford operators map Pauli operators uniformly to other Pauli operators. 

That said, \cref{eq:alternativeexpressioncliffordtwirling} is simply an alternative way to express the twirling operator. While it has the advantage of making the dependence on the Clifford twirling $\Phi_{\cl}^{(k)}(O)$ explicit, it is not particularly informative on its own. However, we can leverage the Clifford-Weingarten calculus to formally compute $\Phi_{t}^{(k)}(O)$. In particular, in the next theorem, we derive the explicit form of the doped twirling operator in terms of doped-Clifford-Weingarten functions, introduced in Ref.~\cite{leone_quantum_2021} and also used in Ref.~\cite{magniAnticoncentrationCliffordCircuits2025}.

\begin{theorem}[$t$-doped twirling operator] The $t$-doped twirling operator $\Phi_{t}^{(k)}(O)$ reads
\be
\Phi_{t}^{(k)}(O)=\sum_{\Omega\in\mathcal{P}}c^{(t)}_{\Omega}(O)\Omega\label{eq:inductionhypo},\quad c^{(t)}_{\Omega}(O)\coloneqq\sum_{\Omega'}[(\mathcal{W}^{-1}\mathcal{T})^{t}\mathcal{W}^{-1}]_{\Omega,\Omega'}\tr(\Omega' O)
\ee
where $\mathcal{W}^{-1}$ are the Clifford-Weingarten functions, $c_{\Omega'}(O)=\sum_{\Omega''}(\mathcal{W}^{-1})_{\Omega'\Omega''}\tr(\Omega''O)$, see \cref{sec:cliffordweingartencalculus}. We defined $\mathcal{T}$ is a $|\mathcal{P}|\times |\mathcal{P}|$ matrix with components $\mathcal{T}_{\Omega,\Omega'}=\tr(\Omega^{\dag}\kappa(\Omega'))$ where we defined $\kappa(\Omega)\coloneqq\int_{\mu_1}\de\mu_1(K)\,K^{\otimes k}\Omega K^{\dag\otimes k}$ the single twirling with measure $\mu_1$. 

\begin{proof}
The proof proceeds by induction on $t$. For $t=0$, we have $\Phi_{t}^{(k)}(O)\equiv \Phi_{\cl}^{(k)}(O)$ by definition, which proves the statement for $t=0$. Let us suppose that $\Phi_{t}^{(k)}(O)$ is indeed of the form in ~\cref{eq:inductionhypo}, and let us show it for $t+1$. By \cref{def:dopedrandomclifford}, we have
\be
\Phi_{t+1}^{(k)}(O)&=\int_{\cl}\de C\int_{\mu_1}\de\mu_1(K) \,C^{\otimes k}K^{\otimes k}\Phi_{t}^{(k)}(O)K^{\dag\otimes k}C^{\dag\otimes k}
=\sum_{\Omega\in\mathcal{P}}c^{(t)}_{\Omega}(O) \int_{\cl}\de C\int_{\mu_1}\de\mu_1(K) \,C^{\otimes k}K^{\otimes k}\Omega K^{\dag\otimes k}C^{\dag\otimes k}\\
&=\sum_{\Omega}c_{\Omega}^{(t)}(O)\sum_{\Omega'\Omega''}(\mathcal{W}^{-1})_{\Omega\Omega''}\tr(\Omega^{\prime\prime\dag}\kappa(\Omega))\Omega'
=\sum_{\Omega,\Omega'}(\mathcal{W}^{-1}\mathcal{T})_{\Omega,\Omega'}c^{(t)}_{\Omega}(O)\Omega'\\
&=\sum_{\Omega,\Omega'}[(\mathcal{W}^{-1}\mathcal{T})^{t+1}]_{\Omega,\Omega'}c_{\Omega}(O)\Omega'
\ee
which proves the statement.
\end{proof}
\end{theorem}
At first glance, the matrix $(\mathcal{W}^{-1}\mathcal{T})^t\mathcal{W}^{-1}$ may appear as a natural generalization of the Clifford-Weingarten functions discussed in \cref{sec:cliffordweingartencalculus}. However, in the following, we present a formulation that is more directly applicable to practical applications, as it explicitly incorporates the dependence on the twirling over the full unitary group.

\begin{lemma}[Convergence towards Haar random]\label{lem:convergenceoftdopedchannel} Let $\mathcal{W}^{-1}$ be the matrix of Clifford-Weingarten functions, and consider the matrix $\mathcal{W}^{-1}\mathcal{T}$. Define $\mathcal{L}$ as a $|\mathcal{P}|\times |\mathcal{P}|$ matrix with components
\be\label{eq:defLmatrix}
\mathcal{L}_{\Omega,\Omega'}\coloneqq\begin{cases}
\sum_{\sigma\in S_k}(\boldsymbol{\Lambda}^{-1})_{\pi\sigma}\mathcal{W}_{\sigma\Omega'},\quad &\pi\in S_k,\\
0 &\Omega\not\in S_k.
\end{cases}
\ee  
Define $\Delta\coloneqq\mathcal{W}^{-1}\mathcal{T}-\mathcal{L}$. The doped-Clifford operator reads
\be
\Phi_{t}^{(k)}(O)=\sum_{\Omega,\Omega'}[\boldsymbol{\Lambda}^{-1}+\Delta^{t}\mathcal{W}^{-1}]_{\Omega,\Omega'}\tr(O\Omega)\Omega'=\Phi_{\haar}^{(k)}(O)+\sum_{\Omega,\Omega'}[\Delta^{t}\mathcal{W}^{-1}]_{\Omega,\Omega'}\tr(O\Omega)\Omega'.
\ee
The doped-Clifford-Weingarten functions are thus defined through the twirling operator as the components of the matrix $[\Delta^t\mathcal{W}^{-1}]_{\Omega,\Omega'}$. 
\begin{proof} For simplicity we denote the non-zero components as $\mathcal{L}_{\pi\Omega}=\sum_{\sigma}(\boldsymbol{\Lambda}^{-1})_{\pi, \sigma}\mathcal{W}_{\sigma\Omega}$,  because the other components of $\mathcal{L}$ are zero by definition. Using that  $\mathcal{T}_{\Omega\pi}=\mathcal{T}_{\pi\Omega}=\mathcal{W}_{\Omega\pi}=\mathcal{W}_{\pi\Omega}$, we now prove $4$ properties of the matrix $\mathcal{L}$.
    \begin{enumerate}[label=(\roman*)]
        \item Let us show that $\mathcal{L}$ is a projector, i.e. $\mathcal{L}^2=\mathcal{L}$:
        \begin{align}(\mathcal{L}^2)_{\Omega\Omega'}&=\sum_{\tau}\mathcal{L}_{\Omega\tau}\mathcal{L}_{\tau\Omega'}=\sum_{\pi,\sigma,\tau}(\boldsymbol{\Lambda}^{-1})_{\Omega\pi}\mathcal{W}_{\pi\tau}(\boldsymbol{\Lambda}^{-1})_{\tau\sigma}\mathcal{W}_{\sigma\Omega'}\\
        &=\sum_{\pi,\sigma,\tau}(\boldsymbol{\Lambda}^{-1})_{\Omega\pi}\boldsymbol{\Lambda}_{\pi\tau}(\boldsymbol{\Lambda}^{-1})_{\tau\sigma}\mathcal{W}_{\sigma\Omega'}=\sum_{\pi,\sigma}\delta_{\pi\sigma}(\boldsymbol{\Lambda}^{-1})_{\Omega\pi}\mathcal{W}_{\sigma\Omega'}=\mathcal{L}
        \end{align}
        where we used the fact that the sum runs only over $\tau\in S_k$ because $\mathcal{L}_{\Omega\Omega'}=0$ for $\Omega\not\in S_k$ and that $\mathcal{W}_{\pi\sigma}=\boldsymbol{\Lambda}_{\pi\sigma}$, i.e. the Gram-matrix $\mathcal{W}$ equals $\Lambda$ restricted to permutation space. 

        \item Let us show that $\mathcal{L}(\mathcal{W}^{-1}\mathcal{T})=\mathcal{L}$:
        \begin{align}
           [\mathcal{L}(\mathcal{W}^{-1}\mathcal{T})]_{\pi\Omega'}&=\sum_{\sigma,\Omega'',\Omega'''} (\boldsymbol{\Lambda}^{-1})_{\pi\sigma}\mathcal{W}_{\pi\tau}(\mathcal{W}^{-1})_{\tau\Omega'''}\mathcal{T}_{\Omega'''\Omega'}\\
           &=\sum_{\sigma,\Omega'''} \delta_{\pi\Omega'''}(\boldsymbol{\Lambda}^{-1})_{\pi\sigma}\mathcal{T}_{\Omega'''\Omega'}=\sum_{\sigma} (\boldsymbol{\Lambda}^{-1})_{\pi\sigma}\mathcal{T}_{\pi\Omega'}\\&=\sum_{\sigma} (\boldsymbol{\Lambda}^{-1})_{\pi\sigma}\mathcal{W}_{\pi\Omega'}=\mathcal{L}_{\pi\Omega'}
        \end{align}
        while $[\mathcal{L}(\mathcal{W}^{-1}\mathcal{T})]_{\Omega\Omega'}=0$ for $\Omega'\not\in S_k$ because $\mathcal{L}_{\Omega\Omega'}=0$  for $\Omega'\not\in S_k$.
        \item Let us show that $(\mathcal{W}^{-1}\mathcal{T})\mathcal{L}=\mathcal{L}$:
        \begin{align}
           [(\mathcal{W}^{-1}\mathcal{T})\mathcal{L}]_{\Omega\Omega'}&= \sum_{\Omega'',\sigma,\tau}(\mathcal{W}^{-1})_{\Omega\Omega''}\mathcal{T}_{\Omega''\sigma}(\boldsymbol{\Lambda}^{-1})_{\sigma\tau}\mathcal{W}_{\tau\Omega'}\\&=\sum_{\Omega'',\sigma,\tau}(\mathcal{W}^{-1})_{\Omega\Omega''}\mathcal{W}_{\Omega''\sigma}(\boldsymbol{\Lambda}^{-1})_{\sigma\tau}\mathcal{W}_{\tau\Omega'}\\
           &=\sum_{\sigma,\tau}\delta_{\Omega\sigma}(\boldsymbol{\Lambda}^{-1})_{\sigma\tau}\mathcal{W}_{\tau\Omega'}=\mathcal{L}_{\Omega\Omega'}
        \end{align}
        where, in the first sum, we restricted the first sum only on $\sigma\in S_k$ because $\mathcal{L}_{\Omega\Omega'}=0$ for $\Omega\not\in S_k$.
        \item Finally, let us show that $\mathcal{L}\mathcal{W}^{-1}=\boldsymbol{\Lambda}^{-1}$ when restricted to permutation space, otherwise zero:
        \begin{align}
            [\mathcal{L}\mathcal{W}^{-1}]_{\pi\Omega}=\sum_{\sigma}(\boldsymbol{\Lambda}^{-1})_{\pi\sigma}\mathcal{W}_{\sigma\Omega'}\mathcal{W}^{-1}_{\Omega'\Omega}=\delta_{\Omega\sigma}(\boldsymbol{\Lambda}^{-1})_{\pi\sigma}
        \end{align}
        while $[\mathcal{L}\mathcal{W}^{-1}]_{\Omega\Omega'}=0$ is zero because $\Lambda_{\Omega\Omega'}=0$ for $\Omega\not\in S_k$. In what follows, we adopt a slight abuse of notation by using $\boldsymbol{\Lambda}$ and $\boldsymbol{\Lambda}^{-1}$ to denote the Gram and Weingarten matrices of the unitary group, viewed as embedded in the Clifford commutant space by padding with zeros outside the permutation subspace.
    \end{enumerate}
By definition, we can write $\mathcal{W}^{-1}\mathcal{T}=\mathcal{L}+\Delta$. However, we notice that $\mathcal{L}\Delta=\mathcal{L}(\mathcal{W}^{-1}\mathcal{T})-\mathcal{L}^2=\mathcal{L}-\mathcal{L}=0$, where we used properties (i) and (ii) above. Similarly, using (i) and (iii), one sees that $\Delta\mathcal{L}=0$. Consequently, we have that $(\mathcal{W}^{-1}\mathcal{T})^t=\mathcal{L}+\Delta^t$. Hence $(\mathcal{W}^{-1}\mathcal{T})^t\mathcal{W}^{-1}=\boldsymbol{\Lambda}^{-1}+\Delta^t\mathcal{W}^{-1}$, where we used property (iv) above (as well as the abuse of notation we pointed out above). We can insert such identity in the expression of the twirling operator, which reads
\begin{align}
    \Phi_t^{(k)}(O)&=\sum_{\Omega\Omega'}[(\mathcal{W}^{-1}\mathcal{T})^t]_{\Omega\Omega'}\tr(\Omega O)\Omega'=\sum_{\Omega\Omega'}[\boldsymbol{\Lambda}^{-1}+\Delta^t\mathcal{W}^{-1}]_{\Omega\Omega'}\tr(\Omega O)\Omega'\\
    &=\sum_{\pi,\sigma}(\boldsymbol{\Lambda}^{-1})_{\pi\sigma}\tr(T_{\pi} O)T_{\sigma}+\sum_{\Omega\Omega'}[\Delta^t\mathcal{W}^{-1}]_{\Omega\Omega'}\tr(\Omega O)\Omega'\\
    &=\Phi_{\haar}^{(k)}(O)+\sum_{\Omega\Omega'}[\Delta^t\mathcal{W}^{-1}]_{\Omega\Omega'}\tr(\Omega O)\Omega'
\end{align}
which concludes the proof. 
\end{proof}
\end{lemma}

\begin{corollary}\label{cor:trivialconvergence}
For any choice of $\mu_1$ for which $\|\Delta\|_{\infty}<1$, then
\be
\lim_{t\rightarrow\infty}\Phi_{t}^{(k)}(O)=\Phi_{\haar}^{(k)}(O).
\ee
\begin{proof}
    The proof immediately follows from \cref{lem:convergenceoftdopedchannel}. 
\end{proof}
\end{corollary}

Now that we have defined the doped-Clifford-Weingarten function, following the same spirit as the Clifford-Weingarten function~\cite{bittel2025completetheorycliffordcommutant}, we aim to study its asymptotic behavior. While a full analytical computation for large $k$ is impractical, we can effectively characterize its behavior in the limit of large $n$.

\subsection{Asymptotics of doped Clifford-Weingarten functions}\label{sec:asymptoticdopedcliffordweingarten}

\begin{lemma}[Properties of $\mathcal{T}$]\label{lem:Tproperties} The $|\mathcal{P}|\times |\mathcal{P}|$ matrix $\mathcal{T}$ with components $\mathcal{T}_{\Omega,\Omega'}\coloneqq\tr(\Omega^{\dag}\kappa(\Omega'))$, obeys the following properties:
\begin{enumerate}[label=(\roman*)]
    \item $\mathcal{T}$ is Hermitian.
    \item $\mathcal{T}_{\Omega,\Omega}=d^{k}\frac{\tr(\omega^{\dag}\kappa(\omega))}{2^k}$, where $\omega$ is defined through $\Omega=\omega^{\otimes n}$.
    \item $\mathcal{T}_{\Omega,\Omega'}\le 2d^{k-1}$ for $\Omega\neq\Omega'$.
\end{enumerate}
\begin{proof}
Item $(i)$ readily follows from the definition. To show $(ii)$ and $(iii)$, we can use the tensor product structure of Pauli monomials see \cref{lem:relevantpropertiespaulimonomials}, i.e.,  $\Omega=\omega^{\otimes n}$. Then
\be
\mathcal{T}_{\Omega,\Omega'}=\tr(\omega^{\dag}\omega')^{n-1}\tr(\omega^{\dag}\kappa(\omega'))=(\mathcal{W}|_{n-1})_{\Omega,\Omega'}\tr(\omega^{\dag}\kappa(\omega'))
\ee
Hence, from \cref{lem:asymptoticweingarten}, we know that $(\mathcal{W}|_{n-1})_{\Omega,\Omega}=(d/2)^k$ and $(\mathcal{W}|_{n-1})_{\Omega,\Omega'}\le(d/2)^{k-1}$. Moreover, it holds that $|\tr(\omega^{\dagger}\kappa(\omega))|\le \|\omega\|_{\infty}\|\kappa(\omega)\|_1\le \|\omega\|_1\le 2^k $, where we have used the data-processing inequality~\cite{nielsen_quantum_2000} in the second-to-last inequality and \cref{lem:relevantpropertiespaulimonomials}.
\end{proof}
\end{lemma}

\begin{lemma}[Properties of $\mathcal{W}^{-1}\mathcal{T}$]\label{lem:propertiesTW} Let $n\ge k^2-3k+13$, and $k\ge 3$. Let $\mathcal{W}^{-1}$ be the matrix of Clifford-Weingarten functions, and consider the matrix $\mathcal{W}^{-1}\mathcal{T}$. Let $\mathcal{L}$ be the matrix defined in \cref{eq:defLmatrix}. Define $\Delta\coloneqq\mathcal{W}^{-1}\mathcal{T}-\mathcal{L}$.  The following facts hold.
\begin{enumerate}[label=(\roman*)]
\item $\Delta_{\Omega,\Omega'}=0$ for $\Omega,\Omega'\in S_k$.
\item $\left|\Delta_{\Omega,\Omega}-\frac{\tr(\omega^{\dag}\kappa(\omega))}{2^k}\right|\le \frac{7|\mathcal{P}|^2}{d}$, for $\Omega\not\in S_k$.
    \item $|\Delta_{\Omega,\Omega'}|\le  \frac{7|\mathcal{P}|^2}{d}$ for $\Omega\neq\Omega'$.
\end{enumerate}
\begin{proof}
First note that $\mathcal{T}_{\pi\Omega}=\mathcal{W}_{\pi\Omega}$ for any $\pi\in S_k$ and $\Omega\in\mathcal{P}$. Hence, item (i) follows immediately by definition. Let us now consider we have $\Delta_{\Omega,\Omega}=(\mathcal{W}^{-1}\mathcal{T})_{\Omega,\Omega}$ for $\Omega\not\in S_k$ and prove item (ii). 
\be
\Delta_{\Omega,\Omega}=(\mathcal{W}^{-1})_{\Omega,\Omega}\mathcal{T}_{\Omega,\Omega}+\sum_{\Omega'\neq \Omega}(\mathcal{W}^{-1})_{\Omega,\Omega'}\mathcal{T}_{\Omega', \Omega}.
\ee
Using \cref{lem:asymptoticweingarten} and \cref{lem:Tproperties}, one can control the second term as 
\be
\left|\sum_{\Omega'\neq \Omega}(\mathcal{W}^{-1})_{\Omega,\Omega'}\mathcal{T}_{\Omega', \Omega}\right|\le |\mathcal{P}|\frac{6|\mathcal{P}|^2}{d^{k+1}}2d^{k-1}=\frac{12|\mathcal{P}|^3}{d^2}.
\ee
It means that we can then use the following chain of inequalities and obtain
\be 
\left|\Delta_{\Omega,\Omega}-\frac{\tr(\omega^{\dag}\kappa(\omega))}{2^k}\right| &\le \left|(\mathcal{W}^{-1})_{\Omega,\Omega}\mathcal{T}_{\Omega,\Omega} - \frac{\tr(\omega^{\dag}\kappa(\omega))}{2^k}\right| + \frac{12|\mathcal{P}|^3}{d^2} \\ 
& = \left|(\mathcal{W}^{-1})_{\Omega,\Omega}d^{k}\frac{\tr(\omega^{\dag}\kappa(\omega))}{2^k} - \frac{\tr(\omega^{\dag}\kappa(\omega))}{2^k}\right| + \frac{12|\mathcal{P}|^3}{d^2} \\ 
& = d^{k}\frac{\tr(\omega^{\dag}\kappa(\omega))}{2^k}\left|\mathcal{W}^{-1}_{\Omega,\Omega} -\frac{1}{d^k}\right|  + \frac{12|\mathcal{P}|^3}{d^2} \\
& \le \frac{6|\mathcal{P}|^2}{d } +  \frac{12|\mathcal{P}|^3}{d^2} \\
& \le \frac{6|\mathcal{P}|^2}{d } +  \frac{12|\mathcal{P}|^3}{d^2} \\
& \le \frac{7|\mathcal{P}|^2}{d},
\ee 
where we have used again \cref{lem:asymptoticweingarten} and \cref{lem:Tproperties}, to prove such a bound, and that $|\mathcal{P}|^2\ge 6|\mathcal{P}|$ for $k\ge 3$, from which item $(ii)$ follows. For item $(iii)$, we can consider $\Delta_{\Omega,\Omega'}$ with $\Omega\neq \Omega'$, $\Omega\in S_k$ and $\Omega'\not\in S_k$ and use the triangle inequality $|\Delta_{\Omega,\Omega'}|\le |(\mathcal{W}^{-1}T)_{\Omega,\Omega'}|+|\mathcal{L}_{\Omega,\Omega'}|$.  Let us bound the two terms separately:
\be\label{eqprova1}
|(\mathcal{W}^{-1}\mathcal T)_{\Omega,\Omega'}|&\le |(\mathcal{W}^{-1})_{\Omega,\Omega}||\mathcal{T}_{\Omega,\Omega'}|+ \left|(\mathcal{W}^{-1})_{\Omega,\Omega'}\right|\left|\mathcal{T}_{\Omega'\Omega'}\right|  +\max_{\Omega}|T_{\Omega,\Omega'}|\sum_{\Omega'\neq \Omega}|(\mathcal{W}^{-1})_{\Omega,\Omega'}|\\ 
&\le \frac{1}{d^k}\left(1+\frac{6|\mathcal{P}^2|}{d}\right)2d^{k-1}+d^{k-1}|\mathcal{P}|\frac{10|\mathcal{P}^2|}{d^{k+1} } + \frac{5 d^k |\mathcal{P}|^2 }{d^{k+1} }\\
\ee
and
\be\label{eqprova2}
|\mathcal{L}_{\Omega,\Omega'}|&\le \left(\max_{\sigma}|\Lambda^{-1}_{\sigma\sigma}|+k!\max_{\sigma\neq \pi}|\Lambda^{-1}_{\sigma\pi}|\right)\max_{\Omega\neq\Omega'}|\mathcal{W}_{\Omega,\Omega'}|\le \left(\frac{1}{d^k}+\frac{k!}{d^{k+1}}\right)d^{k-1}\le \frac{2}{d},
\ee
where we have used the asymptotics of the Weingarten function (see \cref{asymptoticweingarten}) and Clifford-Weingarten functions in \cref{lem:asymptoticweingarten} and that $n\ge k^2-3k+13$ by assumption. Combining \cref{eqprova1} and \cref{eqprova2}, we get
\be
|\Delta_{\Omega,\Omega'}|\le \frac{4}{d}  +  \frac{22|\mathcal{P}^3|}{d^2 } + \frac{ 5|\mathcal{P}^2|}{d} \le  \frac{7|\mathcal{P}|^2}{d}
\ee
where we have made use of the fact that $2|\mathcal{P}|^2\ge 11|\mathcal{P}|+4$ for $k\ge 3$. This concludes the proof.
\end{proof}
\end{lemma}

\begin{theorem}[Asymptotics doped-Clifford-Weingarten functions]\label{th:asymptoticsdopedcliffordweingarten} Assuming that $n\ge\frac{3}{2}(k^2-3k+18+\log\|\Delta\|_{\infty}^{-1})$ and $k\ge 3$, the doped-Clifford-Weingarten functions have the following asymptotics:
\be
|[\Delta^{t}\mathcal{W}^{-1}]_{\Omega,\Omega'}|&\le t\|\Delta\|_{\infty}^{t-1}\frac{18|\mathcal{P}|^2}{d^{k+1}},\quad \Omega,\Omega'\in S_k\\
[\Delta^t\mathcal{W}^{-1}]_{\Omega,\Omega'}&=(\Delta_{\Omega,\Omega})^t(\mathcal{W}^{-1})_{\Omega,\Omega'}\pm t\|\Delta\|_{\infty}^{t-1}\frac{30|\mathcal{P}|^2}{d^{k+1}}\,.
\ee
Moreover, the following facts hold.
\begin{enumerate}[label=(\roman*)]
    \item $\left|(\Delta^{t})_{\Omega,\Omega}-(\Delta_{\Omega,\Omega})^t\right|\le t\frac{7|\mathcal{P}|^3}{d}\|\Delta\|_{\infty}^{t-1}$\,.
    \item $\max |(\Delta^{t})_{\Omega,\Omega'}|\le t\|\Delta\|_{\infty}^{t-1}\frac{7|\mathcal{P}|^2}{d}$ for $\Omega\neq\Omega'$.
\end{enumerate}
\begin{proof}
First, notice that we can write $\mathcal{W}^{-1}\mathcal{T}=\mathcal{L}+\Delta$, and $\mathcal{L}\Delta=0$, while $\mathcal{L}^2=\mathcal{L}$. Therefore, we can express the matrix of the doped-Clifford-Weingarten functions as
\be
(\mathcal{W}^{-1}\mathcal{T})^{t}\mathcal{W}^{-1}=\boldsymbol{\Lambda}^{-1}+\Delta^t\mathcal{W}^{-1}
\ee
where $\boldsymbol{\Lambda}^{-1}$ is only supported on the first $k!\times k!$ block corresponding to permutation. 
Hence we need to characterize $\Delta^t$. Thus, let $\Omega\not\in S_k$. First, the following chain of inequalities holds
\be
(\Delta^t)_{\Omega,\Omega}&\le (\Delta^{t-1})_{\Omega,\Omega}\Delta_{\Omega,\Omega}+\sum_{\Omega'\neq\Omega}|(\Delta^{t-1})_{\Omega,\Omega'}\Delta_{\Omega', \Omega}|\\
&\le (\Delta^{t-1})_{\Omega,\Omega}\Delta_{\Omega,\Omega}+|\mathcal{P}|\max_{\Omega\neq\Omega'}|\Delta_{\Omega,\Omega'}|\|\Delta\|_{\infty}^{t-1}\\
&\le (\Delta^{t-2}\Delta_{\Omega,\Omega}+|\mathcal{P}|\max_{\Omega\neq\Omega'}|\Delta_{\Omega,\Omega'}|\|\Delta\|_{\infty}^{t-2})_{\Omega,\Omega}\Delta_{\Omega,\Omega}+|\mathcal{P}|\max_{\Omega\neq\Omega'}|\Delta_{\Omega,\Omega'}|\|\Delta\|_{\infty}^{t-1}\\
&=(\Delta^{t-2})_{\Omega,\Omega}\Delta_{\Omega,\Omega}^2+2|\mathcal{P}|\max_{\Omega\neq\Omega'}|\Delta_{\Omega,\Omega'}|\|\Delta\|_{\infty}^{t-1}.
\ee
Iterating these inequalities and upper bounding $\max_{\Omega\neq\Omega'}|\Delta_{\Omega,\Omega'}|\le \frac{7|\mathcal{P}^2|}{d}$ as in \cref{lem:propertiesTW}, we finally get
\be
\left|(\Delta^{t})_{\Omega,\Omega}-(\Delta_{\Omega,\Omega})^t\right|\le t\frac{7|\mathcal{P}|^3}{d}\|\Delta\|_{\infty}^{t-1}\,.
\ee
While for $\Omega\neq\Omega'$, we have
\be
(\Delta^{t})_{\Omega,\Omega'}&\le (\Delta^{t-1})_{\Omega,\Omega}\max_{\Omega\neq\Omega'}|\Delta_{\Omega,\Omega'}|+\max_{\Omega\neq\Omega'}|(\Delta^{t-1})_{\Omega,\Omega'}|\sum_{\Omega''\neq\Omega}|\Delta_{\Omega''\Omega'}|\\
&\le \|\Delta\|_{\infty}^{t-1}\max_{\Omega\neq\Omega'}|\Delta_{\Omega,\Omega'}|+\max_{\Omega\neq\Omega'}|(\Delta^{t-1})_{\Omega,\Omega'}|(\|\Delta\|_{\infty}+|\mathcal{P}|\max|\Delta_{\Omega,\Omega'}|).
\ee
Now, let us assume for simplicity that
\be
(\|\Delta\|_{\infty}+|\mathcal{P}|\max|\Delta_{\Omega,\Omega'}|)\le 1,
\ee
which holds whenever $n\ge \frac{3}{2}(k^2-3k+18+\log\|\Delta\|_{\infty}^{-1})$. 
Since it works for any $\Omega\neq\Omega'$, we have the  recurrence relation
\be
\max_{\Omega\neq\Omega'} |(\Delta^{t})_{\Omega,\Omega'}|\le\max_{\Omega\neq\Omega'} |(\Delta^{t-1})_{\Omega,\Omega'}|+\|\Delta\|_{\infty}^{t-1}\max_{\Omega\neq\Omega'}|\Delta_{\Omega,\Omega'}|.
\ee
This gives
\be
\max |(\Delta^{t})_{\Omega,\Omega'}|\le t\|\Delta\|_{\infty}^{t-1}\max_{\Omega\neq\Omega'}|\Delta_{\Omega,\Omega'}|\le  t\|\Delta\|_{\infty}^{t-1}\frac{7|\mathcal{P}|^2}{d},
\ee
where we have used \cref{lem:propertiesTW}.
Now, putting all together, we can prove the result. Notice that $\Delta$ restricted to the first $k!\times k!$ is null. Hence, for $\Omega,\Omega'\in S_k$, we have
\be
|[\Delta^{t}\mathcal{W}^{-1}]_{\Omega,\Omega'}|=\sum_{\Omega''\not \in S_k}(\Delta^{t})_{\Omega\Omega''}(\mathcal{W}^{-1})_{\Omega''\Omega'}\le t\|\Delta\|_{\infty}^{t-1}\frac{49|\mathcal{P}|^5}{d^{k+2}} \le t\|\Delta\|_{\infty}^{t-1}\frac{18|\mathcal{P}|^2}{d^{k+1}}
\ee
where we have used that $n\ge \frac{3}{2}(k^2-3k+13)$. While for $\Omega\not\in S_k$ or $\Omega'\not\in S_k$, we find
\be
[\Delta^{t}\mathcal{W}^{-1}]_{\Omega,\Omega'}&=\sum_{\Omega''}(\Delta^t)_{\Omega\Omega''}(\mathcal{W}^{-1})_{\Omega''\Omega'}
\\
&\le(\Delta_{\Omega,\Omega}^{t})(\mathcal{W}^{-1})_{\Omega,\Omega'}+(\Delta_{\Omega,\Omega'}^{t})(\mathcal{W}^{-1})_{\Omega'\Omega'}+|\mathcal{P}|\max_{\Omega\neq\Omega'}|\mathcal{W}^{-1}|\max_{\Omega\neq\Omega'}\left|\Delta_{\Omega,\Omega'}^{t}\right|\\
&\le \|\Delta\|_{\infty}^t(\mathcal{W}^{-1})_{\Omega,\Omega'}+ t\frac{7|\mathcal{P}|^2}{d^{k+1} }\|\Delta\|_{\infty}^{t-1}\left(1+\frac{6|\mathcal P|^2}{d}\right) + t\frac{49|\mathcal{P}|^5}{d^{k+2} }\|\Delta\|_{\infty}^{t-1} \\
&\le \|\Delta\|_{\infty}^t(\mathcal{W}^{-1})_{\Omega,\Omega'}+ t\frac{7|\mathcal{P}|^2}{d^{k+1} }\|\Delta\|_{\infty}^{t-1}+t\frac{42|\mathcal P|^4}{d^{k+2} }\|\Delta\|_{\infty}^{t-1} + t\frac{18|\mathcal{P}|^2}{d^{k+1} }\|\Delta\|_{\infty}^{t-1} \\
&\le \|\Delta\|_{\infty}^t(\mathcal{W}^{-1})_{\Omega,\Omega'}+ t\frac{25|\mathcal{P}|^2}{d^{k+1} }\|\Delta\|_{\infty}^{t-1}+ t\frac{15|\mathcal{P}|}{d^{k+1} }\|\Delta\|_{\infty}^{t-1} \\
&\le \|\Delta\|_{\infty}^t(\mathcal{W}^{-1})_{\Omega,\Omega'}+ t\frac{30|\mathcal{P}|^2}{d^{k+1} }\|\Delta\|_{\infty}^{t-1},
\ee
where we have used that $|\mathcal{P}|^2\ge 3|\mathcal{P}|$ for $k\ge 3$ and used that $n\ge 3/2(k^2-3k+18)$. This concludes the proof.
\end{proof}
\end{theorem}

Let us informally describe the content of \cref{th:asymptoticsdopedcliffordweingarten}. 
On the one hand, \cref{cor:trivialconvergence} states, quite trivially, that if we dope the Clifford group with any measure described by a perturbation matrix $\Delta$ whose operator norm is strictly less than unity, then in the limit of infinite doping, the twirling operator converges to the one over the full unitary group. This is unsurprising, as the Clifford group doped with any gate outside the Clifford group becomes universal. However, this fact provides the confirmation that, among all the valid measures we could have chosen for $t$-doped Clifford circuits, we selected the one that interpolates between the uniform (Haar) measure over the Clifford group and the uniform (Haar) measure over the full unitary group.
 
On the other hand, \cref{th:asymptoticsdopedcliffordweingarten} provides a more fine-grained structure for the twirling over the doped Clifford group. Specifically, it explicitly separates the dependence on the full unitary group twirling from the doped contributions, which eventually decay as $t$ increases. Informally, we can express the doped-Clifford twirling as
\be\label{eq:informaltwirling}
\Phi_{t}^{(k)}(O) \simeq \Phi_{\haar}^{(k)}(O) + \sum_{\Omega,  \Omega'} (\Delta_{\Omega , \Omega})^t (\mathcal{W}^{-1})_{\Omega,  \Omega'} \tr(\Omega O) \Omega',
\ee
where the Haar twirling appears on the left, and the Clifford twirling on the right, with each $\Omega \not\in S_k$ term being suppressed by a factor of $(\Delta_{\Omega , \Omega})^t\simeq (2^{-k}\|\kappa(\omega)\|_{2}^{2})^{t}$ (see \cref{th:asymptoticsdopedcliffordweingarten}). Naturally, for every $\Omega \not\in S_k$, we have $(\Delta_{\Omega , \Omega})^t \leq \|\Delta\|_{\infty}^t$, which indicates that the largest coefficient is suppressed by the operator norm of the perturbation matrix $\Delta$. 

This finer structural understanding is useful for determining the amount of doping necessary and sufficient to approximate the twirling operator applied to a given operator $O$ of interest over the full unitary group using the twirling operator over the doped Clifford group. As an application of the tools presented above, we analyze this convergence in the next section. 

\section{Convergence of random doped Clifford circuits}\label{sec:mainresults}
In this section, we establish the main result of this work in two complementary ways: first, by analyzing the convergence of the frame potential in \cref{def:framepotential}; as a corollary we provide improved convergence bound of $t$-doped Clifford circuits towards state $k$-designs; second, we analyze the convergence towards relative $\varepsilon$-approximate $k$-design, by providing tight bounds. 

\subsection{Optimal frame potentials convergence}

In this section, we prove tight convergence bounds to the frame potential of $t$-doped Clifford circuits. To this end, we begin by proving some lemmas that are instrumental in deriving the main result.

As a careful reader may infer from \cref{eq:informaltwirling}, the convergence of doped random Clifford circuits towards the Haar measure is governed by the matrix $\Delta$. Specifically, it depends on the sum $\sum_{\Omega}(\Delta_{\Omega,\Omega})^t$. In the following key lemma, we establish a lower bound for this coefficient, which will, in turn, provide upper and lower bounds on the convergence rate of the frame potential. As we will see, such a lower bound depends on the spectral form factor of the doping ensemble $\int\de\mu_1(U)|\tr(U)|^{2k}$, which we characterize later for diagonal single qubit unitaries, as well as Haar random single qubit unitaries.

\begin{lemma}[Lower bounding convergence via spectral form factor]\label{lem:lowerbound} Let $n\ge 3/2(k^{2}-3k + 18)$ and $k\ge 4$. Let $\Delta\coloneqq\mathcal{W}^{-1}\mathcal{T}-\mathcal{L}$, where the matrices $\mathcal{W},\mathcal{T},\mathcal{L}$ are defined in \cref{sec:dopedcliffordhaaraverage}. Then the  bound 
\be
\sum_{\Omega}(\Delta_{\Omega,\Omega})^t\ge(|\mathcal{P}|-k!)\left(\frac{1}{3}\int\de\mu_1(U)\frac{|\tr(U)|^{2k}}{4^k}-\frac{2}{|\mathcal{P}|^{1/10}}\right)^t
\ee
holds.
    \begin{proof}   
First notice that $\Delta_{\Omega,\Omega}=0$ for every $\Omega\in S_k$, i.e.,  for any permutation. Hence, we can restrict the sum over $\Omega\not\in S_k$. Using the asymptotics in \cref{lem:propertiesTW}, we can write the following
       \be
    \sum_{\Omega\not\in S_k}(\Delta_{\Omega,\Omega})^t&\ge \sum_{\Omega\not\in S_k}\left(\frac{\tr(\omega^{\dag}\kappa(\omega))}{2^k}-\frac{7|\mathcal{P}|^2}{d}\right)^t\ge (|\mathcal{P}|-k!)\left(\frac{1}{|\mathcal{P}|-k!}\sum_{\Omega\not\in S_k}\frac{\tr(\omega^{\dag}\kappa(\omega))}{2^k}-\frac{7|\mathcal{P}|^2}{d}\right)^t
    \ee
where we have used Jensen's inequality. We can now add and subtract all the remaining contributions coming from permutations, which leads to
\be
\sum_{\Omega\not\in S_k}(\Delta_{\Omega,\Omega})^t\ge (|\mathcal{P}|-k!)\left(\frac{1}{|\mathcal{P}|-k!}\sum_{\Omega}\frac{\tr(\omega^{\dag}\kappa(\omega))}{2^k}-\frac{k!}{|\mathcal{P}|-k!}-\frac{7|\mathcal{P}|^2}{d}\right)^t.
\ee
We now focus our attention on the first term. First, by \cref{lem:relevantpropertiespaulimonomials} we note that $\omega^{\dag}=\omega^{T}$. Hence, we have the following identity
\be
\sum_{\omega}\frac{\tr(\omega^{\dag}\kappa(\omega))}{2^k}=\sum_{\omega}\tr(\phi_{+}^{\otimes k}\mathbb{1}\otimes \kappa[\omega^{\otimes 2}])=Z_{2}\mathbb{E}_{\sigma}\tr(\phi_{+}^{\otimes k}\mathbb{1}\otimes \kappa[\sigma^{\otimes k}])
\ee
where $\phi_{+}$ is the Bell state between  two copies of a single qubit. The second equality comes from \cref{lem:averagestabstate}, i.e., given a  $2$-qubit stabilizer state $\sigma\in\stab_2$, then $\mathbb{E}_{\sigma}\sigma^{\otimes k}=\frac{1}{Z_2}\sum_{\omega}\omega^{\otimes 2}$, where $Z_2=4\prod_{i=0}^{k-2}(2^i+4)$. Noticing that $\phi_{+}\in\stab_2$, to simplify calculation, we can trivially lower bound the average as
\be
\sum_{\omega}\frac{\tr(\omega^{\dag}\kappa(\omega))}{2^k}&\ge \frac{Z_2}{|\stab_2|}\tr(\phi_{+}^{\otimes k}\mathbb{1}\otimes\kappa[\phi_{+}^{\otimes k}])=\frac{Z_2}{|\stab_2|}\int\de\mu_1(U)\tr(\phi_{+}\mathbb{1}\otimes U\phi_{+}\mathbb{1}\otimes U^{\dag})^k\\&=\frac{Z_2}{|\stab_2|}\int\de\mu_1(U)\frac{|\tr(U)|^{2k}}{4^k}.
\ee
Therefore, the convergence is effectively dictated by the spectral form factor of the ensemble described by the measure $\mu_1$. Before plugging everything together, notice that $|\stab_2|=60$ and that
\be
\frac{Z_2}{|\mathcal{P}|-k!}\ge \frac{Z_2}{|\mathcal{P}|}=4\prod_{i=0}\frac{2^i+4}{2^i+1}=60\frac{4^{k+1}}{(2^k+8)(2^k+4)}.
\ee
Then also notice that 
\begin{equation}
\frac{4^{k+1}}{(2^k+8)(2^k+4)}\ge\frac{1}{3}
\end{equation}
for $k\ge 2$. Lastly, we can bound $\frac{k!}{|\mathcal{P}|-k!}\le\frac{1}{|\mathcal{P}|^{1/10}}$ for every $k\ge 4$.  Finally, we have that $\frac{7|\mathcal{P}|^2}{d}\le \frac{1}{\vert \mathcal P \vert}$ for $n\ge 3/2(k^{2}-3k + 18)$.
Therefore, in summary, we have the following lower bound
\be
\sum_{\Omega\neq S_k}(\Delta_{\Omega,\Omega})^t\ge (|\mathcal{P}|-k!)\left(\frac{1}{3}\int\de\mu_1(U)\frac{|\tr(U)|^{2k}}{4^k}-\frac{2}{|\mathcal{P}|^{1/10}}\right)^t.
\ee
The above equation tells that whenever the average spectral form factor is only polynomially small in $k$, then the convergence is quadratic $t\sim k^2$. 
    \end{proof}
\end{lemma}
In the following lemma, we lower bound spectral form factors for two interesting ensembles: the diagonal ensemble and the Haar random ensemble.
\begin{lemma}[Lower bounding spectral form factors]\label{lem:spectralformfactor} The following lower bounds for the spectral form factors of 
hold for different ensembles.
    \be
&\int\de\mu_1(U)|\tr(U)|^{2k}\ge \frac{2\sqrt{\pi}}{e^2\sqrt{k}},\quad\text{diagonal ensemble}.\\
&\int\de\mu_1(U)|\tr(U)|^{2k}\ge \frac{2\sqrt{\pi}}{e^2\sqrt{k}(k+1)},\quad\text{Haar ensemble}.
    \ee
\begin{proof}
First, notice that the spectral form factor only depends on the difference between the eigenphases
as
\be
|\tr(U)|^2=|1+e^{i(\theta_1-\theta_2)}|=(1+e^{i(\theta_1-\theta_2)})(1+e^{-i\phi})=2+2\cos(\theta_1-\theta_2)=4\cos^2\left(\frac{\theta_1-\theta_2}{2}\right).
\ee
Let us begin when $\mu_1$ is the uniform distribution over diagonal unitary matrices,
\be
\int\de\mu_1(U)\frac{|\tr(U)|^{2k}}{4^k}=\int_{0}^{2\pi}\de\theta\,\cos^{2k}\frac{\theta}{2}=\frac{(2k)!}{(k!)^24^k}\ge\frac{2\sqrt{\pi}}{e^2\sqrt{k}},
\ee
where, for the lower bound, we have used the following on the factorial $\sqrt{2\pi} k^{k+\frac{1}{2}}e^{-k}\le k!\le e k^{k+\frac{1}{2}}e^{-k}$. For the Haar ensemble, we can use the famous eigenvalue probability density on the difference between the eigenphases $p_{\haar}(\theta)=\frac{1}{2\pi}(1-\cos(\theta))$. Therefore, in this case
\be
\int_{0}^{2\pi}\de\theta p_{\haar}(\theta)\cos^{2k}\frac{\theta}{2}=\frac{1}{2\pi}\int_{0}^{2\pi}\de\theta\,(1-\cos(x)) \cos^{2k}\frac{\theta}{2}=\frac{(2k)!}{(k!)^24^k(k+1)}\ge\frac{2\sqrt{\pi}}{e^2\sqrt{k}(k+1)}.
\ee
\end{proof}

\end{lemma}
Therefore, the spectral form factors for the diagonal ensemble and for single-qubit Haar random gates are both polynomially small in \( k \). Consequently, they lead to the same convergence result. Below, we therefore analyze only the case of random diagonal single-qubit gates. The analysis for single-qubit Haar random gates is analogous and, most importantly, leads to the same conclusions.

\begin{corollary}\label{cor:lowerbounddiagonal}
Let $n\ge 3/2(k^{2}-3k + 18)$ and $k\ge 9$. The following bound holds for the diagonal unitary ensemble:
    \be
\sum_{\Omega}(\Delta_{\Omega,\Omega})^t\ge \left(\frac{2\sqrt{\pi}}{3e^2\sqrt{k}}\right)^{t}.
    \ee
\begin{proof}
    By \cref{lem:lowerbound} and \cref{lem:spectralformfactor}, we have
    \be
\sum_{\Omega}(\Delta_{\Omega,\Omega})^t\ge(|\mathcal{P}|-k!)\left(\frac{2\sqrt{\pi}}{3e^2\sqrt{k}}-\frac{2}{|\mathcal{P}|^{1/10}}\right).
    \ee
By \cref{lem:relevantpropertiespaulimonomials}, we have that for any $k\ge 9$, we have that $\frac{2}{|\mathcal{P}|^{1/10}}\le \frac{\sqrt{\pi}}{3e^2\sqrt{k}}$, therefore,  the result follows.
\end{proof}
\end{corollary}

The following lemma establishes upper bounds on the coefficient $\sum_{\Omega}(\Delta_{\Omega,\Omega})^{t}$.
\begin{lemma}[Upper bound~\cite{haferkamp_random_2022}]\label{lem:upperbound} Let $n\ge (k^2-3k+26)$. Let $\mu_1$ be the measure over single qubit diagonal or Haar random unitaries. The following upper bound holds 
\be
\sum_{\Omega}(\Delta_{\Omega,\Omega})^t\le (|\mathcal{P}|-k!)\left(\frac{15}{16}\right)^t .
\ee
\begin{proof}
    Using the asymptotics in \cref{lem:propertiesTW}, we can write
    \be
\sum_{\Omega\not\in S_k}(\Delta_{\Omega,\Omega})^{t}\le \sum_{\omega\not\in S_k}\left(\frac{\tr(\omega^{\dag}\kappa(\omega))}{2^k}+\frac{7|\mathcal{P}|^2}{d}\right)^t\le(|\mathcal{P}|-k!)\left(\max_{\omega\not\in S_k}\frac{\tr(\omega^{\dag}\kappa(\omega))}{2^k}+\frac{7|\mathcal{P}|^2}{d}\right)^t
    \ee
For $\mu_1(U)$ be the measure over the diagonal unitary ensemble, then in Ref.~\cite{haferkamp_random_2022}, the authors have proven that $\max_{\omega\not\in S_k}\frac{\tr(\omega^{\dag}\kappa(\omega))}{2^k}\le \frac{7}{8}$. Moreover, as also noted in Ref.~\cite{haferkamp_random_2022}, the same bound hold for the Haar random ensemble. Moreover we have that $\frac{7|\mathcal{P}|^2}{d}\le \frac{1}{16}$ for every $k\ge 1$ thanks to the hypothesis of the lemma. Therefore, we finally have
\be
\sum_{\Omega\not\in S_k}(\Delta_{\Omega,\Omega})^{t}\le (|\mathcal{P}|-k!)\left(\frac{15}{16}\right)^t
\ee
\end{proof}
    
\end{lemma}

Now we established the key lemmas to show the convergence of doped random circuit in multiple flavours. Let us establish the optimal convergence in frame potential in terms of the coefficient of \cref{lem:lowerbound,lem:upperbound}. 

\begin{theorem}[Frame potential convergence]\label{th:framepotentialconvergence} Let $\mu_1$ be the measure over the diagonal unitary ensemble. Let $n\ge f(k,t)$ where $f(k,t)\coloneqq\frac{7}{4}k^2-6k+19+3t+\log t+t\log k$, and $k\ge 9$. The following upper and lower bounds hold for the difference of the frame potentials
\be
2^{\frac{k^2}{4}-6t-t\log k-1} \le \mathcal{F}_{t}^{(k)}-\mathcal{F}_{\haar}\le 2^{\frac{k^2}{2}-2t\log \frac{16}{15}+1}.
\ee
\begin{proof}
Let us first recall the definition of frame potential, specifically for the doped random Clifford ensemble $\mathcal{C}_t$, described by the measure $\mu_t$ introduced in \cref{def:dopedrandomclifford}:
\be
\mathcal{F}_{t}^{(k)}=\int\de\mu_t(U)\de\mu_t(V)|\tr(U^{\dag}V)|^{2k}=\int\de\mu_{2t}(U)|\tr(U)|^{2k}
\ee
i.e.,  it reduces to the average spectral form factor of the ensemble $\mathcal{C}_{2t}$, where the amount of doping is doubled. We can therefore express it as
\be
\mathcal{F}_{t}^{(k)}=\int\de\mu_{2t}(U)\tr(U^{\otimes k}\otimes U^{\dag\otimes k})=\tr(\Gamma_{2t}^{(k)})
\ee
through the frame operator $\Gamma_{2t}^{(k)}\coloneqq\int\de\mu_t(U) U^{\otimes k}\otimes U^{\dag\otimes k}$. From \cref{lem:convergenceoftdopedchannel}, we can express the frame operator as
\be
\Gamma_{2t}^{(k)}=\Gamma_{\haar}^{(k)}+\sum_{\Omega,\Omega'}[\Delta^{t}\mathcal{W}^{-1}]_{\Omega,\Omega'}\hat{T}\Omega^{\dag}\otimes\Omega'
\ee
where $\hat{T}\prod_{j=1}^{k}T_{k,k+j}$ and $\Gamma_{\haar}^{(k)}=\int\de U\,U^{\otimes k}\otimes U^{\dag\otimes k}$. Given that $\mathcal{F}_t^{(k)}=\tr(\Gamma_{2t}^{(k)})$, below we compute $\tr(\Gamma_{t}^{(k)})$ for simplicity
    \be
\mathcal{F}_{t/2}^{(k)}&=\mathcal{F}_{\haar}^{(k)}+\tr(\hat{T}\sum_{\Omega,\Omega'}[\Delta^{t}\mathcal{W}^{-1}]_{\Omega,\Omega'}\Omega^{\dag}\otimes \Omega')\\
&=\mathcal{F}_{\haar}^{(k)}+\sum_{\Omega,\Omega'}[\Delta^{t}\mathcal{W}^{-1}]_{\Omega,\Omega'}\tr(\hat{T}\Omega^{\dag}\otimes \Omega')\\
&=\mathcal{F}_{\haar}^{(k)}+\sum_{\Omega,\Omega'}[\Delta^{t}\mathcal{W}^{-1}]_{\Omega,\Omega'}\mathcal{W}_{\Omega', \Omega}\\
&=\mathcal{F}_{\haar}^{(k)}+\sum_{\Omega}(\Delta^{t})_{\Omega,\Omega}.
    \ee
Now, by assumption $n\ge f(k,t/2)$ above. In order to apply \cref{th:asymptoticsdopedcliffordweingarten}, a sufficient condition is that $\frac{3}{2}(k^2-3k+18+\log\|\Delta\|_{\infty}^{-1})$. By \cref{lem:upperbound}, if $n\ge (k^2-3k+26)$ then for the diagonal unitary ensemble $\|\Delta\|_{\infty}\le \frac{15}{16}$. Hence, a sufficient condition for applying \cref{th:asymptoticsdopedcliffordweingarten} in this case is $n\ge\frac{3}{2}(k^2-3k+26+\log \frac{16}{15})$. Note that the condition $n\ge f(k,t/2)$ ensures $n\ge\frac{3}{2}(k^2-3k+26+\log \frac{16}{15})$ for any $t,k\ge 1$. Applying \cref{th:asymptoticsdopedcliffordweingarten}, we can express $\sum_{\Omega}(\Delta^t)_{\Omega,\Omega}=\sum_{\Omega}(\Delta_{\Omega,\Omega})^t\pm \frac{7|\mathcal{P}|^4}{d}t\|\Delta\|_{\infty}^{t-1}$. Now, applying \cref{lem:lowerbound,cor:lowerbounddiagonal}, we have
\be
\sum_{\Omega}(\Delta^{t})_{\Omega}\ge  (|\mathcal{P}|-k!)\left(\frac{2\sqrt{\pi}}{3e^2\sqrt{k}}\right)^{t}-\frac{7|\mathcal{P}|^4}{d}t\|\Delta\|_{\infty}^{t-1}.
\ee
To derive a nontrivial lower bound we require that $\frac{7|\mathcal{P}|^4}{d}t\|\Delta\|_{\infty}^{t-1}\le \frac{1}{2}(|\mathcal{P}|-k!)\left(\frac{2\sqrt{\pi}}{3e^2\sqrt{k}}\right)^{t}$. A sufficient condition to ensure it is given by $n\ge f(k,t)$, which explains the reason behind this assumption. To see this, we require $(|\mathcal{P}|-k!)\ge 2^{\frac{k^2}{4}}$ valid for any $k\ge 4$, $|\mathcal{P}|\le 2^{\frac{1}{2}(k^2-3k+6)}$, and $\|\Delta\|_{\infty}\le \frac{15}{16}$, then we require
\be
2^{2(k^2-3k+6)-n+\log 7-(t-1)\log \frac{16}{15}+\log t}\le 2^{\frac{k^2}{4}-t\log_{2}\frac{3e^2\sqrt{k}}{2\sqrt{\pi}}-1}
\ee
from which the condition follows. Hence $\sum_{\Omega}(\Delta^t)_{\Omega,\Omega}\ge 2^{\frac{k^2}{4}-3t-\frac{t}{2}\log k-1}$. The upper bound follows easily. Indeed, we have
\be
\sum_{\Omega,\Omega'}(\Delta^t)_{\Omega,\Omega'}&\le (|\mathcal{P}|-k!)\left(\frac{15}{16}\right)^t+\frac{7|\mathcal{P}|^4}{d}t\left(\frac{15}{16}\right)^{t-1}\\
&\le(|\mathcal{P}|-k!)\left(\frac{15}{16}\right)^t+\frac{1}{2}(|\mathcal{P}|-k!)\left(\frac{2\sqrt{\pi}}{3e^2\sqrt{k}}\right)^{t}\\
&\le2^{\frac{k^2}{2}-t\log \frac{16}{15}}+ 2^{k^2/2-2t-\frac{t}{2}\log k}\le 2^{\frac{k^2}{2}-t\log \frac{16}{15}+1},
\ee
where we have used the condition $n\ge f(k,t/2)$. We are just left to send $t\mapsto 2t$. This concludes the proof.
\end{proof}
\end{theorem}

\begin{remark}
The bounds appearing in \cref{th:framepotentialconvergence} are obtained via a fully analytical, worst-case analysis and should be interpreted as sufficient guarantees rather than tight conditions. In particular, the constraint on the function $f(n,k)$ is likely an artifact of the techniques used to control the non-Haar contributions within the Clifford commutant, and could potentially be relaxed by a more refined analysis. While these sufficient conditions may place the theorem outside the range of current exact classical simulations, we nonetheless expect that the qualitative conclusion of \cref{th:framepotentialconvergence} can be verified numerically for much smaller system sizes.
\end{remark}

The above theorem constitutes one of the main results of this work. In particular, it shows that a necessary and sufficient condition to have the frame potential difference to be smaller than $\varepsilon$ is having the amount of doping to scale \textit{quadratically} with respect to $k$, i.e.,  $t=\tilde{\Theta}(k^2+\log\varepsilon^{-1})$.

\begin{lemma}[Relative state frame potentials and frame potentials are related]\label{lem:relativeframepotentialandframepotentialtdoped} Let $\mathcal{C}_t$ be the $t$-doped Clifford ensemble. Let $\mathcal{S}_t=\{C_t\ket{0}\,:\, C\in\mathcal{C}_t\}$ the ensemble of states induced by $\mathcal{C}_t$. Then it holds that
\be\label{eq:lemmarelativeandframepotentialrelation1}
\mathcal{R}_{\mathcal{S}_t}=\frac{\tr(\Pi_{\sym})}{Z_n}(\mathcal{F}_{t}-\mathcal{F}_{\haar})\,+\frac{\tr(\Pi_{\sym})}{Z_n}\sum_{\Omega\neq\Omega'}(\Delta^{2t})_{\Omega', \Omega}
\ee
where $Z_n=d\prod_{i=0}^{k-2}(d+2^i)$. Moreover, in the hypothesis of \cref{th:framepotentialconvergence}, i.e.,  $n\ge f(k,t)$, then 
\be\label{eq:lemmarelativeandframepotentialrelation2}
\left(\frac{1}{2}-c\right)\frac{\mathcal{F}_{t}^{(k)}-\mathcal{F}_{\haar}}{\mathcal{F}_{\haar}}\le \mathcal{R}_{\mathcal{S}_t}\le \frac{3}{2}\frac{\mathcal{F}_{t}^{(k)}-\mathcal{F}_{\haar}}{\mathcal{F}_{\haar}}
\ee
where $c\le 2^{-f(k,t)+2k}$ for every $t$.
\begin{proof}
Let us denote as $\Psi_{t}^{(k)}$ be the average state of the ensemble $\mathcal{S}_{t}$. Clearly $\Psi_{t}^{(k)}=\Phi_{t}^{(k)}(\sigma^{\otimes k})$ for any $\sigma\in\stab_n$. From \cref{sec:projectivedesigns}, we know that the state frame potential is simply given by the purity of $\Psi_{t}^{(k)}$, which is given by $\pur(\Psi_{t}^{(k)})=\tr(\sigma^{\otimes k}\Phi_{2t}(\sigma^{\otimes k}))$, since the twirling channel $\Phi_{t}(\cdot)$ is self-adjoint. However, it also holds that $\tr(\sigma^{\otimes k}\Phi_{2t}(\sigma^{\otimes k}))=\mathbb{E}_{\sigma}\tr(\sigma^{\otimes k}\Phi_{2t}(\sigma^{\otimes k}))=\frac{1}{Z_n}\sum_{\Omega}\tr(\Omega\Phi_{2t}(\sigma^{\otimes k}))$, see \cref{lem:averagestabstate}. From \cref{lem:relevantpropertiespaulimonomials}, we know that $\tr(\Omega\sigma^{\otimes k})=1$ for any $\sigma\in\stab_n$ and $\Omega\in\mathcal{P}$. Hence, from \cref{lem:convergenceoftdopedchannel}, we have
\be
\pur(\Psi_{t}^{(k)})&=\frac{1}{Z_n}\sum_{\Omega}\tr(\Omega\Phi_{2t}(\sigma^{\otimes k}))=\pur(\Psi_{\haar}^{(k)})+\frac{1}{Z_n}\sum_{\Omega,\Omega',\Omega''}(\Delta^{2t}\mathcal{W}^{-1})_{\Omega',\Omega''}\mathcal{W}_{\Omega'',\Omega}\\
&=\pur(\Psi_{\haar}^{(k)})+\frac{1}{Z_n}\sum_{\Omega,\Omega',\Omega'',\Omega'''}(\Delta^{2t})_{\Omega',\Omega'''}(\mathcal{W}^{-1})_{\Omega''',\Omega''}\mathcal{W}_{\Omega'',\Omega}=\pur(\Psi_{\haar}^{(k)})+\frac{1}{Z_n}\sum_{\Omega,\Omega',\Omega'''}(\Delta^{2t})_{\Omega',\Omega'''}\delta_{\Omega''',\Omega}\\
&=\pur(\Psi_{\haar}^{(k)})+\frac{1}{Z_n}\sum_{\Omega,\Omega'}(\Delta^{2t})_{\Omega', \Omega}=\pur(\Psi_{\haar}^{(k)})+\frac{1}{Z_n}\sum_{\Omega}(\Delta^{2t})_{\Omega,\Omega}+\sum_{\Omega\neq\Omega'}(\Delta^{2t})_{\Omega', \Omega}\\
&=\pur(\Psi_{\haar}^{(k)})+\frac{1}{Z_n}(\mathcal{F}_{t}^{(k)}-\mathcal{F}_{\haar}^{(k)})+\sum_{\Omega\neq\Omega'}(\Delta^{2t})_{\Omega', \Omega}.
\ee
Noticing that $\pur(\Psi_{\haar}^{(k)})=\tr^{-1}(\Pi_{\sym})$, and using \cref{def:relativeframepotential}, \cref{eq:lemmarelativeandframepotentialrelation1} follows. Let us now prove \cref{eq:lemmarelativeandframepotentialrelation2}. First of all, the following bounds hold
\be
\frac{1}{k!}-\frac{4^k}{dk!}\le \frac{\tr(\Pi_{\sym})}{Z_n}\le \frac{1}{k!}\,.
\ee
The lower bound follows from $\frac{Z_n}{d^k}\le 1+\frac{4^k}{d}$ in the hypothesis of \cref{th:framepotentialconvergence} and that $\frac{1}{1+x}\ge 1-x$. The upper bound follows by noticing that $\frac{\tr(\Pi_{\sym})}{Z_n}$ is monotonically increasing with $d$, and therefore, 
\be
\frac{\tr(\Pi_{\sym})}{Z_n}\le \lim_{d}\frac{\tr(\Pi_{\sym})}{Z_n}=\frac{1}{k!}. 
\ee
Moreover, applying \cref{th:asymptoticsdopedcliffordweingarten}, we find $\sum_{\Omega\neq\Omega'}(\Delta^{2t})_{\Omega', \Omega}\le \frac{7|\mathcal{P}|^4}{d}t\|\Delta\|^{t-1}_{\infty}$ and, in the hypothesis of  \cref{th:framepotentialconvergence}, we have $\frac{7|\mathcal{P}|^4}{d}t\|\Delta\|^{t-1}_{\infty}\le \frac{1}{2}(\mathcal{F}_{t}^{(k)}-\mathcal{F}_{\haar})$. 
\end{proof}
    
\end{lemma}

\begin{corollary}[State $k$-design convergence]\label{cor:statekdesignconvergence} Let $\mathcal{S}_t$ be the ensemble of states induced by $\mathcal{C}_t$. Then for $n\ge f(k,t)$ shown in \cref{th:framepotentialconvergence} the bound 
\be
\|\Psi_{\haar}^{(k)}-\Psi_{t}^{(k)}\|_{1}\le 2^{\frac{k^2}{4}-t\log \frac{16}{15}+\frac{1}{2}\log 3}
\ee
holds. Hence, $t=\Omega(k^2+\log\varepsilon^{-1})$ are sufficient for $\mathcal{S}_t$ to be an $\varepsilon$-approximate state $k$-design in trace distance. Similarly, 
\be
\frac{2^{\frac{k^2}{8}-3t-\frac{t}{2}\log k-2}}{\tr(\Pi_{\sym})}\le\|\Psi_{\haar}^{(k)}-\Psi_{t}^{(k)}\|_{\infty},
\ee
and hence $t=\Omega(k^2/\log k+\log\varepsilon^{-1})$ are necessary for $\mathcal{S}_t$ to be a relative $\varepsilon$-approximate state $k$-design.
\begin{proof}
    The proof follows from \cref{eq:bound1stateframepotential}, \cref{lem:relativeframepotentialandframepotentialtdoped,th:framepotentialconvergence}.
\end{proof}

\end{corollary}

\begin{corollary}[Scrambling of doped Clifford circuits]\label{cor:scrambling} Let $\operatorname{OTOC}_k(U)\coloneqq\mathbb{E}_{A_1,\ldots,B_k}\frac{1}{d}\tr(A_1\tilde{B}_1\cdots A_k\tilde{B}_k)$, where $\tilde{B}_i\coloneqq U B_i U^{\dag}$. Then, we define the relative error with respect to the Haar value
\begin{align}
\delta\operatorname{OTOC}_{k}(U)\coloneqq\frac{\operatorname{OTOC}_k(U)-\mathbb{E}_{U\sim \haar}\operatorname{OTOC}_k(U)}{\mathbb{E}_{U\sim \haar}\operatorname{OTOC}_k(U)}
\end{align}
It holds that $\int\de U_t\,\delta\operatorname{OTOC}_{k}(U_t)=\operatorname{negl}(n)$ if and only if $t=\tilde{\Theta}(k^2+\log^{1+c}n)$ where $c>0$ and $\operatorname{negl}(n)$ is a function decreasing faster than any polynomial.
\begin{proof}
   We employ the relation proven in Ref.~\cite{roberts_chaos_2017}: for any ensemble $\mathcal{E}$ of unitaries, it holds that:
    \begin{align}
        \mathbb{E}_{U\sim \mathcal{E}}\operatorname{OTOC}_k(U)=\frac{1}{d^{2k+1}}\mathcal{F}_{\mathcal{E}}
    \end{align}
where $\mathcal{F}_{\mathcal{E}}$ is the frame potential. Considering the relative ratio, and applying the result of \cref{th:framepotentialconvergence} we find that fixing $t=O(k^2+\log^{1+c}n)$ for some $c>0$, then $(\mathcal{F}_{t}-\mathcal{F}_{\haar})/\mathcal{F}_{\haar}=\exp(O(\log^{1+c}n))$. Conversely, we have that fixing $t=\Omega(k^2/\log k+\log^{1+c}n)$ $(\mathcal{F}_{t}-\mathcal{F}_{\haar})/\mathcal{F}_{\haar}=\exp(\Omega(\log^{1+c}n))$. Noting that $\exp(\log^{1+c}n)=\operatorname{negl}(n)$ proves the desired result.
\end{proof}
\end{corollary}

\subsection{Optimal convergence of $t$-doped Clifford circuit towards relative unitary designs}
In this section, we establish the optimal convergence of the ensemble of $t$-doped Clifford circuits through approximate $k$-design in relative error. The key ingredient for the proof, is the lower bound, as for the upper bound we can import the main result shown in Ref.~\cite{haferkamp_random_2022}.

Let us first establish the following key lemma. 
\begin{lemma}[Lower bounding the largest eigenvalue of the $t$-doped state]\label{lem:lowerboundinfinitynormdopedstate}
    Let $\mu_1$ be any measure over the single qubit unitary group. Let $\Psi_{t}^{(k)}\coloneqq\Phi_{t}^{(k)}(\sigma^{\otimes k})$ for some $\sigma\in\stab_n$. Then
    \be
\|\Psi_{t}^{(k)}\|_{\infty}\ge B_{\mu_1}^{t}\frac{d^{k/2}}{2Z_n},
    \ee 
where $B_{\mu_1}\coloneqq\int\de\vec{\alpha}\delta(|\vec{\alpha}|\le\frac{1}{8kt})$ is the volume of a ball of radius $\frac{1}{8kt}$ with the measure $\de\vec{\alpha}$, see \cref{eq:alternativeexpressioncliffordtwirling}.

\begin{proof}
Our task is to lower bound the infinity norm of $\Psi_t^{(k)}$. Let us first find tight bounds on $\|\Psi_{\cl}^{(k)}\|_{\infty}$. Let $\Omega\in\mathcal{P}_P$ a projective Pauli monomial with $m=k/2$. We know that $\Omega^2=d^m\Omega$, hence $\Omega/d^m$ is a projector and $\|\Omega\|_{\infty}=d^m$. On the other hand, we also have $\|\Omega\|_{1}=\tr(\Omega)=d^{k-m}$. For the choice $m=k/2$, we have that $\Omega$ is proportional to a statevector, and that $\|\Omega\|_{\infty}=\|\Omega\|_1=d^{k/2}$. We remark that a projective Pauli monomial with $m=k/2$ always exists for $k=0\mod 8$, while it is not guaranteed to exist for any $k$. For example, for $k=4$ it does not exist. However, a projective Pauli monomial with $m=\frac{k}{2}-1$ for $k$ even always exists, see~\cite{bittel2025completetheorycliffordcommutant} for more details. We use $m=k/2$ for simplicity. We can lower bound 
\be
\|\Psi_{\cl}^{(k)}\|_{\infty}\ge \frac{1}{\|\Omega\|_1}\tr(\Psi_{\cl}^{(k)}\Omega)=\frac{1}{Z_n\|\Omega\|_1}\sum_{\Omega'\in\mathcal{P}}\tr(\Omega\Omega')\ge \frac{1}{Z_n\|\Omega\|_1}\tr(\Omega^2)=\frac{d^k}{Z_n\|\Omega\|_1}=\frac{d^{k/2}}{Z_n}
\ee
where we used that $\Psi_{\cl}^{(k)}\propto\sum_{\Omega\in\mathcal{P}}\Omega$ (see~\cref{sec:cliffordweingartencalculus}), that $\tr(\Omega\Omega')\ge 1$, and that $\tr(\Omega^2)=d^k$. 

We are now ready to bound $\|\Psi_{t}^{(k)}\|_{\infty}$. Let $\ket{v}$ the normalized eigenstate corresponding to the maximum eigenvalue $\|\Psi_{\cl}^{(k)}\|_{\infty}$. We can bound $\|\Psi_{t}^{(k)}\|_{\infty}\ge \langle v|\Psi_{t}^{(k)}|v\rangle$. Writing $\Psi_{t}^{(k)}=\Phi_{t}^{(k)}(\Psi_{\cl}^{(k)})$, we can lower bound it as
\be
\tr(\Phi_{t}^{(k)}(\Psi_{\cl}^{(k)})\ketbra{v})\ge \mathbb{E}_{\vec{P}_i}\int\prod_{i}\de\vec{\alpha_i}\delta(|\vec{\alpha}|\le\varepsilon)\tr(\ketbra{v}\left[\prod_{i}U_{i}^{\otimes k}(\vec{\alpha}_i)\right]^{\dag}\Psi_{\cl}^{(k)}\left[\prod_{i}U_{i}^{\otimes k}(\vec{\alpha}_i)\right]^{\dag}).
\ee
Now, notice that for every $i$, we have
\be
\|U(\vec{\alpha})-\mathbb{1}\|_{\infty}=\sum_{l=1}^{\infty}\frac{1}{l!}\|(i\vec{\alpha}\cdot\vec{P})^l\|_{\infty}\le\sum_{l=1}^{\infty}\frac{1}{l!}|\vec{\alpha}|^{l}=e^{|\vec{\alpha}|}-1\le 2\varepsilon
\ee
Since, we restricted the domain of integration up to $|\vec{\alpha}|\le \varepsilon$. The above inequality holds whenever $\varepsilon\le 1$. It easily follows that $\|U^{\otimes k}(\vec{\alpha})-\mathbb{1}\|_{\infty}\le 2k\varepsilon$. Therefore, by denoting $V\coloneqq\left[\prod_{i}U_{i}^{\otimes k}(\vec{\alpha}_i)\right]^{\dag}$ for readability, we have
\be
\tr(\ketbra{v}\left[\prod_{i}U_{i}^{\otimes k}(\vec{\alpha}_i)\right]^{\dag}\Psi_{\cl}^{(k)}\left[\prod_{i}U_{i}^{\otimes k}(\vec{\alpha}_i)\right]^{\dag})\ge \tr(\ketbra{v}\Psi_{\cl}^{(k)})-\left|\tr(\ketbra{v}V\Psi_{\cl}^{(k)}V^{\dag})-\tr(\ketbra{v}\Psi_{\cl}^{(k)})\right|
\ee
Now, notice that
\be
\left|\tr(\ketbra{v}V\Psi_{\cl}^{(k)}V^{\dag})-\tr(\ketbra{v}\Psi_{\cl}^{(k)})\right|&=
\left|\tr(\ketbra{v}V\Psi_{\cl}^{(k)}[V^{\dag}-\mathbb{1}])+\tr(\Psi_{\cl}^{(k)}\ketbra{v}[V-\mathbb{1}])\right|\\
&\le \|V\|_{\infty}\|\Psi_{\cl}^{(k)}[V^{\dag}-\mathbb{1}]\ketbra{v}\|_{1}+\|V-\mathbb{1}\|_{\infty}\|\Psi_{\cl}^{(k)}\ketbra{v}\|_{1}\\
&\le \|\Psi_{\cl}^{(k)}\|_{\infty}\|V^{\dag}-\mathbb{1}\|_{\infty}\|\ketbra{v}\|_1+\|V-\mathbb{1}\|_{\infty}\langle v|\Psi_{\cl}^{(k)}|v\rangle\\
&=2\|\Psi_{\cl}^{(k)}\|_{\infty}\|V-\mathbb{1}\|_{\infty}\,.
\ee
Noticing that $\|V-\mathbb{1}\|_{\infty}\le t\|U^{\otimes k}(\vec{\alpha})-\mathbb{1}\|_{\infty}\le 4tk\varepsilon$, we obtain
\be
\tr(\ketbra{v}\left[\prod_{i}U_{i}^{\otimes k}(\vec{\alpha}_i)\right]^{\dag}\Psi_{\cl}^{(k)}\left[\prod_{i}U_{i}^{\otimes k}(\vec{\alpha}_i)\right]^{\dag})\ge \|\Psi_{\cl}^{(k)}\|_{\infty}(1-4tk\varepsilon)
\ee
Choosing $\varepsilon=\frac{1}{8kt}$, we arrive at
\be
\|\Psi_{t}^{(k)}\|_{\infty}\ge \frac{1}{2}B_{\mu_1}^t\|\Psi_{\cl}^{(k)}\|_{\infty} \ge B_{\mu_1}^{t}\frac{d^{k/2}}{2Z_n}
\ee
concluding the proof.
\end{proof}

\end{lemma}
The next theorem establishes the key lower bound to convergence to relative $\varepsilon$-approximate $k$-design.
\begin{theorem}[Lower bound to relative unitary designs]\label{th:lowerboundrelativeerrorunitarydesigns}
    Let $\delta>0$. Let $\mathcal{C}_t$ be the ensemble of $t$-doped Clifford circuits. For any $n\ge 2\delta^{-1}k$ and $0\le\varepsilon\le d^{\frac{k}{2}(1-\delta)}$, then $\mathcal{C}_t$ cannot be a relative $\varepsilon$-approximate unitary design unless 
    \be
t\ge \frac{nk}{6\delta^{-1}\log B_{\mu_1}^{-1}}\,.
    \ee
    \begin{proof}
According to \cref{def:unitarydesignrelativeerror}, a necessary condition for $\mathcal{C}_t$ being a relative design is the channel $\delta\Phi=(1+ \varepsilon)\Phi_{\haar}^{(k)}-\Phi_{t}^{(k)}$ to be a completely positive quantum channel. A necessary condition for complete positivity is simply positivity, i.e.,  the channel $\delta\Phi$ to output positive matrices when applied on states. We can then apply $\delta\Phi$ on $\sigma^{\otimes k}$ for $\sigma\in\stab_n$, and get
\be
\delta\Phi(\sigma^{\otimes k})=(1+\varepsilon)\Psi_{\haar}^{(k)}-\Psi_{t}^{(k)}.
\ee
A necessary condition for positivity of the matrix $\delta\Phi(\sigma^{\otimes k})$ is the smallest eigenvalue to be positive. Denoting $\lambda(\delta\Phi(\sigma^{\otimes k}))$ the smallest eigenvalue of $\delta\Phi(\sigma^{\otimes k})$, we can upper bound it as $\lambda(\delta\Phi(\sigma^{\otimes k}))\le (1+\varepsilon)\frac{1}{\tr(\Pi_{\sym})}-B_{\mu_1}^t\frac{d^{k/2}}{2Z_n}$, because $[\Psi_{\haar}^{(k)},\Psi_t^{(k)}]=0$. Imposing it larger than $0$, from \cref{lem:lowerboundinfinitynormdopedstate}, we arrive at the condition
\be
t\ge \frac{1}{\log B_{\mu_1}^{-1}}\left(\frac{nk}{2}-\log(1+\varepsilon)+\log \frac{\tr(\Pi_{\sym})} {2 Z_n}\right).
\ee
Then, for any $n\ge k-1$, we can further lower bound $\log\frac{\tr\Pi_{\sym}}{Z_n}$ using $\log\tr\Pi_{\sym}\ge nk-k\log k$, $\log Z_n\le nk+\sum_{i=0}^{k-2}\le nk+1-\frac{1}{d}$, implying $\log\frac{\tr\Pi_{\sym}}{Z_n}\ge -k\log k-1+\frac{1}{d}$. We arrive at 
\be
t\ge \frac{1}{\log B_{\mu_1}^{-1}}\left(\frac{nk}{2}-\log(1+\varepsilon)-k\log k-2\right).
\ee
Let us now discriminate two cases. If $\varepsilon\le 1$, we have $\log(1+\varepsilon)\le 1$, hence 
\be
t\ge \frac{1}{\log B_{\mu_1}^{-1}}\left(\frac{nk}{2}-k\log k-3\right).
\ee
Whereas, for any $1<\varepsilon\le d^{\frac{k}{2}(1-\delta)}$ and $0<\delta\le1$, we have $\log(1+\varepsilon)\le \log\varepsilon+\frac{\log e}{\varepsilon}\le \frac{nk(1-\delta)}{2}+\log e$, which leads to
\be\label{eq1th22}
t\ge \frac{1}{\log B_{\mu_1}^{-1}}\left(\frac{nk\delta}{2}-k\log k-2-\log e\right).
\ee
Therefore, for any $\varepsilon$ in the range $0\le \varepsilon\le d^{\frac{k(1-\delta)}{2}}$ with $0<\delta<1$, the lower bound in \cref{eq1th22} applies being the lowest lower bound. Finally, we have that $\log k+(2+\log e)/k\le \frac{2}{3}\frac{n}{2}\delta$ for any $n\ge 2\delta^{-1}k$ and $k>4$, which leads to $t\ge \frac{nk}{6\delta^{-1}\log B_{\mu_1}^{-1}}$ as desired. 
    \end{proof}
\end{theorem}

In the following, we import the result from Ref.~\cite{haferkamp_random_2022} to prove an upper bound to relative error unitary designs that matches the lower bound of \cref{th:lowerboundrelativeerrorunitarydesigns}. We then also consider the measure $\mu_1$ used in Ref.~\cite{haferkamp_random_2022}, which is the sum of three delta-functions on $U$, $U^{\dag}$ and $I$ where $U$ is a generic single qubit non-Clifford gate. 
\begin{corollary}[Lower bounding the number of non-Clifford resources]  Let $\delta>0$. Let $\mathcal{C}_t$ be the ensemble of $t$-doped Clifford circuits and let $\mu_1$ be a single-qubit measure over one of the following ensembles: the discrete set $\{U, U^{\dagger}, I\}$, the unitary ensemble, or the diagonal unitary ensemble. For any $n\ge 2\delta^{-1}k$ and $0\le\varepsilon\le d^{\frac{k}{2}(1-\delta)}$, then the following lower bounds on the number of non-Clifford resources $t$ hold.
\be
t& \ge \frac{nk\delta}{18} \quad &\{U,U^{\dagger},I\}. \\ 
t& \ge \frac{nk\delta}{6\log(nk\delta)(17 + 6\log k)} \quad &\emph{unitary ensemble}. \\
t& \ge \frac{nk\delta}{6\log(nk\delta)(2+\log k)} \quad &\emph{diagonal unitary ensemble}.
\ee

\begin{proof}
Let $\mu_1$ be the discrete measure that samples over $U,U^{\dag},I$. Then for any $U$ such that $\|U-I\|_{\infty}\ge\Omega(1)$, i.e.,  that does not depend on $k,t$, then we simply have $B_{\mu_1}=\frac{1}{3}$, hence we have $t\ge \frac{nk\delta}{18}$. 

In the second case, we want to compute $B_{\mu_1}=\int \de \vec{\alpha} \delta(\alpha \le (8kt)^{-1})$ for the Haar measure. Let us start by considering the unitary decomposition $U=\exp(i \vec{\alpha} \cdot \vec{P})$.
 For a single-qubit unitary, the proposed decomposition can also be written as
\be 
U = \begin{pmatrix}
    \cos(\vert \vec{\alpha} \vert) + i \frac{\alpha_3}{\vert \vec{\alpha} \vert} \sin(\vert \vec{\alpha}\vert) & i \frac{\alpha_1}{\vert \vec{\alpha} \vert} \sin(\vert \vec{\alpha}\vert) +\frac{\alpha_2}{\vert \vec{\alpha} \vert} \sin(\vert \vec{\alpha}\vert) \\
    i \frac{\alpha_1}{\vert \vec{\alpha} \vert} \sin(\vert \vec{\alpha}\vert) - \frac{\alpha_2}{\vert \vec{\alpha} \vert} \sin(\vert \vec{\alpha}\vert) & \cos(\vert \vec{\alpha}\vert) - i \frac{\alpha_3}{\vert \vec{\alpha} \vert} \sin(\vert \vec{\alpha}\vert)
\end{pmatrix}.
\ee 
Following Ref.~\cite{Baker2002}, 
we get that the measure over the single-qubit unitary group can be expressed as 
\be 
\int_{|\vec{\alpha}|\le c}\prod_{i}\de\vec{\alpha_i} = \frac{1}{\pi^2}\int_{0}^{2\pi} \de \phi\int_{0}^{\pi} \de\theta\int_{0}^{\sin^2 c} \de r \frac{1}{\sqrt{1-r^2}}r^2\sin\theta\,.
\ee 
The conversion is straightforward: first converting to spinorial coordinates, then to spherical ones. Note that for $c=2\pi$ the integral returns $1$, being the measure normalized. 
The integral is then equal to 
\be 
\int_{|\vec{\alpha}|\le c}\prod_{i}\de\vec{\alpha_i} = \frac{1}{\pi}\left(2\arctan\left(\frac{\sqrt{2}\sin^2(c)}{\cos(c)\sqrt{3-\cos(2c)}}\right)-\sqrt{2}\sin^2(c)\cos(c)\sqrt{3-\cos(2c)}\right).
\ee 
By considering $c=\frac{1}{8kt}$, 
we get 
\be 
\int_{|\vec{\alpha}|\le \frac{1}{8kt}}\prod_{i}\de\vec{\alpha_i} \ge \frac{4}{3 \pi (8kt)^6 }\left(1- \frac{1}{(8kt)^2}\right) \ge \frac{63}{48\pi(8kt)^6},
\ee 
from which we obtain $t\log t\ge \frac{nk\delta}{6(17+6\log k)}$, and, therefore, 
\be
t\ge \frac{nk}{6\log(nk\delta)(17+6\log k)} 
\ee
is a necessary condition.
Let us conclude by computing the same but considering diagonal unitary matrices. We have then 
\be 
B_{\mu_1}=\int_{-\frac{1}{8kt}}^{\frac{1}{8kt}} \de \phi = \frac{1}{4kt} ,
\ee 
from which we obtain $t \log t \ge \frac{nk\delta}{6(2+\log k)}$ and, therefore, 
\be
t\ge\frac{nk\delta}{6\log(nk\delta)(2+\log k)} ,
\ee
concluding the proof.
\end{proof}
\end{corollary}

\begin{proposition}[Upper bound to relative unitary designs, Corollary 1 in Ref.~\cite{haferkamp_random_2022}]\label{prop:haferkamp} Let $\varepsilon>0$. Let $\mathcal{C}_t$ be the ensemble of $t$-doped Clifford circuits. Let $U$ be a single qubit non-Clifford gate. Let $\mu_1$ be the uniform measure over $U,U^{\dag},I$. Then  there exist two constants $C_1(U),C_2(U)$ such that for $t\ge C_1(U)\log^2(k)(2nk+\log\varepsilon^{-1})$ then $\mathcal{C}_t$ is a relative $\varepsilon$-approximate $k$-design, whenever $n\ge C_{2}(U)k^2$. If instead $\mu_1$ is the single-qubit Haar measure, analogous conclusions can be reached.
\end{proposition}

Therefore, combining \cref{th:lowerboundrelativeerrorunitarydesigns} and \cref{prop:haferkamp}, we conclude that, for $k=O(\sqrt{n})$, $t$-doped Clifford circuits are relative $\varepsilon$-approximate $k$-design iff $t=\tilde{\Theta}(kn)$. 

The convergence lower bound in \cref{th:lowerboundrelativeerrorunitarydesigns} holds whenever $k=O(n)$. Since a $k+1$-design must be at least a $k$-design, we have that for any $k=\Omega(n)$ the number of non-Clifford gates to be applied is at least $t=\Omega(n^2)$, underlying an extremely high cost in terms of non-Clifford resources in producing $k$-design with large $k$. In the following proposition, we also cover the regime $k=\Omega(n)$, using the same proof technique of \cref{th:lowerboundrelativeerrorunitarydesigns} and a different version of \cref{lem:lowerboundinfinitynormdopedstate}.

\begin{theorem}[Lower bound to relative unitary $k$-design for $k=\Omega(n)$]\label{th:alternativelowerbound} Let $\delta>0$. Let $\mathcal{C}_t$ be the ensemble of $t$-doped Clifford circuit and $U=e^{i\frac{\pi}{8}Z}$ being the $T$-gate. For any $8n\delta^{-1}\le k\le 2^{\frac{n\delta}{2}}$ and $0\le \varepsilon\le d^{\frac{k}{2}(1-\delta)}$, then $\mathcal{C}_t$ cannot be a relative $\varepsilon$-approximate unitary $k$-design unless
\be
t\ge \frac{k\delta}{8}.
\ee

\begin{proof}
Consider $\Psi_t^{(k)} \coloneqq \Phi_t^{(k)}(\sigma^{\otimes k})$ for $\sigma \in \stab_n$. We then have the bound 
\be
\|\Psi_t^{(k)}\|_2 \le \sqrt{\rank(\Psi_t^{(k)})} \cdot \|\Psi_t^{(k)}\|_{\infty}. 
\ee
The 2-norm $\|\Psi_t^{(k)}\|_2$ corresponds to (the square root of) the purity of the state vector $\Psi_t^{(k)}$. For an ensemble of pure states $\mathcal{E} = \{ \ket{\psi_i} \}$, the average over $k$-fold tensor powers is given by $\Psi_{\mathcal{E}}^{(k)} = \frac{1}{|\mathcal{E}|} \sum_{\ket{\psi} \in \mathcal{E}} \ketbra{\psi}^{\otimes k}$, whose purity can be lower bounded
\be
\tr((\Psi_{\mathcal{E}}^{(k)})^2)=\frac{1}{|\mathcal{E}|} + \frac{1}{|\mathcal{E}|^2} \sum_{\ket{\psi} \neq \ket{\phi} \in \mathcal{E}} |\langle \psi | \phi \rangle|^{2} \ge \frac{1}{|\mathcal{E}|}.
\ee
This implies that the 2-norm is lower bounded by the inverse square root of the size of the ensemble, i.e., $\|\Psi_{\mathcal{E}}^{(k)}\|_2 \ge |\mathcal{E}|^{-1/2}$. Choosing $\mathcal{E} = \mathcal{C}_t$, the ensemble of $t$-doped Clifford circuits, we can crudely upper bound the ensemble size by $|\mathcal{C}_t| \le |\mathcal{C}_n| \cdot 4^{nt}$, where $|\mathcal{C}_n| \le 2^{2n^2 + n}$ is the size of the Clifford group, and $4^{nt}$ accounts for the number of possible rotations $e^{i\frac{\pi}{8}P}$ with $P \in \mathbb{P}_n$. Therefore, we obtain the lower bound
\be
\|\Psi_t^{(k)}\|_{\infty} \ge \frac{2^{-nt - n^2 - n/2}}{  \sqrt{\tr(\Pi_{\sym})}},
\ee
where we have used that $\rank(\Psi_{t}^{(k)})=\tr(\Pi_{\sym})$. Now, using the same strategy as in \cref{th:lowerboundrelativeerrorunitarydesigns}, we can upper bound the smallest eigenvalue of $\delta\Phi(\sigma^{\otimes k}) = (1+\varepsilon)\Psi_{\text{Haar}}^{(k)} - \Psi_t^{(k)}$ as
\be
\lambda_{\min}(\delta\Phi(\sigma^{\otimes k})) \le \frac{1+\varepsilon}{\tr(\Pi_{\sym})} - \frac{2^{-nt - n^2 - n/2}}{\sqrt{\tr(\Pi_{\sym})}}.
\ee
Requiring this quantity to be non-negative and applying the bounds from \cref{th:lowerboundrelativeerrorunitarydesigns}, we arrive at
\be
t \ge \frac{k}{2} - n - \frac{k \log k}{2n} - \frac{\log(1+\varepsilon)}{n} - \frac{1}{2}.
\ee
For any $\varepsilon \le d^{\frac{k}{2}(1 - \delta)}$ with $0 < \delta \le 1$, we obtain
\be
t \ge \frac{k\delta}{2} - n - \frac{k \log k}{2n} - \frac{\log e}{n} - \frac{1}{2},
\ee
which implies $t \ge \frac{k\delta}{8}$, for any $8n\delta^{-1} \le k \le 2^{n\delta/2}$, thus concluding the proof.

\end{proof}
\end{theorem}

\subsection{The non-Clifford cost of generating pseudo-random unitaries}\label{sec:pseudo-randomess}
In this section, we provide a proof of \cref{prop1:pseudo} by synthesizing results from Refs.~\cite{grewal_efficient_2023,schuster2025randomunitariesextremelylow}. A similar proof is also presented in Ref.~\cite{grewal2024pseudoentanglementaintcheap}. First, Ref.~\cite{schuster2025randomunitariesextremelylow} shows that it is possible to construct a pseudo-random unitary ensemble using a shallow quantum circuit of depth \( O(\operatorname{poly} \log \log n) \), assuming the hardness of the 
\emph{learning with errors} (LWE) problem for quantum computers. We refer the reader to Ref.~\cite{schuster2025randomunitariesextremelylow} for full technical details.
A circuit of depth \( D \) on \( n \) qubits contains at most \( O(nD) \) gates, and each gate can be implemented using \( O(1) \) single-qubit non-Clifford gates~\cite{nielsen_quantum_2000}. Hence, the construction in Ref.~\cite{schuster2025randomunitariesextremelylow} requires at most \( O(n \operatorname{poly} \log \log n) \) non-Clifford gates, which yields the upper bound in \cref{prop1:pseudo}.

To establish the lower bound, consider a quantum state vector \( \ket{\psi} \) on \( n \) qubits. Its associated stabilizer group is defined as
\be
G_{\psi} \coloneqq \{ P \in \mathbb{P}_n \,:\, \langle \psi | P | \psi \rangle = \pm 1 \}.
\ee
According to Refs.~\cite{grewal_efficient_2023,leone_learning_2022}, if \( \ket{\psi} \) is prepared using \( t \) single-qubit non-Clifford gates, then \( \dim G_{\psi} \geq n - 2t \). Furthermore, Ref.~\cite{grewal_efficient_2023} introduced a polynomial-time quantum algorithm that, given a state vector \( \ket{\psi} \) and parameters \( k \in [n] \) $\varepsilon=\Omega(\operatorname{poly}^{-1}n)$, can determine whether
\begin{itemize}
    \item[(A)] \( \dim G_{\psi} \ge n - k \), or
    \item[(B)] \( \ket{\psi} \) is \( \varepsilon \)-far in trace distance from any state satisfying (A).
\end{itemize}

Now, assume for contradiction that there exists a pseudo-random ensemble \( \mathcal{U} \) generated using only \( \le \alpha n \) with $\alpha<1$ non-Clifford gates (with high probability). Then, we can build a polynomial-time distinguisher as follows: apply the algorithm from Ref.~\cite{grewal_efficient_2023} to states generated (from $\ket{0}^{\otimes n}$) by unitaries from \( \mathcal{U} \) and from the Haar measure. For unitaries in \( \mathcal{U} \), the algorithm will output case (A). We now show that for Haar-random unitaries, the algorithm almost surely outputs case (B).

To do so, it suffices to prove that a Haar-random state is far in trace distance from any state with stabilizer group dimension \( \ge n - k \), for \( k \le \alpha n \). By the compression theorem~\cite{leone_learning_2022}, any such state vector can be expressed as \( \ket{\psi} = C(\ket{0}^{\otimes n-k} \otimes \ket{\phi_k}) \), where \( \ket{\phi_k} \) is a \( k \)-qubit state vector and \( C \) is a Clifford circuit. Thus, the number of such states is at most \( |\mathcal{C}_n| \cdot |S_{k,\varepsilon}| \), where \( S_{k,\varepsilon} \) is an \( \varepsilon \)-net covering of the \( k \)-qubit pure state space.

Applying Lévy's lemma and a 
union bound, we obtain
\be\label{eqprovapseudo}
\Pr_{\ket{\psi} \sim \operatorname{Haar}}\left( \exists \ket{\psi_{n-k}} \text{ with } \dim G_{\psi_{n-k}} \ge n - k \text{ and } \tr(\psi \psi_{n-k}) \ge d^{-1/3} \right) 
\le |\mathcal{C}_n| \cdot |S_{k,\varepsilon}| \cdot e^{-\Omega(2^{n/3})}=\operatorname{negl}\,(n).
\ee
The last step of \cref{eqprovapseudo} follows from the fact that $|C_n|\le2^{2n^2+n}$, while 
\be
S_{k,\varepsilon}=\left(\frac{1}{\varepsilon}\right)^{O(2^k)}=2^{O(2^{n\alpha})}
\ee
for any $\varepsilon=\Theta(1)$. Hence, fixing $\varepsilon=1/4$ and choosing $\alpha$ appropriately, \cref{eqprovapseudo} follows. Hence,
\cref{eqprovapseudo} proves that, with overwhelming probability, a Haar-random state is at least \( \sqrt{1 - d^{-1/3}} \)-far in trace distance from any state vector \( \ket{\phi} = C(\ket{0} \otimes \ket{\phi_k}) \) with \( \ket{\phi_k} \in S_{k,1/4} \). Since \( S_{k,1/4} \) is an \( \varepsilon \)-covering net with $\varepsilon=1/4$, this further implies that the Haar-random state is at least \( \sqrt{1 - d^{-1/3}} - 1/4\ge 0.2\)-far from any state with \( \dim G \ge n - k \).
We conclude that the algorithm from Ref.~\cite{grewal_efficient_2023} serves as a valid distinguisher unless \( t > \alpha n=\Omega(n) \), completing the proof of \cref{prop1:pseudo}. 

\subsection{Implications for $\varepsilon$-nets}
\label{sec:nets}

In this work, we have  quantified 
in what precise sense the 
ensemble of doped Clifford circuits is spread within the unitary group. We here explore finally a different reading of that notion, i.e., to what extent the quantum circuits prepared by doped Clifford circuits approximate all quantum states, in the sense of constituting an $\epsilon$-net, defined below.

\begin{definition}[$\varepsilon$-net] Let $\mathcal{E}$ be a discrete ensemble of unitary operators. $\mathcal{E}$ is a $\varepsilon$ covering net iff for any $U\in\mathcal{U}_n$, there exists $U'\in\mathcal{E}$ such that $\|U-U'\|_{\diamond}\le \varepsilon$, where $\|\cdot\|_{\diamond}$ is the diamond norm.
\end{definition}
Below, we show that, as one would expect, creating an $\varepsilon$-net requires an exponential amount of non-Clifford resources, as it also requires an exponential gate complexity. 
\begin{proposition}[$\varepsilon$-net via $t$-doped Clifford circuits] Let $\mathcal{C}_t$ be the ensemble of $t$-doped Clifford circuits doped with single qubit $T$ gates. Then $\mathcal{C}_t$ cannot form a $\varepsilon$-net with respect to the diamond norm on unitaries unless $t=\Omega(4^n/(n\log\varepsilon^{-1}))$.
\begin{proof}
The proof is done via a simple counting argument, and using some steps of the proof of \cref{th:alternativelowerbound}. It is known~\cite[Proposition 7]{szarek1982nets} (see also 
Ref.\ \cite[Lemma 9]{zhao_learning_2023}) that for an ensemble of unitaries $\mathcal{E}$ to form a $\varepsilon$-net with respect to the diamond norm the size $|\mathcal{E}|$ must be lower bounded by
\be
|\mathcal{E}|\ge \left(\frac{c_{\diamond}}{\varepsilon}\right)^{d^2},
\ee
with $c_{\diamond}>0$ being a universal constant. Hence, 
a necessary condition for $\mathcal{C}_t$ to form a $\varepsilon$-net is 
\be
|\mathcal{C}_t|\ge \left(\frac{c_{\diamond}}{d}\right)^{d^2}.
\ee
We can upper bound $|\mathcal{C}_t|\le 2^{2n^2+n+4nt}$, which immediately gives the necessary condition on the doping parameter $t$ as
\be
t\ge \frac{d^{2}}{n}\log\frac{c_{\diamond}}{\varepsilon}-n-1
\ee
proving the statement.
\end{proof}
    
\end{proposition}
These statements provide further evidence that exploring the full set of unitaries---for example, by constructing an $\epsilon$-net or by generating unitaries at random---is extremely costly in terms of non-Clifford resources.

\stoptoc

\section*{Acknowledgements}

This work has been inspired by the Ph.D. thesis of L.L.~\cite{leone_clifford_2023}. We thank Toshihiro Yada for having identified a flaw in an earlier version of the manuscript. This work has been supported by the DFG (CRC 183, FOR 2724), by the BMBF (Hybrid++, QuSol), the BMWK (EniQmA), the Munich Quantum Valley (K-8), the QuantERA (HQCC),
the Alexander-von-Humboldt Foundation and Berlin Quantum. This work has also been funded by the DFG under Germany's Excellence Strategy – The Berlin Mathematics Research Center MATH+ (EXC-2046/1, project ID: 390685689). S.F.E.O. acknowledges support from PNRR MUR project PE0000023-NQSTI. AH acknowledges support from the PNRR MUR project PE0000023-NQSTI and the PNRR MUR project CN 00000013-ICSC.


\begin{thebibliography}{106}%
\makeatletter
\providecommand \@ifxundefined [1]{%
 \@ifx{#1\undefined}
}%
\providecommand \@ifnum [1]{%
 \ifnum #1\expandafter \@firstoftwo
 \else \expandafter \@secondoftwo
 \fi
}%
\providecommand \@ifx [1]{%
 \ifx #1\expandafter \@firstoftwo
 \else \expandafter \@secondoftwo
 \fi
}%
\providecommand \natexlab [1]{#1}%
\providecommand \enquote  [1]{``#1''}%
\providecommand \bibnamefont  [1]{#1}%
\providecommand \bibfnamefont [1]{#1}%
\providecommand \citenamefont [1]{#1}%
\providecommand \href@noop [0]{\@secondoftwo}%
\providecommand \href [0]{\begingroup \@sanitize@url \@href}%
\providecommand \@href[1]{\@@startlink{#1}\@@href}%
\providecommand \@@href[1]{\endgroup#1\@@endlink}%
\providecommand \@sanitize@url [0]{\catcode `\\12\catcode `\$12\catcode `\&12\catcode `\#12\catcode `\^12\catcode `\_12\catcode `\%12\relax}%
\providecommand \@@startlink[1]{}%
\providecommand \@@endlink[0]{}%
\providecommand \url  [0]{\begingroup\@sanitize@url \@url }%
\providecommand \@url [1]{\endgroup\@href {#1}{\urlprefix }}%
\providecommand \urlprefix  [0]{URL }%
\providecommand \Eprint [0]{\href }%
\providecommand \doibase [0]{https://doi.org/}%
\providecommand \selectlanguage [0]{\@gobble}%
\providecommand \bibinfo  [0]{\@secondoftwo}%
\providecommand \bibfield  [0]{\@secondoftwo}%
\providecommand \translation [1]{[#1]}%
\providecommand \BibitemOpen [0]{}%
\providecommand \bibitemStop [0]{}%
\providecommand \bibitemNoStop [0]{.\EOS\space}%
\providecommand \EOS [0]{\spacefactor3000\relax}%
\providecommand \BibitemShut  [1]{\csname bibitem#1\endcsname}%
\let\auto@bib@innerbib\@empty
\bibitem [{\citenamefont {Ambainis}\ and\ \citenamefont {Smith}(2004)}]{ambainis_smaapp2004}%
  \BibitemOpen
  \bibfield  {author} {\bibinfo {author} {\bibfnamefont {A.}~\bibnamefont {Ambainis}}\ and\ \bibinfo {author} {\bibfnamefont {A.}~\bibnamefont {Smith}},\ }\bibfield  {title} {\bibinfo {title} {Small pseudo-random families of matrices: Derandomizing approximate quantum encryption},\ }in\ \href {https://doi.org/I:10.1007/978-3-540-27821-4_23} {\emph {\bibinfo {booktitle} {Proc. RANDOM’04, Lecture Notes in Computer Science}}},\ Vol.\ \bibinfo {volume} {3122}\ (\bibinfo  {publisher} {Springer},\ \bibinfo {year} {2004})\ pp.\ \bibinfo {pages} {249--260}\BibitemShut {NoStop}%
\bibitem [{\citenamefont {Hayden}\ \emph {et~al.}(2004)\citenamefont {Hayden}, \citenamefont {Leung}, \citenamefont {Shor},\ and\ \citenamefont {Winter}}]{hayden_ransta2004}%
  \BibitemOpen
  \bibfield  {author} {\bibinfo {author} {\bibfnamefont {P.}~\bibnamefont {Hayden}}, \bibinfo {author} {\bibfnamefont {D.}~\bibnamefont {Leung}}, \bibinfo {author} {\bibfnamefont {P.~W.}\ \bibnamefont {Shor}},\ and\ \bibinfo {author} {\bibfnamefont {A.}~\bibnamefont {Winter}},\ }\bibfield  {title} {\bibinfo {title} {Randomizing quantum states: Constructions and applications},\ }\href {https://doi.org/10.1007/s00220-004-1087-6} {\bibfield  {journal} {\bibinfo  {journal} {Commun. Math. Phys.}\ }\textbf {\bibinfo {volume} {250}},\ \bibinfo {pages} {371} (\bibinfo {year} {2004})}\BibitemShut {NoStop}%
\bibitem [{\citenamefont {Kretschmer}(2021)}]{kretschmer_quapru2021}%
  \BibitemOpen
  \bibfield  {author} {\bibinfo {author} {\bibfnamefont {W.}~\bibnamefont {Kretschmer}},\ }\bibfield  {title} {\bibinfo {title} {Quantum pseudorandomness and classical complexity},\ }in\ \href {https://doi.org/https://doi.org/10.4230/LIPIcs.TQC.2021.2} {\emph {\bibinfo {booktitle} {TQC 2021, Vol. 197}}}\ (\bibinfo {year} {2021})\ pp.\ \bibinfo {pages} {2:1--2:20}\BibitemShut {NoStop}%
\bibitem [{\citenamefont {Morimae}\ and\ \citenamefont {Yamakawa}(2022)}]{morimae_quacom2022}%
  \BibitemOpen
  \bibfield  {author} {\bibinfo {author} {\bibfnamefont {T.}~\bibnamefont {Morimae}}\ and\ \bibinfo {author} {\bibfnamefont {T.}~\bibnamefont {Yamakawa}},\ }\bibfield  {title} {\bibinfo {title} {Quantum commitments and signatures without one-way functions},\ }in\ \href {https://doi.org/https://doi.org/10.1007/978-3-031-15802-5_10} {\emph {\bibinfo {booktitle} {CRYPTO 2022}}}\ (\bibinfo {year} {2022})\ pp.\ \bibinfo {pages} {269--295}\BibitemShut {NoStop}%
\bibitem [{\citenamefont {Ananth}\ \emph {et~al.}(2022)\citenamefont {Ananth}, \citenamefont {Qian},\ and\ \citenamefont {Yuen}}]{ananth_cryfro2022}%
  \BibitemOpen
  \bibfield  {author} {\bibinfo {author} {\bibfnamefont {P.}~\bibnamefont {Ananth}}, \bibinfo {author} {\bibfnamefont {L.}~\bibnamefont {Qian}},\ and\ \bibinfo {author} {\bibfnamefont {H.}~\bibnamefont {Yuen}},\ }\bibfield  {title} {\bibinfo {title} {Cryptography from pseudorandom quantum states},\ }in\ \href {https://doi.org/0.48550/arXiv.2112.10020} {\emph {\bibinfo {booktitle} {CRYPTO 2022}}}\ (\bibinfo {year} {2022})\ pp.\ \bibinfo {pages} {208--236}\BibitemShut {NoStop}%
\bibitem [{\citenamefont {Sen}(2006)}]{sen_ranmea2006}%
  \BibitemOpen
  \bibfield  {author} {\bibinfo {author} {\bibfnamefont {P.}~\bibnamefont {Sen}},\ }\bibfield  {title} {\bibinfo {title} {Random measurement bases, quantum state distinction and applications to the hidden subgroup problem},\ }in\ \href {https://doi.org/10.48550/arXiv.quant-ph/0512085} {\emph {\bibinfo {booktitle} {21st Annual IEEE Conference on Computational Complexity (CCC’06)}}}\ (\bibinfo {year} {2006})\ pp.\ \bibinfo {pages} {14--287}\BibitemShut {NoStop}%
\bibitem [{\citenamefont {Brand{\~a}o}\ and\ \citenamefont {Horodecki}(2013)}]{brandao_expspe2013}%
  \BibitemOpen
  \bibfield  {author} {\bibinfo {author} {\bibfnamefont {F.~G. S.~L.}\ \bibnamefont {Brand{\~a}o}}\ and\ \bibinfo {author} {\bibfnamefont {M.}~\bibnamefont {Horodecki}},\ }\bibfield  {title} {\bibinfo {title} {Exponential quantum speed-ups are generic},\ }\href {https://doi.org/10.26421/QIC13.11-12-1} {\bibfield  {journal} {\bibinfo  {journal} {Quant. Inf. Comp.}\ }\textbf {\bibinfo {volume} {13}},\ \bibinfo {pages} {901} (\bibinfo {year} {2013})}\BibitemShut {NoStop}%
\bibitem [{\citenamefont {Boixo}\ \emph {et~al.}(2018)\citenamefont {Boixo}, \citenamefont {Isakov}, \citenamefont {Smelyanskiy}, \citenamefont {Babbush}, \citenamefont {Ding}, \citenamefont {Jiang}, \citenamefont {Bremner}, \citenamefont {Martinis},\ and\ \citenamefont {Neven}}]{boixo_characterizing_2018}%
  \BibitemOpen
  \bibfield  {author} {\bibinfo {author} {\bibfnamefont {S.}~\bibnamefont {Boixo}}, \bibinfo {author} {\bibfnamefont {S.~V.}\ \bibnamefont {Isakov}}, \bibinfo {author} {\bibfnamefont {V.~N.}\ \bibnamefont {Smelyanskiy}}, \bibinfo {author} {\bibfnamefont {R.}~\bibnamefont {Babbush}}, \bibinfo {author} {\bibfnamefont {N.}~\bibnamefont {Ding}}, \bibinfo {author} {\bibfnamefont {Z.}~\bibnamefont {Jiang}}, \bibinfo {author} {\bibfnamefont {M.~J.}\ \bibnamefont {Bremner}}, \bibinfo {author} {\bibfnamefont {J.~M.}\ \bibnamefont {Martinis}},\ and\ \bibinfo {author} {\bibfnamefont {H.}~\bibnamefont {Neven}},\ }\bibfield  {title} {\bibinfo {title} {Characterizing quantum supremacy in near-term devices},\ }\href {https://doi.org/10.1038/s41567-018-0124-x} {\bibfield  {journal} {\bibinfo  {journal} {Nature Phys.}\ }\textbf {\bibinfo {volume} {14}},\ \bibinfo {pages} {595} (\bibinfo {year} {2018})}\BibitemShut {NoStop}%
\bibitem [{\citenamefont {Arute}\ \emph {et~al.}(2019)\citenamefont {Arute} \emph {et~al.}}]{arute_quasup2019}%
  \BibitemOpen
  \bibfield  {author} {\bibinfo {author} {\bibfnamefont {F.}~\bibnamefont {Arute}} \emph {et~al.},\ }\bibfield  {title} {\bibinfo {title} {Quantum supremacy using a programmable superconducting processor},\ }\href {https://doi.org/10.1038/s41586-019-1666-5} {\bibfield  {journal} {\bibinfo  {journal} {Nature}\ }\textbf {\bibinfo {volume} {574}},\ \bibinfo {pages} {505} (\bibinfo {year} {2019})}\BibitemShut {NoStop}%
\bibitem [{\citenamefont {Bouland}\ \emph {et~al.}(2019)\citenamefont {Bouland}, \citenamefont {Fefferman}, \citenamefont {Nirkhe},\ and\ \citenamefont {Vazirani}}]{bouland_complexity_2019}%
  \BibitemOpen
  \bibfield  {author} {\bibinfo {author} {\bibfnamefont {A.}~\bibnamefont {Bouland}}, \bibinfo {author} {\bibfnamefont {B.}~\bibnamefont {Fefferman}}, \bibinfo {author} {\bibfnamefont {C.}~\bibnamefont {Nirkhe}},\ and\ \bibinfo {author} {\bibfnamefont {U.}~\bibnamefont {Vazirani}},\ }\bibfield  {title} {\bibinfo {title} {On the complexity and verification of quantum random circuit sampling},\ }\href {https://doi.org/10.1038/s41567-018-0318-2} {\bibfield  {journal} {\bibinfo  {journal} {Nature Phys.}\ }\textbf {\bibinfo {volume} {15}},\ \bibinfo {pages} {159} (\bibinfo {year} {2019})}\BibitemShut {NoStop}%
\bibitem [{\citenamefont {Kueng}\ \emph {et~al.}(2017)\citenamefont {Kueng}, \citenamefont {Rauhut},\ and\ \citenamefont {Terstiege}}]{kueng_lowran2017}%
  \BibitemOpen
  \bibfield  {author} {\bibinfo {author} {\bibfnamefont {R.}~\bibnamefont {Kueng}}, \bibinfo {author} {\bibfnamefont {H.}~\bibnamefont {Rauhut}},\ and\ \bibinfo {author} {\bibfnamefont {U.}~\bibnamefont {Terstiege}},\ }\bibfield  {title} {\bibinfo {title} {Low rank matrix recovery from rank one measurements},\ }\href {https://doi.org/10.1016/j.acha.2015.07.007} {\bibfield  {journal} {\bibinfo  {journal} {Appl. Comput. Harmon. Anal.}\ }\textbf {\bibinfo {volume} {42}},\ \bibinfo {pages} {88} (\bibinfo {year} {2017})}\BibitemShut {NoStop}%
\bibitem [{\citenamefont {Kimmel}\ and\ \citenamefont {Liu}(2017)}]{kimmel_pharet2017}%
  \BibitemOpen
  \bibfield  {author} {\bibinfo {author} {\bibfnamefont {S.}~\bibnamefont {Kimmel}}\ and\ \bibinfo {author} {\bibfnamefont {Y.}~\bibnamefont {Liu}},\ }\bibfield  {title} {\bibinfo {title} {Phase retrieval using unitary 2-designs},\ }in\ \href {https://doi.org/10.1109/SAMPTA.2017.8024414} {\emph {\bibinfo {booktitle} {2017 International Conference on Sampling Theory and Applications (SampTA)}}}\ (\bibinfo {year} {2017})\ pp.\ \bibinfo {pages} {345--349}\BibitemShut {NoStop}%
\bibitem [{\citenamefont {Kueng}\ \emph {et~al.}(2016)\citenamefont {Kueng}, \citenamefont {Zhu},\ and\ \citenamefont {Gross}}]{kueng_distinguishing_2016}%
  \BibitemOpen
  \bibfield  {author} {\bibinfo {author} {\bibfnamefont {R.}~\bibnamefont {Kueng}}, \bibinfo {author} {\bibfnamefont {H.}~\bibnamefont {Zhu}},\ and\ \bibinfo {author} {\bibfnamefont {D.}~\bibnamefont {Gross}},\ }\href {https://arxiv.org/abs/1609.08595} {\bibinfo {title} {{Distinguishing quantum states using Clifford orbits}}} (\bibinfo {year} {2016}),\ \Eprint {https://arxiv.org/abs/1609.08595} {arXiv:1609.08595} \BibitemShut {NoStop}%
\bibitem [{\citenamefont {Oszmaniec}\ \emph {et~al.}(2016)\citenamefont {Oszmaniec} \emph {et~al.}}]{oszmaniec_ranbos2016}%
  \BibitemOpen
  \bibfield  {author} {\bibinfo {author} {\bibfnamefont {M.}~\bibnamefont {Oszmaniec}} \emph {et~al.},\ }\bibfield  {title} {\bibinfo {title} {Random bosonic states for robust quantum metrology},\ }\href {https://doi.org/10.1103/PhysRevX.6.041044} {\bibfield  {journal} {\bibinfo  {journal} {Phys. Rev. X}\ }\textbf {\bibinfo {volume} {6}},\ \bibinfo {pages} {041044} (\bibinfo {year} {2016})}\BibitemShut {NoStop}%
\bibitem [{\citenamefont {Eisert}\ \emph {et~al.}(2020)\citenamefont {Eisert}, \citenamefont {Hangleiter}, \citenamefont {Walk}, \citenamefont {Roth}, \citenamefont {Markham}, \citenamefont {Parekh}, \citenamefont {Chabaud},\ and\ \citenamefont {Kashefi}}]{Eisert_2020}%
  \BibitemOpen
  \bibfield  {author} {\bibinfo {author} {\bibfnamefont {J.}~\bibnamefont {Eisert}}, \bibinfo {author} {\bibfnamefont {D.}~\bibnamefont {Hangleiter}}, \bibinfo {author} {\bibfnamefont {N.}~\bibnamefont {Walk}}, \bibinfo {author} {\bibfnamefont {I.}~\bibnamefont {Roth}}, \bibinfo {author} {\bibfnamefont {D.}~\bibnamefont {Markham}}, \bibinfo {author} {\bibfnamefont {R.}~\bibnamefont {Parekh}}, \bibinfo {author} {\bibfnamefont {U.}~\bibnamefont {Chabaud}},\ and\ \bibinfo {author} {\bibfnamefont {E.}~\bibnamefont {Kashefi}},\ }\bibfield  {title} {\bibinfo {title} {Quantum certification and benchmarking},\ }\href {https://doi.org/10.1038/s42254-020-0186-4} {\bibfield  {journal} {\bibinfo  {journal} {Nature Rev. Phys.}\ }\textbf {\bibinfo {volume} {2}},\ \bibinfo {pages} {382–390} (\bibinfo {year} {2020})}\BibitemShut {NoStop}%
\bibitem [{\citenamefont {Dankert}\ \emph {et~al.}(2009)\citenamefont {Dankert}, \citenamefont {Cleve}, \citenamefont {Emerson},\ and\ \citenamefont {Livine}}]{dankert_exact_2009}%
  \BibitemOpen
  \bibfield  {author} {\bibinfo {author} {\bibfnamefont {C.}~\bibnamefont {Dankert}}, \bibinfo {author} {\bibfnamefont {R.}~\bibnamefont {Cleve}}, \bibinfo {author} {\bibfnamefont {J.}~\bibnamefont {Emerson}},\ and\ \bibinfo {author} {\bibfnamefont {E.}~\bibnamefont {Livine}},\ }\bibfield  {title} {\bibinfo {title} {Exact and approximate unitary 2-designs and their application to fidelity estimation},\ }\href {https://doi.org/10.1103/PhysRevA.80.012304} {\bibfield  {journal} {\bibinfo  {journal} {Phys. Rev. A}\ }\textbf {\bibinfo {volume} {80}},\ \bibinfo {pages} {012304} (\bibinfo {year} {2009})}\BibitemShut {NoStop}%
\bibitem [{\citenamefont {Magesan}\ \emph {et~al.}(2011)\citenamefont {Magesan}, \citenamefont {Gambetta},\ and\ \citenamefont {Emerson}}]{Magesan_2011}%
  \BibitemOpen
  \bibfield  {author} {\bibinfo {author} {\bibfnamefont {E.}~\bibnamefont {Magesan}}, \bibinfo {author} {\bibfnamefont {J.~M.}\ \bibnamefont {Gambetta}},\ and\ \bibinfo {author} {\bibfnamefont {J.}~\bibnamefont {Emerson}},\ }\bibfield  {title} {\bibinfo {title} {Scalable and robust randomized benchmarking of quantum processes},\ }\href {https://doi.org/10.1103/physrevlett.106.180504} {\bibfield  {journal} {\bibinfo  {journal} {Phys. Rev. Lett.}\ }\textbf {\bibinfo {volume} {106}},\ \bibinfo {pages} {180504} (\bibinfo {year} {2011})}\BibitemShut {NoStop}%
\bibitem [{\citenamefont {Magni}\ \emph {et~al.}(2025)\citenamefont {Magni}, \citenamefont {Christopoulos}, \citenamefont {Luca},\ and\ \citenamefont {Turkeshi}}]{magniAnticoncentrationCliffordCircuits2025}%
  \BibitemOpen
  \bibfield  {author} {\bibinfo {author} {\bibfnamefont {B.}~\bibnamefont {Magni}}, \bibinfo {author} {\bibfnamefont {A.}~\bibnamefont {Christopoulos}}, \bibinfo {author} {\bibfnamefont {A.~D.}\ \bibnamefont {Luca}},\ and\ \bibinfo {author} {\bibfnamefont {X.}~\bibnamefont {Turkeshi}},\ }\href {https://doi.org/10.48550/arXiv.2502.20455} {\bibinfo {title} {Anticoncentration in {{Clifford circuits}} and {{beyond}}: {{From random tensor networks}} to {{pseudo-magic states}}}} (\bibinfo {year} {2025}),\ \Eprint {https://arxiv.org/abs/2502.20455} {arXiv:2502.20455} \BibitemShut {NoStop}%
\bibitem [{\citenamefont {Devetak}(2005)}]{devetak_thepri2005}%
  \BibitemOpen
  \bibfield  {author} {\bibinfo {author} {\bibfnamefont {I.}~\bibnamefont {Devetak}},\ }\bibfield  {title} {\bibinfo {title} {The private classical capacity and quantum capacity of a quantum channel},\ }\href {https://doi.org/10.1109/TIT.2004.839515} {\bibfield  {journal} {\bibinfo  {journal} {IEEE Trans. Inf. Th.}\ }\textbf {\bibinfo {volume} {51}},\ \bibinfo {pages} {44} (\bibinfo {year} {2005})}\BibitemShut {NoStop}%
\bibitem [{\citenamefont {Devetak}\ and\ \citenamefont {Winter}(2004)}]{devetak_relqua2004}%
  \BibitemOpen
  \bibfield  {author} {\bibinfo {author} {\bibfnamefont {I.}~\bibnamefont {Devetak}}\ and\ \bibinfo {author} {\bibfnamefont {A.}~\bibnamefont {Winter}},\ }\bibfield  {title} {\bibinfo {title} {Relating quantum privacy and quantum coherence: An operational approach},\ }\href {https://doi.org/10.1103/PhysRevLett.93.080501} {\bibfield  {journal} {\bibinfo  {journal} {Phys. Rev. Lett.}\ }\textbf {\bibinfo {volume} {93}},\ \bibinfo {pages} {080501} (\bibinfo {year} {2004})}\BibitemShut {NoStop}%
\bibitem [{\citenamefont {Groisman}\ \emph {et~al.}(2005)\citenamefont {Groisman}, \citenamefont {Popescu},\ and\ \citenamefont {Winter}}]{groisman_quacla2005}%
  \BibitemOpen
  \bibfield  {author} {\bibinfo {author} {\bibfnamefont {B.}~\bibnamefont {Groisman}}, \bibinfo {author} {\bibfnamefont {S.}~\bibnamefont {Popescu}},\ and\ \bibinfo {author} {\bibfnamefont {A.}~\bibnamefont {Winter}},\ }\bibfield  {title} {\bibinfo {title} {Quantum, classical, and total amount of correlations in a quantum state},\ }\href {https://doi.org/10.1103/PhysRevA.72.032317} {\bibfield  {journal} {\bibinfo  {journal} {Phys. Rev. A}\ }\textbf {\bibinfo {volume} {72}},\ \bibinfo {pages} {032317} (\bibinfo {year} {2005})}\BibitemShut {NoStop}%
\bibitem [{\citenamefont {Abeyesinghe}\ \emph {et~al.}(2009)\citenamefont {Abeyesinghe} \emph {et~al.}}]{abeyesinghe_themot2009}%
  \BibitemOpen
  \bibfield  {author} {\bibinfo {author} {\bibfnamefont {A.}~\bibnamefont {Abeyesinghe}} \emph {et~al.},\ }\bibfield  {title} {\bibinfo {title} {The mother of all protocols: Restructuring quantum information’s family tree},\ }\href {https://doi.org/10.1098/rspa.2009.0202} {\bibfield  {journal} {\bibinfo  {journal} {Proc. R. Soc. A}\ }\textbf {\bibinfo {volume} {465}},\ \bibinfo {pages} {2537} (\bibinfo {year} {2009})}\BibitemShut {NoStop}%
\bibitem [{\citenamefont {Dupuis}\ \emph {et~al.}(2014)\citenamefont {Dupuis} \emph {et~al.}}]{dupuis_oneshot2014}%
  \BibitemOpen
  \bibfield  {author} {\bibinfo {author} {\bibfnamefont {F.}~\bibnamefont {Dupuis}} \emph {et~al.},\ }\bibfield  {title} {\bibinfo {title} {One-shot decoupling},\ }\href {https://doi.org/10.1007/s00220-014-1990-4} {\bibfield  {journal} {\bibinfo  {journal} {Commun. Math. Phys.}\ }\textbf {\bibinfo {volume} {328}},\ \bibinfo {pages} {251} (\bibinfo {year} {2014})}\BibitemShut {NoStop}%
\bibitem [{\citenamefont {Szehr}\ \emph {et~al.}(2013)\citenamefont {Szehr} \emph {et~al.}}]{szehr_decuni2013}%
  \BibitemOpen
  \bibfield  {author} {\bibinfo {author} {\bibfnamefont {O.}~\bibnamefont {Szehr}} \emph {et~al.},\ }\bibfield  {title} {\bibinfo {title} {Decoupling with unitary approximate two-designs},\ }\href {https://doi.org/10.1088/1367-2630/15/5/053022} {\bibfield  {journal} {\bibinfo  {journal} {New J. Phys.}\ }\textbf {\bibinfo {volume} {15}},\ \bibinfo {pages} {053022} (\bibinfo {year} {2013})}\BibitemShut {NoStop}%
\bibitem [{\citenamefont {Horodecki}\ \emph {et~al.}(2005)\citenamefont {Horodecki}, \citenamefont {Oppenheim},\ and\ \citenamefont {Winter}}]{horodecki_parqua2005}%
  \BibitemOpen
  \bibfield  {author} {\bibinfo {author} {\bibfnamefont {M.}~\bibnamefont {Horodecki}}, \bibinfo {author} {\bibfnamefont {J.}~\bibnamefont {Oppenheim}},\ and\ \bibinfo {author} {\bibfnamefont {A.}~\bibnamefont {Winter}},\ }\bibfield  {title} {\bibinfo {title} {Partial quantum information},\ }\href {https://doi.org/10.1038/nature03909} {\bibfield  {journal} {\bibinfo  {journal} {Nature}\ }\textbf {\bibinfo {volume} {436}},\ \bibinfo {pages} {673–676} (\bibinfo {year} {2005})}\BibitemShut {NoStop}%
\bibitem [{\citenamefont {Horodecki}\ \emph {et~al.}(2006)\citenamefont {Horodecki}, \citenamefont {Oppenheim},\ and\ \citenamefont {Winter}}]{horodecki_quasta2007}%
  \BibitemOpen
  \bibfield  {author} {\bibinfo {author} {\bibfnamefont {M.}~\bibnamefont {Horodecki}}, \bibinfo {author} {\bibfnamefont {J.}~\bibnamefont {Oppenheim}},\ and\ \bibinfo {author} {\bibfnamefont {A.}~\bibnamefont {Winter}},\ }\bibfield  {title} {\bibinfo {title} {Quantum state merging and negative information},\ }\href {https://doi.org/10.1007/s00220-006-0118-x} {\bibfield  {journal} {\bibinfo  {journal} {Comm. Math. Phys.}\ }\textbf {\bibinfo {volume} {269}},\ \bibinfo {pages} {107–136} (\bibinfo {year} {2006})}\BibitemShut {NoStop}%
\bibitem [{\citenamefont {Nakata}\ \emph {et~al.}(2021)\citenamefont {Nakata}, \citenamefont {Wakakuwa},\ and\ \citenamefont {Yamasaki}}]{nakata_onesho2021}%
  \BibitemOpen
  \bibfield  {author} {\bibinfo {author} {\bibfnamefont {Y.}~\bibnamefont {Nakata}}, \bibinfo {author} {\bibfnamefont {E.}~\bibnamefont {Wakakuwa}},\ and\ \bibinfo {author} {\bibfnamefont {H.}~\bibnamefont {Yamasaki}},\ }\bibfield  {title} {\bibinfo {title} {One-shot quantum error correction of classical and quantum information},\ }\href {https://doi.org/10.1103/PhysRevA.104.012408} {\bibfield  {journal} {\bibinfo  {journal} {Phys. Rev. A}\ }\textbf {\bibinfo {volume} {104}},\ \bibinfo {pages} {012408} (\bibinfo {year} {2021})}\BibitemShut {NoStop}%
\bibitem [{\citenamefont {Wakakuwa}\ and\ \citenamefont {Nakata}(2023)}]{wakakuwa_onesho2023}%
  \BibitemOpen
  \bibfield  {author} {\bibinfo {author} {\bibfnamefont {E.}~\bibnamefont {Wakakuwa}}\ and\ \bibinfo {author} {\bibfnamefont {Y.}~\bibnamefont {Nakata}},\ }\bibfield  {title} {\bibinfo {title} {One-shot triple-resource trade-off in quantum channel coding},\ }\href {https://doi.org/10.1109/TIT.2022.3222775} {\bibfield  {journal} {\bibinfo  {journal} {IEEE Trans. Inf. Th.}\ }\textbf {\bibinfo {volume} {69}},\ \bibinfo {pages} {2400} (\bibinfo {year} {2023})}\BibitemShut {NoStop}%
\bibitem [{\citenamefont {Popescu}\ \emph {et~al.}(2006)\citenamefont {Popescu}, \citenamefont {Short},\ and\ \citenamefont {Winter}}]{popescu_entanglement_2006}%
  \BibitemOpen
  \bibfield  {author} {\bibinfo {author} {\bibfnamefont {S.}~\bibnamefont {Popescu}}, \bibinfo {author} {\bibfnamefont {A.~J.}\ \bibnamefont {Short}},\ and\ \bibinfo {author} {\bibfnamefont {A.}~\bibnamefont {Winter}},\ }\bibfield  {title} {\bibinfo {title} {Entanglement and the foundations of statistical mechanics},\ }\href {https://doi.org/10.1038/nphys444} {\bibfield  {journal} {\bibinfo  {journal} {Nature Phys.}\ }\textbf {\bibinfo {volume} {2}},\ \bibinfo {pages} {754} (\bibinfo {year} {2006})}\BibitemShut {NoStop}%
\bibitem [{\citenamefont {Linden}\ \emph {et~al.}(2009)\citenamefont {Linden}, \citenamefont {Popescu}, \citenamefont {Short},\ and\ \citenamefont {Winter}}]{linden_quantum_2009}%
  \BibitemOpen
  \bibfield  {author} {\bibinfo {author} {\bibfnamefont {N.}~\bibnamefont {Linden}}, \bibinfo {author} {\bibfnamefont {S.}~\bibnamefont {Popescu}}, \bibinfo {author} {\bibfnamefont {A.~J.}\ \bibnamefont {Short}},\ and\ \bibinfo {author} {\bibfnamefont {A.}~\bibnamefont {Winter}},\ }\bibfield  {title} {\bibinfo {title} {Quantum mechanical evolution towards thermal equilibrium},\ }\href {https://doi.org/10.1103/PhysRevE.79.061103} {\bibfield  {journal} {\bibinfo  {journal} {Phys. Rev. E}\ }\textbf {\bibinfo {volume} {79}},\ \bibinfo {pages} {061103} (\bibinfo {year} {2009})}\BibitemShut {NoStop}%
\bibitem [{\citenamefont {del Rio}\ \emph {et~al.}(2016)\citenamefont {del Rio} \emph {et~al.}}]{delrio_relthe2016}%
  \BibitemOpen
  \bibfield  {author} {\bibinfo {author} {\bibfnamefont {L.}~\bibnamefont {del Rio}} \emph {et~al.},\ }\bibfield  {title} {\bibinfo {title} {Relative thermalization},\ }\href {https://doi.org/10.1103/PhysRevE.94.022104} {\bibfield  {journal} {\bibinfo  {journal} {Phys. Rev. E}\ }\textbf {\bibinfo {volume} {94}},\ \bibinfo {pages} {022104} (\bibinfo {year} {2016})}\BibitemShut {NoStop}%
\bibitem [{\citenamefont {Kaneko}\ \emph {et~al.}(2020)\citenamefont {Kaneko} \emph {et~al.}}]{kaneko_chaofm2020}%
  \BibitemOpen
  \bibfield  {author} {\bibinfo {author} {\bibfnamefont {K.}~\bibnamefont {Kaneko}} \emph {et~al.},\ }\bibfield  {title} {\bibinfo {title} {Characterizing complexity of many-body quantum dynamics by higher-order eigenstate thermalization},\ }\href {https://doi.org/10.1103/PhysRevA.101.042126} {\bibfield  {journal} {\bibinfo  {journal} {Phys. Rev. A}\ }\textbf {\bibinfo {volume} {101}},\ \bibinfo {pages} {042126} (\bibinfo {year} {2020})}\BibitemShut {NoStop}%
\bibitem [{\citenamefont {Gogolin}\ and\ \citenamefont {Eisert}(2016)}]{christian_review}%
  \BibitemOpen
  \bibfield  {author} {\bibinfo {author} {\bibfnamefont {C.}~\bibnamefont {Gogolin}}\ and\ \bibinfo {author} {\bibfnamefont {J.}~\bibnamefont {Eisert}},\ }\bibfield  {title} {\bibinfo {title} {Equilibration, thermalisation, and the emergence of statistical mechanics in closed quantum systems},\ }\href {https://doi.org/10.1088/0034-4885/79/5/056001} {\bibfield  {journal} {\bibinfo  {journal} {Rep. Prog. Phys.}\ }\textbf {\bibinfo {volume} {79}},\ \bibinfo {pages} {56001} (\bibinfo {year} {2016})}\BibitemShut {NoStop}%
\bibitem [{\citenamefont {Ippoliti}\ and\ \citenamefont {Ho}(2022)}]{ippoliti_solmod2022}%
  \BibitemOpen
  \bibfield  {author} {\bibinfo {author} {\bibfnamefont {M.}~\bibnamefont {Ippoliti}}\ and\ \bibinfo {author} {\bibfnamefont {W.~W.}\ \bibnamefont {Ho}},\ }\bibfield  {title} {\bibinfo {title} {Solvable model of deep thermalization with distinct design times},\ }\href {https://doi.org/10.22331/q-2022-12-29-886} {\bibfield  {journal} {\bibinfo  {journal} {Quantum}\ }\textbf {\bibinfo {volume} {6}},\ \bibinfo {pages} {886} (\bibinfo {year} {2022})}\BibitemShut {NoStop}%
\bibitem [{\citenamefont {Hayden}\ and\ \citenamefont {Preskill}(2007)}]{hayden_black_2007}%
  \BibitemOpen
  \bibfield  {author} {\bibinfo {author} {\bibfnamefont {P.}~\bibnamefont {Hayden}}\ and\ \bibinfo {author} {\bibfnamefont {J.}~\bibnamefont {Preskill}},\ }\bibfield  {title} {\bibinfo {title} {Black holes as mirrors: {{Quantum}} information in random subsystems},\ }\href {https://doi.org/10.1088/1126-6708/2007/09/120} {\bibfield  {journal} {\bibinfo  {journal} {J. High En. Phys.}\ }\textbf {\bibinfo {volume} {2007}},\ \bibinfo {pages} {120} (\bibinfo {year} {2007})}\BibitemShut {NoStop}%
\bibitem [{\citenamefont {Nakata}\ \emph {et~al.}(2023)\citenamefont {Nakata} \emph {et~al.}}]{nakata_blaasc2023}%
  \BibitemOpen
  \bibfield  {author} {\bibinfo {author} {\bibfnamefont {Y.}~\bibnamefont {Nakata}} \emph {et~al.},\ }\bibfield  {title} {\bibinfo {title} {{Black holes as clouded mirrors: the Hayden-Preskill protocol with symmetry}},\ }\href {https://doi.org/10.22331/q-2023-02-21-928} {\bibfield  {journal} {\bibinfo  {journal} {Quantum}\ }\textbf {\bibinfo {volume} {7}},\ \bibinfo {pages} {928} (\bibinfo {year} {2023})}\BibitemShut {NoStop}%
\bibitem [{\citenamefont {Onorati}\ \emph {et~al.}(2017)\citenamefont {Onorati}, \citenamefont {Buerschaper}, \citenamefont {Kliesch}, \citenamefont {Brown}, \citenamefont {Werner},\ and\ \citenamefont {Eisert}}]{RandomHamiltonians}%
  \BibitemOpen
  \bibfield  {author} {\bibinfo {author} {\bibfnamefont {E.}~\bibnamefont {Onorati}}, \bibinfo {author} {\bibfnamefont {O.}~\bibnamefont {Buerschaper}}, \bibinfo {author} {\bibfnamefont {M.}~\bibnamefont {Kliesch}}, \bibinfo {author} {\bibfnamefont {W.}~\bibnamefont {Brown}}, \bibinfo {author} {\bibfnamefont {A.~H.}\ \bibnamefont {Werner}},\ and\ \bibinfo {author} {\bibfnamefont {J.}~\bibnamefont {Eisert}},\ }\bibfield  {title} {\bibinfo {title} {{Mixing properties of stochastic quantum Hamiltonians}},\ }\href {https://doi.org/10.1007/s00220-017-2950-6} {\bibfield  {journal} {\bibinfo  {journal} {Commun. Math. Phys.}\ }\textbf {\bibinfo {volume} {355}},\ \bibinfo {pages} {905} (\bibinfo {year} {2017})}\BibitemShut {NoStop}%
\bibitem [{\citenamefont {Nakata}\ \emph {et~al.}(2024)\citenamefont {Nakata}, \citenamefont {Takeuchi}, \citenamefont {Kliesch},\ and\ \citenamefont {Darmawan}}]{nakata2025computationalcomplexityunitarystate}%
  \BibitemOpen
  \bibfield  {author} {\bibinfo {author} {\bibfnamefont {Y.}~\bibnamefont {Nakata}}, \bibinfo {author} {\bibfnamefont {Y.}~\bibnamefont {Takeuchi}}, \bibinfo {author} {\bibfnamefont {M.}~\bibnamefont {Kliesch}},\ and\ \bibinfo {author} {\bibfnamefont {A.}~\bibnamefont {Darmawan}},\ }\href {https://arxiv.org/abs/2410.23353} {\bibinfo {title} {On computational complexity of unitary and state design properties}} (\bibinfo {year} {2024}),\ \Eprint {https://arxiv.org/abs/2410.23353} {arXiv:2410.23353} \BibitemShut {NoStop}%
\bibitem [{\citenamefont {Oliviero}\ \emph {et~al.}(2024)\citenamefont {Oliviero}, \citenamefont {Leone}, \citenamefont {Lloyd},\ and\ \citenamefont {Hamma}}]{oliviero_unscrambling_2024}%
  \BibitemOpen
  \bibfield  {author} {\bibinfo {author} {\bibfnamefont {S.~F.~E.}\ \bibnamefont {Oliviero}}, \bibinfo {author} {\bibfnamefont {L.}~\bibnamefont {Leone}}, \bibinfo {author} {\bibfnamefont {S.}~\bibnamefont {Lloyd}},\ and\ \bibinfo {author} {\bibfnamefont {A.}~\bibnamefont {Hamma}},\ }\bibfield  {title} {\bibinfo {title} {Unscrambling {{quantum information}} with {{Clifford decoders}}},\ }\href {https://doi.org/10.1103/PhysRevLett.132.080402} {\bibfield  {journal} {\bibinfo  {journal} {Phys. Rev. Lett.}\ }\textbf {\bibinfo {volume} {132}},\ \bibinfo {pages} {080402} (\bibinfo {year} {2024})}\BibitemShut {NoStop}%
\bibitem [{\citenamefont {Leone}\ \emph {et~al.}(2024)\citenamefont {Leone}, \citenamefont {Oliviero}, \citenamefont {Lloyd},\ and\ \citenamefont {Hamma}}]{leone_learning_2024}%
  \BibitemOpen
  \bibfield  {author} {\bibinfo {author} {\bibfnamefont {L.}~\bibnamefont {Leone}}, \bibinfo {author} {\bibfnamefont {S.~F.~E.}\ \bibnamefont {Oliviero}}, \bibinfo {author} {\bibfnamefont {S.}~\bibnamefont {Lloyd}},\ and\ \bibinfo {author} {\bibfnamefont {A.}~\bibnamefont {Hamma}},\ }\bibfield  {title} {\bibinfo {title} {Learning efficient decoders for quasichaotic quantum scramblers},\ }\href {https://doi.org/10.1103/PhysRevA.109.022429} {\bibfield  {journal} {\bibinfo  {journal} {Phys. Rev. A}\ }\textbf {\bibinfo {volume} {109}},\ \bibinfo {pages} {022429} (\bibinfo {year} {2024})}\BibitemShut {NoStop}%
\bibitem [{\citenamefont {Sekino}\ and\ \citenamefont {Susskind}(2008)}]{sekino_fast_2008}%
  \BibitemOpen
  \bibfield  {author} {\bibinfo {author} {\bibfnamefont {Y.}~\bibnamefont {Sekino}}\ and\ \bibinfo {author} {\bibfnamefont {L.}~\bibnamefont {Susskind}},\ }\bibfield  {title} {\bibinfo {title} {Fast scramblers},\ }\href {https://doi.org/10.1088/1126-6708/2008/10/065} {\bibfield  {journal} {\bibinfo  {journal} {J. High En. Phys.}\ }\textbf {\bibinfo {volume} {2008}},\ \bibinfo {pages} {065} (\bibinfo {year} {2008})}\BibitemShut {NoStop}%
\bibitem [{\citenamefont {Lashkari}\ \emph {et~al.}(2013)\citenamefont {Lashkari}, \citenamefont {Stanford}, \citenamefont {Hastings}, \citenamefont {Osborne},\ and\ \citenamefont {Hayden}}]{lashkari_fast_2013}%
  \BibitemOpen
  \bibfield  {author} {\bibinfo {author} {\bibfnamefont {N.}~\bibnamefont {Lashkari}}, \bibinfo {author} {\bibfnamefont {D.}~\bibnamefont {Stanford}}, \bibinfo {author} {\bibfnamefont {M.}~\bibnamefont {Hastings}}, \bibinfo {author} {\bibfnamefont {T.}~\bibnamefont {Osborne}},\ and\ \bibinfo {author} {\bibfnamefont {P.}~\bibnamefont {Hayden}},\ }\bibfield  {title} {\bibinfo {title} {Towards the fast scrambling conjecture},\ }\href {https://doi.org/10.1007/JHEP04(2013)022} {\bibfield  {journal} {\bibinfo  {journal} {J. High En. Phys.}\ }\textbf {\bibinfo {volume} {2013}},\ \bibinfo {pages} {22} (\bibinfo {year} {2013})}\BibitemShut {NoStop}%
\bibitem [{\citenamefont {Maldacena}\ \emph {et~al.}(2016)\citenamefont {Maldacena}, \citenamefont {Shenker},\ and\ \citenamefont {Stanford}}]{maldacena_bound_2016}%
  \BibitemOpen
  \bibfield  {author} {\bibinfo {author} {\bibfnamefont {J.}~\bibnamefont {Maldacena}}, \bibinfo {author} {\bibfnamefont {S.~H.}\ \bibnamefont {Shenker}},\ and\ \bibinfo {author} {\bibfnamefont {D.}~\bibnamefont {Stanford}},\ }\bibfield  {title} {\bibinfo {title} {A bound on chaos},\ }\href {https://doi.org/10.1007/JHEP08(2016)106} {\bibfield  {journal} {\bibinfo  {journal} {J. High En. Phys.}\ }\textbf {\bibinfo {volume} {2016}},\ \bibinfo {pages} {106} (\bibinfo {year} {2016})}\BibitemShut {NoStop}%
\bibitem [{\citenamefont {Roberts}\ and\ \citenamefont {Yoshida}(2017)}]{roberts_chaos_2017}%
  \BibitemOpen
  \bibfield  {author} {\bibinfo {author} {\bibfnamefont {D.~A.}\ \bibnamefont {Roberts}}\ and\ \bibinfo {author} {\bibfnamefont {B.}~\bibnamefont {Yoshida}},\ }\bibfield  {title} {\bibinfo {title} {Chaos and complexity by design},\ }\href {https://doi.org/10.1007/JHEP04(2017)121} {\bibfield  {journal} {\bibinfo  {journal} {J. High En. Phys.}\ }\textbf {\bibinfo {volume} {2017}},\ \bibinfo {pages} {121} (\bibinfo {year} {2017})}\BibitemShut {NoStop}%
\bibitem [{\citenamefont {Cotler}\ \emph {et~al.}(2023)\citenamefont {Cotler} \emph {et~al.}}]{cotler_emequa2023}%
  \BibitemOpen
  \bibfield  {author} {\bibinfo {author} {\bibfnamefont {J.~S.}\ \bibnamefont {Cotler}} \emph {et~al.},\ }\bibfield  {title} {\bibinfo {title} {Emergent quantum state designs from individual many-body wave functions},\ }\href {https://doi.org/10.1103/PRXQuantum.4.010311} {\bibfield  {journal} {\bibinfo  {journal} {PRX Quantum}\ }\textbf {\bibinfo {volume} {4}},\ \bibinfo {pages} {010311} (\bibinfo {year} {2023})}\BibitemShut {NoStop}%
\bibitem [{\citenamefont {Mark}\ \emph {et~al.}(2024)\citenamefont {Mark} \emph {et~al.}}]{mark_maxent2024}%
  \BibitemOpen
  \bibfield  {author} {\bibinfo {author} {\bibfnamefont {D.~K.}\ \bibnamefont {Mark}} \emph {et~al.},\ }\bibfield  {title} {\bibinfo {title} {{Maximum entropy principle in deep thermalization and in Hilbert-space ergodicity}},\ }\href {https://doi.org/10.1103/PhysRevX.14.041051} {\bibfield  {journal} {\bibinfo  {journal} {Phys. Rev. X}\ }\textbf {\bibinfo {volume} {14}},\ \bibinfo {pages} {041051} (\bibinfo {year} {2024})}\BibitemShut {NoStop}%
\bibitem [{\citenamefont {Pilatowsky-Cameo}\ \emph {et~al.}(2024)\citenamefont {Pilatowsky-Cameo} \emph {et~al.}}]{pilatowsky_hilerg2024}%
  \BibitemOpen
  \bibfield  {author} {\bibinfo {author} {\bibfnamefont {S.}~\bibnamefont {Pilatowsky-Cameo}} \emph {et~al.},\ }\bibfield  {title} {\bibinfo {title} {Hilbert-space ergodicity in driven quantum systems: Obstructions and designs},\ }\href {https://doi.org/10.1103/PhysRevX.14.041059.} {\bibfield  {journal} {\bibinfo  {journal} {Phys. Rev. X}\ }\textbf {\bibinfo {volume} {14}},\ \bibinfo {pages} {041059} (\bibinfo {year} {2024})}\BibitemShut {NoStop}%
\bibitem [{\citenamefont {Munson}\ \emph {et~al.}(2025)\citenamefont {Munson}, \citenamefont {Kothakonda}, \citenamefont {Haferkamp}, \citenamefont {Yunger~Halpern}, \citenamefont {Eisert},\ and\ \citenamefont {Faist}}]{Munson2025Mar}%
  \BibitemOpen
  \bibfield  {author} {\bibinfo {author} {\bibfnamefont {A.}~\bibnamefont {Munson}}, \bibinfo {author} {\bibfnamefont {N.~B.~T.}\ \bibnamefont {Kothakonda}}, \bibinfo {author} {\bibfnamefont {J.}~\bibnamefont {Haferkamp}}, \bibinfo {author} {\bibfnamefont {N.}~\bibnamefont {Yunger~Halpern}}, \bibinfo {author} {\bibfnamefont {J.}~\bibnamefont {Eisert}},\ and\ \bibinfo {author} {\bibfnamefont {P.}~\bibnamefont {Faist}},\ }\bibfield  {title} {\bibinfo {title} {{Complexity-Constrained Quantum Thermodynamics}},\ }\href {https://doi.org/10.1103/PRXQuantum.6.010346} {\bibfield  {journal} {\bibinfo  {journal} {PRX Quantum}\ }\textbf {\bibinfo {volume} {6}},\ \bibinfo {pages} {010346} (\bibinfo {year} {2025})}\BibitemShut {NoStop}%
\bibitem [{\citenamefont {Emerson}\ \emph {et~al.}(2003)\citenamefont {Emerson}, \citenamefont {Weinstein}, \citenamefont {Saraceno}, \citenamefont {Lloyd},\ and\ \citenamefont {Cory}}]{emerson_pseudorandom_2003}%
  \BibitemOpen
  \bibfield  {author} {\bibinfo {author} {\bibfnamefont {J.}~\bibnamefont {Emerson}}, \bibinfo {author} {\bibfnamefont {Y.~S.}\ \bibnamefont {Weinstein}}, \bibinfo {author} {\bibfnamefont {M.}~\bibnamefont {Saraceno}}, \bibinfo {author} {\bibfnamefont {S.}~\bibnamefont {Lloyd}},\ and\ \bibinfo {author} {\bibfnamefont {D.~G.}\ \bibnamefont {Cory}},\ }\bibfield  {title} {\bibinfo {title} {Pseudo-{{random unitary operators}} for {{Quant. Inf. Proc.}}},\ }\href {https://doi.org/10.1126/science.1090790} {\bibfield  {journal} {\bibinfo  {journal} {Science}\ }\textbf {\bibinfo {volume} {302}},\ \bibinfo {pages} {2098} (\bibinfo {year} {2003})}\BibitemShut {NoStop}%
\bibitem [{\citenamefont {Gross}\ \emph {et~al.}(2007)\citenamefont {Gross}, \citenamefont {Audenaert},\ and\ \citenamefont {Eisert}}]{Gross_2007}%
  \BibitemOpen
  \bibfield  {author} {\bibinfo {author} {\bibfnamefont {D.}~\bibnamefont {Gross}}, \bibinfo {author} {\bibfnamefont {K.}~\bibnamefont {Audenaert}},\ and\ \bibinfo {author} {\bibfnamefont {J.}~\bibnamefont {Eisert}},\ }\bibfield  {title} {\bibinfo {title} {Evenly distributed unitaries: On the structure of unitary designs},\ }\href {https://doi.org/10.1063/1.2716992} {\bibfield  {journal} {\bibinfo  {journal} {J. Math. Phys.}\ }\textbf {\bibinfo {volume} {48}},\ \bibinfo {pages} {052104} (\bibinfo {year} {2007})}\BibitemShut {NoStop}%
\bibitem [{\citenamefont {Haferkamp}\ \emph {et~al.}(2023)\citenamefont {Haferkamp}, \citenamefont {{Montealegre-Mora}}, \citenamefont {Heinrich}, \citenamefont {Eisert}, \citenamefont {Gross},\ and\ \citenamefont {Roth}}]{haferkamp2020QuantumHomeopathyWorks}%
  \BibitemOpen
  \bibfield  {author} {\bibinfo {author} {\bibfnamefont {J.}~\bibnamefont {Haferkamp}}, \bibinfo {author} {\bibfnamefont {F.}~\bibnamefont {{Montealegre-Mora}}}, \bibinfo {author} {\bibfnamefont {M.}~\bibnamefont {Heinrich}}, \bibinfo {author} {\bibfnamefont {J.}~\bibnamefont {Eisert}}, \bibinfo {author} {\bibfnamefont {D.}~\bibnamefont {Gross}},\ and\ \bibinfo {author} {\bibfnamefont {I.}~\bibnamefont {Roth}},\ }\bibfield  {title} {\bibinfo {title} {Efficient {{unitary designs}} with a {{system-size independent number}} of {{non-Clifford gates}}},\ }\href {https://doi.org/10.1007/s00220-022-04507-6} {\bibfield  {journal} {\bibinfo  {journal} {Commun. Math. Phys.}\ }\textbf {\bibinfo {volume} {397}},\ \bibinfo {pages} {995} (\bibinfo {year} {2023})}\BibitemShut {NoStop}%
\bibitem [{\citenamefont {Brand{\~a}o}\ \emph {et~al.}(2016)\citenamefont {Brand{\~a}o}, \citenamefont {Harrow},\ and\ \citenamefont {Horodecki}}]{brandao_local_2016}%
  \BibitemOpen
  \bibfield  {author} {\bibinfo {author} {\bibfnamefont {F.~G. S.~L.}\ \bibnamefont {Brand{\~a}o}}, \bibinfo {author} {\bibfnamefont {A.~W.}\ \bibnamefont {Harrow}},\ and\ \bibinfo {author} {\bibfnamefont {M.}~\bibnamefont {Horodecki}},\ }\bibfield  {title} {\bibinfo {title} {Local {{random quantum circuits}} are {{approximate polynomial-designs}}},\ }\href {https://doi.org/10.1007/s00220-016-2706-8} {\bibfield  {journal} {\bibinfo  {journal} {Comm. Math. Phys.}\ }\textbf {\bibinfo {volume} {346}},\ \bibinfo {pages} {397} (\bibinfo {year} {2016})}\BibitemShut {NoStop}%
\bibitem [{\citenamefont {Harrow}\ and\ \citenamefont {Low}(2009)}]{Harrow2009Random2-designs}%
  \BibitemOpen
  \bibfield  {author} {\bibinfo {author} {\bibfnamefont {A.~W.}\ \bibnamefont {Harrow}}\ and\ \bibinfo {author} {\bibfnamefont {R.~A.}\ \bibnamefont {Low}},\ }\bibfield  {title} {\bibinfo {title} {{Random quantum circuits are approximate 2-designs}},\ }\href {https://link.springer.com/article/10.1007/s00220-009-0873-6} {\bibfield  {journal} {\bibinfo  {journal} {Comm. Math. Phys.}\ }\textbf {\bibinfo {volume} {291}},\ \bibinfo {pages} {302} (\bibinfo {year} {2009})}\BibitemShut {NoStop}%
\bibitem [{\citenamefont {Harrow}\ and\ \citenamefont {Mehraban}(2023)}]{harrow_approximate_2018}%
  \BibitemOpen
  \bibfield  {author} {\bibinfo {author} {\bibfnamefont {A.~W.}\ \bibnamefont {Harrow}}\ and\ \bibinfo {author} {\bibfnamefont {S.}~\bibnamefont {Mehraban}},\ }\bibfield  {title} {\bibinfo {title} {Approximate unitary t-designs by short random quantum circuits using nearest-neighbor and long-range gates},\ }\href {https://doi.org/10.1007/s00220-023-04675-z} {\bibfield  {journal} {\bibinfo  {journal} {Comm. Math. Phys.}\ }\textbf {\bibinfo {volume} {401}},\ \bibinfo {pages} {1531–1626} (\bibinfo {year} {2023})}\BibitemShut {NoStop}%
\bibitem [{\citenamefont {LaRacuente}\ and\ \citenamefont {Leditzky}(2024)}]{laracuente2024approximateunitarykdesignsshallow}%
  \BibitemOpen
  \bibfield  {author} {\bibinfo {author} {\bibfnamefont {N.}~\bibnamefont {LaRacuente}}\ and\ \bibinfo {author} {\bibfnamefont {F.}~\bibnamefont {Leditzky}},\ }\href {https://arxiv.org/abs/2407.07876} {\bibinfo {title} {Approximate unitary $k$-designs from shallow, low-communication circuits}} (\bibinfo {year} {2024}),\ \Eprint {https://arxiv.org/abs/2407.07876} {arXiv:2407.07876} \BibitemShut {NoStop}%
\bibitem [{\citenamefont {Ji}\ \emph {et~al.}(2018{\natexlab{a}})\citenamefont {Ji}, \citenamefont {Liu},\ and\ \citenamefont {Song}}]{Ji_2018}%
  \BibitemOpen
  \bibfield  {author} {\bibinfo {author} {\bibfnamefont {Z.}~\bibnamefont {Ji}}, \bibinfo {author} {\bibfnamefont {Y.-K.}\ \bibnamefont {Liu}},\ and\ \bibinfo {author} {\bibfnamefont {F.}~\bibnamefont {Song}},\ }\bibinfo {title} {Pseudorandom quantum states},\ in\ \href {https://doi.org/10.1007/978-3-319-96878-0_5} {\emph {\bibinfo {booktitle} {Advances in Cryptology – CRYPTO 2018}}}\ (\bibinfo  {publisher} {Springer International Publishing},\ \bibinfo {year} {2018})\ p.\ \bibinfo {pages} {126–152}\BibitemShut {NoStop}%
\bibitem [{\citenamefont {Metger}\ \emph {et~al.}(2024)\citenamefont {Metger}, \citenamefont {Poremba}, \citenamefont {Sinha},\ and\ \citenamefont {Yuen}}]{metger2024pseudorandomunitariesnonadaptivesecurity}%
  \BibitemOpen
  \bibfield  {author} {\bibinfo {author} {\bibfnamefont {T.}~\bibnamefont {Metger}}, \bibinfo {author} {\bibfnamefont {A.}~\bibnamefont {Poremba}}, \bibinfo {author} {\bibfnamefont {M.}~\bibnamefont {Sinha}},\ and\ \bibinfo {author} {\bibfnamefont {H.}~\bibnamefont {Yuen}},\ }\href {https://arxiv.org/abs/2402.14803} {\bibinfo {title} {Pseudorandom unitaries with non-adaptive security}} (\bibinfo {year} {2024}),\ \Eprint {https://arxiv.org/abs/2402.14803} {arXiv:2402.14803} \BibitemShut {NoStop}%
\bibitem [{\citenamefont {Schuster}\ \emph {et~al.}(2025)\citenamefont {Schuster}, \citenamefont {Haferkamp},\ and\ \citenamefont {Huang}}]{schuster2025randomunitariesextremelylow}%
  \BibitemOpen
  \bibfield  {author} {\bibinfo {author} {\bibfnamefont {T.}~\bibnamefont {Schuster}}, \bibinfo {author} {\bibfnamefont {J.}~\bibnamefont {Haferkamp}},\ and\ \bibinfo {author} {\bibfnamefont {H.-Y.}\ \bibnamefont {Huang}},\ }\href {https://arxiv.org/abs/2407.07754} {\bibinfo {title} {Random unitaries in extremely low depth}} (\bibinfo {year} {2025}),\ \Eprint {https://arxiv.org/abs/2407.07754} {arXiv:2407.07754} \BibitemShut {NoStop}%
\bibitem [{\citenamefont {Ma}\ and\ \citenamefont {Huang}(2024)}]{ma2024constructrandomunitaries}%
  \BibitemOpen
  \bibfield  {author} {\bibinfo {author} {\bibfnamefont {F.}~\bibnamefont {Ma}}\ and\ \bibinfo {author} {\bibfnamefont {H.-Y.}\ \bibnamefont {Huang}},\ }\href {https://arxiv.org/abs/2410.10116} {\bibinfo {title} {How to construct random unitaries}} (\bibinfo {year} {2024}),\ \Eprint {https://arxiv.org/abs/2410.10116} {arXiv:2410.10116} \BibitemShut {NoStop}%
\bibitem [{\citenamefont {Haug}\ \emph {et~al.}(2024)\citenamefont {Haug}, \citenamefont {Bharti},\ and\ \citenamefont {Koh}}]{haug2024pseudorandomunitariesrealsparse}%
  \BibitemOpen
  \bibfield  {author} {\bibinfo {author} {\bibfnamefont {T.}~\bibnamefont {Haug}}, \bibinfo {author} {\bibfnamefont {K.}~\bibnamefont {Bharti}},\ and\ \bibinfo {author} {\bibfnamefont {D.~E.}\ \bibnamefont {Koh}},\ }\href {https://arxiv.org/abs/2306.11677} {\bibinfo {title} {Pseudorandom unitaries are neither real nor sparse nor noise-robust}} (\bibinfo {year} {2024}),\ \Eprint {https://arxiv.org/abs/2306.11677} {arXiv:2306.11677} \BibitemShut {NoStop}%
\bibitem [{\citenamefont {Eastin}\ and\ \citenamefont {Knill}(2009)}]{Eastin_2009}%
  \BibitemOpen
  \bibfield  {author} {\bibinfo {author} {\bibfnamefont {B.}~\bibnamefont {Eastin}}\ and\ \bibinfo {author} {\bibfnamefont {E.}~\bibnamefont {Knill}},\ }\bibfield  {title} {\bibinfo {title} {Restrictions on transversal encoded quantum gate sets},\ }\href {https://doi.org/10.1103/PhysRevLett.102.110502} {\bibfield  {journal} {\bibinfo  {journal} {Phys. Rev. Lett.}\ }\textbf {\bibinfo {volume} {102}},\ \bibinfo {pages} {110502} (\bibinfo {year} {2009})}\BibitemShut {NoStop}%
\bibitem [{\citenamefont {Gottesman}(1997)}]{gottesman_stabilizer_1997}%
  \BibitemOpen
  \bibfield  {author} {\bibinfo {author} {\bibfnamefont {D.}~\bibnamefont {Gottesman}},\ }\emph {\bibinfo {title} {Stabilizer codes and quantum error correction}},\ \href {https://doi.org/10.48550/arXiv.quant-ph/9705052} {Ph.D. thesis},\ \bibinfo  {school} {California Institute of Technology} (\bibinfo {year} {1997})\BibitemShut {NoStop}%
\bibitem [{\citenamefont {Webb}(2016)}]{webb_clifford_2016}%
  \BibitemOpen
  \bibfield  {author} {\bibinfo {author} {\bibfnamefont {Z.}~\bibnamefont {Webb}},\ }\bibfield  {title} {\bibinfo {title} {The {{Clifford}} group forms a unitary 3-design},\ }\href {https://doi.org/10.26421/QIC16.15-16-8} {\bibfield  {journal} {\bibinfo  {journal} {QIC}\ }\textbf {\bibinfo {volume} {16}},\ \bibinfo {pages} {1379} (\bibinfo {year} {2016})}\BibitemShut {NoStop}%
\bibitem [{\citenamefont {Zhu}(2017)}]{zhu_multiqubit_2017}%
  \BibitemOpen
  \bibfield  {author} {\bibinfo {author} {\bibfnamefont {H.}~\bibnamefont {Zhu}},\ }\bibfield  {title} {\bibinfo {title} {Multiqubit {{Clifford}} groups are unitary 3-designs},\ }\href {https://doi.org/10.1103/PhysRevA.96.062336} {\bibfield  {journal} {\bibinfo  {journal} {Phys. Rev. A}\ }\textbf {\bibinfo {volume} {96}},\ \bibinfo {pages} {062336} (\bibinfo {year} {2017})}\BibitemShut {NoStop}%
\bibitem [{\citenamefont {Scott}(2008)}]{scott_optimizing_2008}%
  \BibitemOpen
  \bibfield  {author} {\bibinfo {author} {\bibfnamefont {A.~J.}\ \bibnamefont {Scott}},\ }\bibfield  {title} {\bibinfo {title} {Optimizing quantum process tomography with unitary 2-designs},\ }\href {https://doi.org/10.1088/1751-8113/41/5/055308} {\bibfield  {journal} {\bibinfo  {journal} {J. Phys. A}\ }\textbf {\bibinfo {volume} {41}},\ \bibinfo {pages} {055308} (\bibinfo {year} {2008})}\BibitemShut {NoStop}%
\bibitem [{\citenamefont {Roth}\ \emph {et~al.}(2018)\citenamefont {Roth}, \citenamefont {Kueng}, \citenamefont {Kimmel}, \citenamefont {Liu}, \citenamefont {Gross}, \citenamefont {Eisert},\ and\ \citenamefont {Kliesch}}]{roth_recovering_2018}%
  \BibitemOpen
  \bibfield  {author} {\bibinfo {author} {\bibfnamefont {I.}~\bibnamefont {Roth}}, \bibinfo {author} {\bibfnamefont {R.}~\bibnamefont {Kueng}}, \bibinfo {author} {\bibfnamefont {S.}~\bibnamefont {Kimmel}}, \bibinfo {author} {\bibfnamefont {Y.-K.}\ \bibnamefont {Liu}}, \bibinfo {author} {\bibfnamefont {D.}~\bibnamefont {Gross}}, \bibinfo {author} {\bibfnamefont {J.}~\bibnamefont {Eisert}},\ and\ \bibinfo {author} {\bibfnamefont {M.}~\bibnamefont {Kliesch}},\ }\bibfield  {title} {\bibinfo {title} {Recovering {{quantum gates}} from {{few average gate fidelities}}},\ }\href {https://doi.org/10.1103/PhysRevLett.121.170502} {\bibfield  {journal} {\bibinfo  {journal} {Phys. Rev. Lett.}\ }\textbf {\bibinfo {volume} {121}},\ \bibinfo {pages} {170502} (\bibinfo {year} {2018})}\BibitemShut {NoStop}%
\bibitem [{\citenamefont {Wallman}\ and\ \citenamefont {Flammia}(2014)}]{wallman_randomized_2014}%
  \BibitemOpen
  \bibfield  {author} {\bibinfo {author} {\bibfnamefont {J.~J.}\ \bibnamefont {Wallman}}\ and\ \bibinfo {author} {\bibfnamefont {S.~T.}\ \bibnamefont {Flammia}},\ }\bibfield  {title} {\bibinfo {title} {Randomized benchmarking with confidence},\ }\href {https://doi.org/10.1088/1367-2630/16/10/103032} {\bibfield  {journal} {\bibinfo  {journal} {New J. Phys.}\ }\textbf {\bibinfo {volume} {16}},\ \bibinfo {pages} {103032} (\bibinfo {year} {2014})}\BibitemShut {NoStop}%
\bibitem [{\citenamefont {Wallman}(2018)}]{wallman_randomized_2018}%
  \BibitemOpen
  \bibfield  {author} {\bibinfo {author} {\bibfnamefont {J.~J.}\ \bibnamefont {Wallman}},\ }\bibfield  {title} {\bibinfo {title} {Randomized benchmarking with gate-dependent noise},\ }\href {https://doi.org/10.22331/q-2018-01-29-47} {\bibfield  {journal} {\bibinfo  {journal} {Quantum}\ }\textbf {\bibinfo {volume} {2}},\ \bibinfo {pages} {47} (\bibinfo {year} {2018})}\BibitemShut {NoStop}%
\bibitem [{\citenamefont {Zhu}\ \emph {et~al.}(2016)\citenamefont {Zhu}, \citenamefont {Kueng}, \citenamefont {Grassl},\ and\ \citenamefont {Gross}}]{zhu2016cliffordgroupfailsgracefully}%
  \BibitemOpen
  \bibfield  {author} {\bibinfo {author} {\bibfnamefont {H.}~\bibnamefont {Zhu}}, \bibinfo {author} {\bibfnamefont {R.}~\bibnamefont {Kueng}}, \bibinfo {author} {\bibfnamefont {M.}~\bibnamefont {Grassl}},\ and\ \bibinfo {author} {\bibfnamefont {D.}~\bibnamefont {Gross}},\ }\href {https://arxiv.org/abs/1609.08172} {\bibinfo {title} {{The Clifford group fails gracefully to be a unitary 4-design}}} (\bibinfo {year} {2016}),\ \Eprint {https://arxiv.org/abs/1609.08172} {arXiv:1609.08172} \BibitemShut {NoStop}%
\bibitem [{\citenamefont {Deshpande}\ \emph {et~al.}(2025)\citenamefont {Deshpande}, \citenamefont {Fefferman}, \citenamefont {Ghosh}, \citenamefont {Gullans},\ and\ \citenamefont {Hangleiter}}]{deshpande2025peakedquantumadvantageusing}%
  \BibitemOpen
  \bibfield  {author} {\bibinfo {author} {\bibfnamefont {A.}~\bibnamefont {Deshpande}}, \bibinfo {author} {\bibfnamefont {B.}~\bibnamefont {Fefferman}}, \bibinfo {author} {\bibfnamefont {S.}~\bibnamefont {Ghosh}}, \bibinfo {author} {\bibfnamefont {M.}~\bibnamefont {Gullans}},\ and\ \bibinfo {author} {\bibfnamefont {D.}~\bibnamefont {Hangleiter}},\ }\href {https://arxiv.org/abs/2510.05262} {\bibinfo {title} {Peaked quantum advantage using error correction}} (\bibinfo {year} {2025}),\ \Eprint {https://arxiv.org/abs/2510.05262} {arXiv:2510.05262 [quant-ph]} \BibitemShut {NoStop}%
\bibitem [{\citenamefont {Zhang}\ \emph {et~al.}(2025)\citenamefont {Zhang}, \citenamefont {Vijay}, \citenamefont {Gu},\ and\ \citenamefont {Bao}}]{zhang2025designsmagicaugmentedcliffordcircuits}%
  \BibitemOpen
  \bibfield  {author} {\bibinfo {author} {\bibfnamefont {Y.}~\bibnamefont {Zhang}}, \bibinfo {author} {\bibfnamefont {S.}~\bibnamefont {Vijay}}, \bibinfo {author} {\bibfnamefont {Y.}~\bibnamefont {Gu}},\ and\ \bibinfo {author} {\bibfnamefont {Y.}~\bibnamefont {Bao}},\ }\href {https://arxiv.org/abs/2507.02828} {\bibinfo {title} {Designs from magic-augmented clifford circuits}} (\bibinfo {year} {2025}),\ \Eprint {https://arxiv.org/abs/2507.02828} {arXiv:2507.02828 [quant-ph]} \BibitemShut {NoStop}%
\bibitem [{\citenamefont {Bittel}\ \emph {et~al.}(2025)\citenamefont {Bittel}, \citenamefont {Eisert}, \citenamefont {Leone}, \citenamefont {Mele},\ and\ \citenamefont {Oliviero}}]{bittel2025completetheorycliffordcommutant}%
  \BibitemOpen
  \bibfield  {author} {\bibinfo {author} {\bibfnamefont {L.}~\bibnamefont {Bittel}}, \bibinfo {author} {\bibfnamefont {J.}~\bibnamefont {Eisert}}, \bibinfo {author} {\bibfnamefont {L.}~\bibnamefont {Leone}}, \bibinfo {author} {\bibfnamefont {A.~A.}\ \bibnamefont {Mele}},\ and\ \bibinfo {author} {\bibfnamefont {S.~F.~E.}\ \bibnamefont {Oliviero}},\ }\href {https://arxiv.org/abs/2504.12263} {\bibinfo {title} {{A complete theory of the Clifford commutant}}} (\bibinfo {year} {2025}),\ \Eprint {https://arxiv.org/abs/2504.12263} {arXiv:2504.12263} \BibitemShut {NoStop}%
\bibitem [{\citenamefont {Haferkamp}(2022)}]{haferkamp_random_2022}%
  \BibitemOpen
  \bibfield  {author} {\bibinfo {author} {\bibfnamefont {J.}~\bibnamefont {Haferkamp}},\ }\bibfield  {title} {\bibinfo {title} {{Random quantum circuits are approximate unitary $t$-designs in depth $O(nt^{5+o(1)})$}},\ }\href {https://doi.org/10.22331/q-2022-09-08-795} {\bibfield  {journal} {\bibinfo  {journal} {Quantum}\ }\textbf {\bibinfo {volume} {6}},\ \bibinfo {pages} {795} (\bibinfo {year} {2022})}\BibitemShut {NoStop}%
\bibitem [{\citenamefont {Jian}\ \emph {et~al.}(2022)\citenamefont {Jian}, \citenamefont {Bentsen},\ and\ \citenamefont {Swingle}}]{jian2022lineargrowthcircuitcomplexity}%
  \BibitemOpen
  \bibfield  {author} {\bibinfo {author} {\bibfnamefont {S.-K.}\ \bibnamefont {Jian}}, \bibinfo {author} {\bibfnamefont {G.}~\bibnamefont {Bentsen}},\ and\ \bibinfo {author} {\bibfnamefont {B.}~\bibnamefont {Swingle}},\ }\href {https://arxiv.org/abs/2206.14205} {\bibinfo {title} {{Linear growth of circuit complexity from Brownian dynamics}}} (\bibinfo {year} {2022}),\ \Eprint {https://arxiv.org/abs/2206.14205} {arXiv:2206.14205} \BibitemShut {NoStop}%
\bibitem [{\citenamefont {Haferkamp}\ \emph {et~al.}(2022)\citenamefont {Haferkamp}, \citenamefont {Faist}, \citenamefont {Kothakonda}, \citenamefont {Eisert},\ and\ \citenamefont {Yunger~Halpern}}]{Haferkamp2022May}%
  \BibitemOpen
  \bibfield  {author} {\bibinfo {author} {\bibfnamefont {J.}~\bibnamefont {Haferkamp}}, \bibinfo {author} {\bibfnamefont {P.}~\bibnamefont {Faist}}, \bibinfo {author} {\bibfnamefont {N.~B.~T.}\ \bibnamefont {Kothakonda}}, \bibinfo {author} {\bibfnamefont {J.}~\bibnamefont {Eisert}},\ and\ \bibinfo {author} {\bibfnamefont {N.}~\bibnamefont {Yunger~Halpern}},\ }\bibfield  {title} {\bibinfo {title} {{Linear growth of quantum circuit complexity}},\ }\href {https://doi.org/10.1038/s41567-022-01539-6} {\bibfield  {journal} {\bibinfo  {journal} {Nat. Phys.}\ }\textbf {\bibinfo {volume} {18}},\ \bibinfo {pages} {528} (\bibinfo {year} {2022})}\BibitemShut {NoStop}%
\bibitem [{\citenamefont {Kitaev}(2014)}]{kitaev_hidden_2014}%
  \BibitemOpen
  \bibfield  {author} {\bibinfo {author} {\bibfnamefont {A.}~\bibnamefont {Kitaev}},\ }\bibfield  {title} {\bibinfo {title} {Hidden correlations in the {{Hawking}} radiation and thermal noise},\ }in\ \href@noop {} {\emph {\bibinfo {booktitle} {Talk given at the {{Fundamental Physics Prize Symposium}}}}},\ Vol.~\bibinfo {volume} {10}\ (\bibinfo {year} {2014})\BibitemShut {NoStop}%
\bibitem [{\citenamefont {Leone}\ \emph {et~al.}(2021{\natexlab{a}})\citenamefont {Leone}, \citenamefont {Oliviero}, \citenamefont {Zhou},\ and\ \citenamefont {Hamma}}]{leone_quantum_2021}%
  \BibitemOpen
  \bibfield  {author} {\bibinfo {author} {\bibfnamefont {L.}~\bibnamefont {Leone}}, \bibinfo {author} {\bibfnamefont {S.~F.~E.}\ \bibnamefont {Oliviero}}, \bibinfo {author} {\bibfnamefont {Y.}~\bibnamefont {Zhou}},\ and\ \bibinfo {author} {\bibfnamefont {A.}~\bibnamefont {Hamma}},\ }\bibfield  {title} {\bibinfo {title} {Quantum {{chaos}} is {{quantum}}},\ }\href {https://doi.org/10.22331/q-2021-05-04-453} {\bibfield  {journal} {\bibinfo  {journal} {Quantum}\ }\textbf {\bibinfo {volume} {5}},\ \bibinfo {pages} {453} (\bibinfo {year} {2021}{\natexlab{a}})}\BibitemShut {NoStop}%
\bibitem [{\citenamefont {Oliviero}\ \emph {et~al.}(2021{\natexlab{a}})\citenamefont {Oliviero}, \citenamefont {Leone},\ and\ \citenamefont {Hamma}}]{oliviero_transitions_2021}%
  \BibitemOpen
  \bibfield  {author} {\bibinfo {author} {\bibfnamefont {S.~F.~E.}\ \bibnamefont {Oliviero}}, \bibinfo {author} {\bibfnamefont {L.}~\bibnamefont {Leone}},\ and\ \bibinfo {author} {\bibfnamefont {A.}~\bibnamefont {Hamma}},\ }\bibfield  {title} {\bibinfo {title} {Transitions in entanglement complexity in random quantum circuits by measurements},\ }\href {https://doi.org/10.1016/j.physleta.2021.127721} {\bibfield  {journal} {\bibinfo  {journal} {Phys. Lett. A}\ }\textbf {\bibinfo {volume} {418}},\ \bibinfo {pages} {127721} (\bibinfo {year} {2021}{\natexlab{a}})}\BibitemShut {NoStop}%
\bibitem [{\citenamefont {Bhatia}(1997)}]{bhatia_matrix_1997}%
  \BibitemOpen
  \bibfield  {author} {\bibinfo {author} {\bibfnamefont {R.}~\bibnamefont {Bhatia}},\ }\href {https://doi.org/10.1007/978-1-4612-0653-8} {\emph {\bibinfo {title} {Matrix {{analysis}}}}},\ Vol.\ \bibinfo {volume} {169}\ (\bibinfo  {publisher} {Springer New York},\ \bibinfo {year} {1997})\BibitemShut {NoStop}%
\bibitem [{\citenamefont {Cotler}\ \emph {et~al.}(2017)\citenamefont {Cotler}, \citenamefont {{Hunter-Jones}}, \citenamefont {Liu},\ and\ \citenamefont {Yoshida}}]{cotler_chaos_2017}%
  \BibitemOpen
  \bibfield  {author} {\bibinfo {author} {\bibfnamefont {J.}~\bibnamefont {Cotler}}, \bibinfo {author} {\bibfnamefont {N.}~\bibnamefont {{Hunter-Jones}}}, \bibinfo {author} {\bibfnamefont {J.}~\bibnamefont {Liu}},\ and\ \bibinfo {author} {\bibfnamefont {B.}~\bibnamefont {Yoshida}},\ }\bibfield  {title} {\bibinfo {title} {Chaos, complexity, and random matrices},\ }\href {https://doi.org/10.1007/JHEP11(2017)048} {\bibfield  {journal} {\bibinfo  {journal} {J. High En. Phys.}\ }\textbf {\bibinfo {volume} {2017}},\ \bibinfo {pages} {48} (\bibinfo {year} {2017})}\BibitemShut {NoStop}%
\bibitem [{\citenamefont {Liu}(2018)}]{liu_spectral_2018}%
  \BibitemOpen
  \bibfield  {author} {\bibinfo {author} {\bibfnamefont {J.}~\bibnamefont {Liu}},\ }\bibfield  {title} {\bibinfo {title} {Spectral form factors and late time quantum chaos},\ }\href {https://doi.org/10.1103/PhysRevD.98.086026} {\bibfield  {journal} {\bibinfo  {journal} {Phys. Rev. D}\ }\textbf {\bibinfo {volume} {98}},\ \bibinfo {pages} {086026} (\bibinfo {year} {2018})}\BibitemShut {NoStop}%
\bibitem [{\citenamefont {Leone}\ \emph {et~al.}(2021{\natexlab{b}})\citenamefont {Leone}, \citenamefont {Oliviero},\ and\ \citenamefont {Hamma}}]{leone_isospectral_2021}%
  \BibitemOpen
  \bibfield  {author} {\bibinfo {author} {\bibfnamefont {L.}~\bibnamefont {Leone}}, \bibinfo {author} {\bibfnamefont {S.~F.~E.}\ \bibnamefont {Oliviero}},\ and\ \bibinfo {author} {\bibfnamefont {A.}~\bibnamefont {Hamma}},\ }\bibfield  {title} {\bibinfo {title} {Isospectral {{twirling}} and {{quantum chaos}}},\ }\href {https://doi.org/10.3390/e23081073} {\bibfield  {journal} {\bibinfo  {journal} {Entropy}\ }\textbf {\bibinfo {volume} {23}},\ \bibinfo {pages} {1073} (\bibinfo {year} {2021}{\natexlab{b}})}\BibitemShut {NoStop}%
\bibitem [{\citenamefont {Oliviero}\ \emph {et~al.}(2021{\natexlab{b}})\citenamefont {Oliviero}, \citenamefont {Leone}, \citenamefont {Caravelli},\ and\ \citenamefont {Hamma}}]{oliviero_random_2021}%
  \BibitemOpen
  \bibfield  {author} {\bibinfo {author} {\bibfnamefont {S.~F.~E.}\ \bibnamefont {Oliviero}}, \bibinfo {author} {\bibfnamefont {L.}~\bibnamefont {Leone}}, \bibinfo {author} {\bibfnamefont {F.}~\bibnamefont {Caravelli}},\ and\ \bibinfo {author} {\bibfnamefont {A.}~\bibnamefont {Hamma}},\ }\bibfield  {title} {\bibinfo {title} {Random {{matrix theory}} of the {{isospectral}} twirling},\ }\href {https://doi.org/10.21468/SciPostPhys.10.3.076} {\bibfield  {journal} {\bibinfo  {journal} {SciPost Phys.}\ }\textbf {\bibinfo {volume} {10}},\ \bibinfo {pages} {76} (\bibinfo {year} {2021}{\natexlab{b}})}\BibitemShut {NoStop}%
\bibitem [{\citenamefont {Gharibyan}\ \emph {et~al.}(2018)\citenamefont {Gharibyan}, \citenamefont {Hanada}, \citenamefont {Shenker},\ and\ \citenamefont {Tezuka}}]{gharibyan_onset_2018}%
  \BibitemOpen
  \bibfield  {author} {\bibinfo {author} {\bibfnamefont {H.}~\bibnamefont {Gharibyan}}, \bibinfo {author} {\bibfnamefont {M.}~\bibnamefont {Hanada}}, \bibinfo {author} {\bibfnamefont {S.~H.}\ \bibnamefont {Shenker}},\ and\ \bibinfo {author} {\bibfnamefont {M.}~\bibnamefont {Tezuka}},\ }\bibfield  {title} {\bibinfo {title} {Onset of random matrix behavior in scrambling systems},\ }\href {https://doi.org/10.1007/JHEP07(2018)124} {\bibfield  {journal} {\bibinfo  {journal} {J. High En. Phys.}\ }\textbf {\bibinfo {volume} {2018}},\ \bibinfo {pages} {124} (\bibinfo {year} {2018})}\BibitemShut {NoStop}%
\bibitem [{\citenamefont {Winer}\ and\ \citenamefont {Swingle}(2023)}]{winer2023reappearancethermalizationdynamicslatetime}%
  \BibitemOpen
  \bibfield  {author} {\bibinfo {author} {\bibfnamefont {M.}~\bibnamefont {Winer}}\ and\ \bibinfo {author} {\bibfnamefont {B.}~\bibnamefont {Swingle}},\ }\href {https://arxiv.org/abs/2307.14415} {\bibinfo {title} {Reappearance of thermalization dynamics in the late-time spectral form factor}} (\bibinfo {year} {2023}),\ \Eprint {https://arxiv.org/abs/2307.14415} {arXiv:2307.14415} \BibitemShut {NoStop}%
\bibitem [{\citenamefont {Keating}\ and\ \citenamefont {Snaith}(2000)}]{Keating2000RandomMT}%
  \BibitemOpen
  \bibfield  {author} {\bibinfo {author} {\bibfnamefont {J.~P.}\ \bibnamefont {Keating}}\ and\ \bibinfo {author} {\bibfnamefont {N.~C.}\ \bibnamefont {Snaith}},\ }\bibfield  {title} {\bibinfo {title} {Random matrix theory and $\zeta$(1/2+it)},\ }\href {https://api.semanticscholar.org/CorpusID:11095649} {\bibfield  {journal} {\bibinfo  {journal} {Comm. Math. Phys.}\ }\textbf {\bibinfo {volume} {214}},\ \bibinfo {pages} {57} (\bibinfo {year} {2000})}\BibitemShut {NoStop}%
\bibitem [{\citenamefont {Imry}(1986)}]{Imry1986ActiveTC}%
  \BibitemOpen
  \bibfield  {author} {\bibinfo {author} {\bibfnamefont {Y.}~\bibnamefont {Imry}},\ }\bibfield  {title} {\bibinfo {title} {Active transmission channels and universal conductance fluctuations},\ }\href {https://api.semanticscholar.org/CorpusID:56328297} {\bibfield  {journal} {\bibinfo  {journal} {Eur. Phys. Lett.}\ }\textbf {\bibinfo {volume} {1}},\ \bibinfo {pages} {249} (\bibinfo {year} {1986})}\BibitemShut {NoStop}%
\bibitem [{\citenamefont {Bravyi}\ and\ \citenamefont {Gosset}(2016)}]{bravyi_improved_2016}%
  \BibitemOpen
  \bibfield  {author} {\bibinfo {author} {\bibfnamefont {S.}~\bibnamefont {Bravyi}}\ and\ \bibinfo {author} {\bibfnamefont {D.}~\bibnamefont {Gosset}},\ }\bibfield  {title} {\bibinfo {title} {Improved {{Classical simulation}} of {{quantum circuits dominated}} by {{Clifford gates}}},\ }\href {https://doi.org/10.1103/PhysRevLett.116.250501} {\bibfield  {journal} {\bibinfo  {journal} {Phys. Rev. Lett.}\ }\textbf {\bibinfo {volume} {116}},\ \bibinfo {pages} {250501} (\bibinfo {year} {2016})}\BibitemShut {NoStop}%
\bibitem [{\citenamefont {Gu}\ \emph {et~al.}(2025)\citenamefont {Gu}, \citenamefont {Oliviero},\ and\ \citenamefont {Leone}}]{guMagic2025}%
  \BibitemOpen
  \bibfield  {author} {\bibinfo {author} {\bibfnamefont {A.}~\bibnamefont {Gu}}, \bibinfo {author} {\bibfnamefont {S.~F.}\ \bibnamefont {Oliviero}},\ and\ \bibinfo {author} {\bibfnamefont {L.}~\bibnamefont {Leone}},\ }\bibfield  {title} {\bibinfo {title} {Magic-induced computational separation in entanglement theory},\ }\href {https://doi.org/10.1103/PRXQuantum.6.020324} {\bibfield  {journal} {\bibinfo  {journal} {PRX Quantum}\ }\textbf {\bibinfo {volume} {6}},\ \bibinfo {pages} {020324} (\bibinfo {year} {2025})}\BibitemShut {NoStop}%
\bibitem [{\citenamefont {Ji}\ \emph {et~al.}(2018{\natexlab{b}})\citenamefont {Ji}, \citenamefont {Liu},\ and\ \citenamefont {Song}}]{ji_pseudorandom_2018}%
  \BibitemOpen
  \bibfield  {author} {\bibinfo {author} {\bibfnamefont {Z.}~\bibnamefont {Ji}}, \bibinfo {author} {\bibfnamefont {Y.-K.}\ \bibnamefont {Liu}},\ and\ \bibinfo {author} {\bibfnamefont {F.}~\bibnamefont {Song}},\ }\bibfield  {title} {\bibinfo {title} {Pseudorandom {{quantum states}}},\ }in\ \href {https://doi.org/10.1007/978-3-319-96878-0_5} {\emph {\bibinfo {booktitle} {Advances in {{Cryptology}} -- {{CRYPTO}} 2018}}},\ \bibinfo {series and number} {Lecture {{Notes}} in {{Computer Science}}},\ \bibinfo {editor} {edited by\ \bibinfo {editor} {\bibfnamefont {H.}~\bibnamefont {Shacham}}\ and\ \bibinfo {editor} {\bibfnamefont {A.}~\bibnamefont {Boldyreva}}}\ (\bibinfo  {publisher} {Springer International Publishing},\ \bibinfo {address} {Cham},\ \bibinfo {year} {2018})\ pp.\ \bibinfo {pages} {126--152}\BibitemShut {NoStop}%
\bibitem [{\citenamefont {Grewal}\ \emph {et~al.}(2023)\citenamefont {Grewal}, \citenamefont {Iyer}, \citenamefont {Kretschmer},\ and\ \citenamefont {Liang}}]{grewal_efficient_2023}%
  \BibitemOpen
  \bibfield  {author} {\bibinfo {author} {\bibfnamefont {S.}~\bibnamefont {Grewal}}, \bibinfo {author} {\bibfnamefont {V.}~\bibnamefont {Iyer}}, \bibinfo {author} {\bibfnamefont {W.}~\bibnamefont {Kretschmer}},\ and\ \bibinfo {author} {\bibfnamefont {D.}~\bibnamefont {Liang}},\ }\href {https://doi.org/10.48550/arXiv.2305.13409} {\bibinfo {title} {Efficient {{learning}} of {{quantum states prepared with few non-Clifford gates}}}} (\bibinfo {year} {2023}),\ \Eprint {https://arxiv.org/abs/2305.13409} {arXiv:2305.13409} \BibitemShut {NoStop}%
\bibitem [{\citenamefont {Skinner}\ \emph {et~al.}(2019)\citenamefont {Skinner}, \citenamefont {Ruhman},\ and\ \citenamefont {Nahum}}]{PhysRevX.9.031009}%
  \BibitemOpen
  \bibfield  {author} {\bibinfo {author} {\bibfnamefont {B.}~\bibnamefont {Skinner}}, \bibinfo {author} {\bibfnamefont {J.}~\bibnamefont {Ruhman}},\ and\ \bibinfo {author} {\bibfnamefont {A.}~\bibnamefont {Nahum}},\ }\bibfield  {title} {\bibinfo {title} {Measurement-induced phase transitions in the dynamics of entanglement},\ }\href {https://doi.org/10.1103/PhysRevX.9.031009} {\bibfield  {journal} {\bibinfo  {journal} {Phys. Rev. X}\ }\textbf {\bibinfo {volume} {9}},\ \bibinfo {pages} {031009} (\bibinfo {year} {2019})}\BibitemShut {NoStop}%
\bibitem [{\citenamefont {Mello}\ \emph {et~al.}(2025)\citenamefont {Mello}, \citenamefont {Santini}, \citenamefont {Lami}, \citenamefont {De~Nardis},\ and\ \citenamefont {Collura}}]{PhysRevLett.134.150403}%
  \BibitemOpen
  \bibfield  {author} {\bibinfo {author} {\bibfnamefont {A.~F.}\ \bibnamefont {Mello}}, \bibinfo {author} {\bibfnamefont {A.}~\bibnamefont {Santini}}, \bibinfo {author} {\bibfnamefont {G.}~\bibnamefont {Lami}}, \bibinfo {author} {\bibfnamefont {J.}~\bibnamefont {De~Nardis}},\ and\ \bibinfo {author} {\bibfnamefont {M.}~\bibnamefont {Collura}},\ }\bibfield  {title} {\bibinfo {title} {Clifford dressed time-dependent variational principle},\ }\href {https://doi.org/10.1103/PhysRevLett.134.150403} {\bibfield  {journal} {\bibinfo  {journal} {Phys. Rev. Lett.}\ }\textbf {\bibinfo {volume} {134}},\ \bibinfo {pages} {150403} (\bibinfo {year} {2025})}\BibitemShut {NoStop}%
\bibitem [{\citenamefont {Weingarten}(1978)}]{weingarten_asymptotic_1978}%
  \BibitemOpen
  \bibfield  {author} {\bibinfo {author} {\bibfnamefont {D.}~\bibnamefont {Weingarten}},\ }\bibfield  {title} {\bibinfo {title} {Asymptotic behavior of group integrals in the limit of infinite rank},\ }\href {https://doi.org/10.1063/1.523807} {\bibfield  {journal} {\bibinfo  {journal} {J. Math. Phys.}\ }\textbf {\bibinfo {volume} {19}},\ \bibinfo {pages} {999} (\bibinfo {year} {1978})}\BibitemShut {NoStop}%
\bibitem [{\citenamefont {Liu}\ \emph {et~al.}(2018)\citenamefont {Liu}, \citenamefont {Lloyd}, \citenamefont {Zhu},\ and\ \citenamefont {Zhu}}]{liu_entanglement_2018}%
  \BibitemOpen
  \bibfield  {author} {\bibinfo {author} {\bibfnamefont {Z.-W.}\ \bibnamefont {Liu}}, \bibinfo {author} {\bibfnamefont {S.}~\bibnamefont {Lloyd}}, \bibinfo {author} {\bibfnamefont {E.}~\bibnamefont {Zhu}},\ and\ \bibinfo {author} {\bibfnamefont {H.}~\bibnamefont {Zhu}},\ }\bibfield  {title} {\bibinfo {title} {Entanglement, quantum randomness, and complexity beyond scrambling},\ }\href {https://doi.org/10.1007/JHEP07(2018)041} {\bibfield  {journal} {\bibinfo  {journal} {J. High En. Phys.}\ }\textbf {\bibinfo {volume} {2018}},\ \bibinfo {pages} {41} (\bibinfo {year} {2018})}\BibitemShut {NoStop}%
\bibitem [{\citenamefont {Low}(2010)}]{low_pseudorandomness_2010}%
  \BibitemOpen
  \bibfield  {author} {\bibinfo {author} {\bibfnamefont {R.~A.}\ \bibnamefont {Low}},\ }\href {https://arxiv.org/abs/1006.5227} {\bibinfo {title} {Pseudo-randomness and learning in quantum computation}} (\bibinfo {year} {2010}),\ \Eprint {https://arxiv.org/abs/1006.5227} {arXiv:1006.5227 [quant-ph]} \BibitemShut {NoStop}%
\bibitem [{\citenamefont {Oszmaniec}\ \emph {et~al.}(2022)\citenamefont {Oszmaniec}, \citenamefont {Sawicki},\ and\ \citenamefont {Horodecki}}]{EpsilonNet}%
  \BibitemOpen
  \bibfield  {author} {\bibinfo {author} {\bibfnamefont {M.}~\bibnamefont {Oszmaniec}}, \bibinfo {author} {\bibfnamefont {A.}~\bibnamefont {Sawicki}},\ and\ \bibinfo {author} {\bibfnamefont {M.}~\bibnamefont {Horodecki}},\ }\bibfield  {title} {\bibinfo {title} {Epsilon-nets, unitary designs and random quantum circuits},\ }\href {https://doi.org/10.1109/TIT.2021.3128110} {\bibfield  {journal} {\bibinfo  {journal} {IEEE Trans. Inf. Th.}\ }\textbf {\bibinfo {volume} {68}},\ \bibinfo {pages} {989} (\bibinfo {year} {2022})}\BibitemShut {NoStop}%
\bibitem [{\citenamefont {Lami}\ \emph {et~al.}(2025)\citenamefont {Lami}, \citenamefont {Haug},\ and\ \citenamefont {De~Nardis}}]{lamiQuantumStateDesigns2025}%
  \BibitemOpen
  \bibfield  {author} {\bibinfo {author} {\bibfnamefont {G.}~\bibnamefont {Lami}}, \bibinfo {author} {\bibfnamefont {T.}~\bibnamefont {Haug}},\ and\ \bibinfo {author} {\bibfnamefont {J.}~\bibnamefont {De~Nardis}},\ }\bibfield  {title} {\bibinfo {title} {Quantum {{state designs}} with {{Clifford-enhanced matrix product states}}},\ }\href {https://doi.org/10.1103/PRXQuantum.6.010345} {\bibfield  {journal} {\bibinfo  {journal} {PRX Quantum}\ }\textbf {\bibinfo {volume} {6}},\ \bibinfo {pages} {010345} (\bibinfo {year} {2025})}\BibitemShut {NoStop}%
\bibitem [{\citenamefont {Gross}\ \emph {et~al.}(2021)\citenamefont {Gross}, \citenamefont {Nezami},\ and\ \citenamefont {Walter}}]{gross_schurweyl_2019}%
  \BibitemOpen
  \bibfield  {author} {\bibinfo {author} {\bibfnamefont {D.}~\bibnamefont {Gross}}, \bibinfo {author} {\bibfnamefont {S.}~\bibnamefont {Nezami}},\ and\ \bibinfo {author} {\bibfnamefont {M.}~\bibnamefont {Walter}},\ }\bibfield  {title} {\bibinfo {title} {{Schur–Weyl duality for the Clifford group with applications: Property testing, a robust Hudson theorem, and de Finetti representations}},\ }\href {https://doi.org/10.1007/s00220-021-04118-7} {\bibfield  {journal} {\bibinfo  {journal} {Comm. Math. Phys.}\ }\textbf {\bibinfo {volume} {385}},\ \bibinfo {pages} {1325–1393} (\bibinfo {year} {2021})}\BibitemShut {NoStop}%
\bibitem [{\citenamefont {Nielsen}\ and\ \citenamefont {Chuang}(2000)}]{nielsen_quantum_2000}%
  \BibitemOpen
  \bibfield  {author} {\bibinfo {author} {\bibfnamefont {M.~A.}\ \bibnamefont {Nielsen}}\ and\ \bibinfo {author} {\bibfnamefont {I.~L.}\ \bibnamefont {Chuang}},\ }\href {https://doi.org/10.1017/CBO9780511976667} {\emph {\bibinfo {title} {Quantum {{computation}} and {{quantum information}}}}}\ (\bibinfo  {publisher} {Cambridge University Press},\ \bibinfo {year} {2000})\BibitemShut {NoStop}%
\bibitem [{\citenamefont {Baker}(2002)}]{Baker2002}%
  \BibitemOpen
  \bibfield  {author} {\bibinfo {author} {\bibfnamefont {A.}~\bibnamefont {Baker}},\ }\bibfield  {title} {\bibinfo {title} {{Lie groups}},\ }in\ \href {https://doi.org/10.1007/978-1-4471-0183-3_7} {\emph {\bibinfo {booktitle} {{Matrix groups}}}}\ (\bibinfo  {publisher} {Springer},\ \bibinfo {address} {London, England, UK},\ \bibinfo {year} {2002})\ pp.\ \bibinfo {pages} {181--209}\BibitemShut {NoStop}%
\bibitem [{\citenamefont {Grewal}\ \emph {et~al.}(2024)\citenamefont {Grewal}, \citenamefont {Iyer}, \citenamefont {Kretschmer},\ and\ \citenamefont {Liang}}]{grewal2024pseudoentanglementaintcheap}%
  \BibitemOpen
  \bibfield  {author} {\bibinfo {author} {\bibfnamefont {S.}~\bibnamefont {Grewal}}, \bibinfo {author} {\bibfnamefont {V.}~\bibnamefont {Iyer}}, \bibinfo {author} {\bibfnamefont {W.}~\bibnamefont {Kretschmer}},\ and\ \bibinfo {author} {\bibfnamefont {D.}~\bibnamefont {Liang}},\ }\href {https://arxiv.org/abs/2404.00126} {\bibinfo {title} {Pseudoentanglement ain't cheap}} (\bibinfo {year} {2024}),\ \Eprint {https://arxiv.org/abs/2404.00126} {arXiv:2404.00126 [quant-ph]} \BibitemShut {NoStop}%
\bibitem [{\citenamefont {Leone}\ \emph {et~al.}(2022)\citenamefont {Leone}, \citenamefont {Oliviero}, \citenamefont {Lloyd},\ and\ \citenamefont {Hamma}}]{leone_learning_2022}%
  \BibitemOpen
  \bibfield  {author} {\bibinfo {author} {\bibfnamefont {L.}~\bibnamefont {Leone}}, \bibinfo {author} {\bibfnamefont {S.~F.~E.}\ \bibnamefont {Oliviero}}, \bibinfo {author} {\bibfnamefont {S.}~\bibnamefont {Lloyd}},\ and\ \bibinfo {author} {\bibfnamefont {A.}~\bibnamefont {Hamma}},\ }\href {https://doi.org/10.48550/arXiv.2212.11338} {\bibinfo {title} {Learning efficient decoders for quasi-chaotic quantum scramblers}} (\bibinfo {year} {2022}),\ \Eprint {https://arxiv.org/abs/2212.11338} {arXiv:2212.11338} \BibitemShut {NoStop}%
\bibitem [{\citenamefont {Szarek}(1982)}]{szarek1982nets}%
  \BibitemOpen
  \bibfield  {author} {\bibinfo {author} {\bibfnamefont {S.~J.}\ \bibnamefont {Szarek}},\ }\bibfield  {title} {\bibinfo {title} {{Nets of Grassmann manifold and orthogonal groups}},\ }in\ \href@noop {} {\emph {\bibinfo {booktitle} {Proceedings of Research Workshop on Banach Space Theory}}}\ (\bibinfo  {publisher} {University of Iowa Press},\ \bibinfo {address} {Iowa City, Iowa},\ \bibinfo {year} {1982})\ pp.\ \bibinfo {pages} {169--185}\BibitemShut {NoStop}%
\bibitem [{\citenamefont {Zhao}\ \emph {et~al.}(2024)\citenamefont {Zhao}, \citenamefont {Lewis}, \citenamefont {Kannan}, \citenamefont {Quek}, \citenamefont {Huang},\ and\ \citenamefont {Caro}}]{zhao_learning_2023}%
  \BibitemOpen
  \bibfield  {author} {\bibinfo {author} {\bibfnamefont {H.}~\bibnamefont {Zhao}}, \bibinfo {author} {\bibfnamefont {L.}~\bibnamefont {Lewis}}, \bibinfo {author} {\bibfnamefont {I.}~\bibnamefont {Kannan}}, \bibinfo {author} {\bibfnamefont {Y.}~\bibnamefont {Quek}}, \bibinfo {author} {\bibfnamefont {H.-Y.}\ \bibnamefont {Huang}},\ and\ \bibinfo {author} {\bibfnamefont {M.~C.}\ \bibnamefont {Caro}},\ }\bibfield  {title} {\bibinfo {title} {{Learning Quantum States and Unitaries of Bounded Gate Complexity}},\ }\href {https://doi.org/10.1103/PRXQuantum.5.040306} {\bibfield  {journal} {\bibinfo  {journal} {PRX Quantum}\ }\textbf {\bibinfo {volume} {5}},\ \bibinfo {pages} {040306} (\bibinfo {year} {2024})}\BibitemShut {NoStop}%
\bibitem [{\citenamefont {Leone}(2023)}]{leone_clifford_2023}%
  \BibitemOpen
  \bibfield  {author} {\bibinfo {author} {\bibfnamefont {L.}~\bibnamefont {Leone}},\ }\emph {\bibinfo {title} {Clifford Group and beyond: {{Theory}} and Applications in Quantum Information}},\ \href@noop {} {Ph.D. thesis},\ \bibinfo  {school} {University of Massachusetts Boston} (\bibinfo {year} {2023})\BibitemShut {NoStop}%
\end{thebibliography}
\end{document}